\documentclass[10pt,aps,pra,twocolumn,groupedaddress,reprint,amsmath,amssymb,showkeys,superscriptaddress]{revtex4-2}
\usepackage{bookmark}
\usepackage[T1]{fontenc}     
\usepackage[utf8]{inputenc}  
\usepackage{lmodern}         
\usepackage{graphicx}
\usepackage{mathtools}
\usepackage{physics}
\usepackage{comment}
\usepackage{hyperref}
\hypersetup{
    colorlinks,
    linkcolor={red!50!black},
    citecolor={blue!50!black},
    urlcolor={blue!80!black}
}

\newcommand{\be}{{\varepsilon_{0}}}

\newcommand{\mB}{{m_\mathrm{B}}}
\newcommand{\mF}{{m_\mathrm{F}}}
\newcommand{\nB}{{n_\mathrm{B}}}
\newcommand{\nF}{{n_\mathrm{F}}}
\newcommand{\NB}{{N_\mathrm{B}}}
\newcommand{\NF}{{N_\mathrm{F}}}

\newcommand{\muB}{{\mu_\mathrm{B}}}
\newcommand{\muF}{{\mu_\mathrm{F}}}

\newcommand{\eF}{{\epsilon_\mathrm{F}}}
\newcommand{\kF}{{k_\mathrm{F}}}
\newcommand{\til}{~}

\IfFileExists{newtxtext.sty}
    {\usepackage{newtxtext,newtxmath}}
    {\IfFileExists{stix.sty}
       {\usepackage{stix}}
       {\IfFileExists{mathptmx.sty}
       {\usepackage{mathptmx}}{} } }

\begin{document}

\title{Stability of Bose-Fermi mixtures in two dimensions: a lowest-order constrained variational approach}

\author{Pietro Cordioli}
\affiliation{Dipartimento di Fisica e Astronomia “Augusto Righi”, Universit\`a di Bologna, Via Irnerio 46, I-40126, Bologna, Italy} 
\author{Leonardo Pisani}%
\email[]{leonardo.pisani2@unibo.it}
\affiliation{Dipartimento di Fisica e Astronomia “Augusto Righi”, Universit\`a di Bologna, Via Irnerio 46, I-40126, Bologna, Italy} 
\affiliation{INFN, Sezione di Bologna, Viale Berti Pichat 6/2, I-40127, Bologna, Italy}
\author{Pierbiagio Pieri}%
\email[]{pierbiagio.pieri@unibo.it}
\affiliation{Dipartimento di Fisica e Astronomia “Augusto Righi”, Universit\`a di Bologna, Via Irnerio 46, I-40126, Bologna, Italy} 
\affiliation{INFN, Sezione di Bologna, Viale Berti Pichat 6/2, I-40127, Bologna, Italy}

\date{\today}

\begin{abstract}
We investigate the problem of mechanical stability in two-dimensional Bose–Fermi mixtures at zero temperature, focusing on systems with a tunable Bose–Fermi (BF) interaction and a weak but finite boson–boson (BB) repulsion. The analysis is carried out within the framework of the lowest-order constrained variational (LOCV) approach, which allows for a non-perturbative treatment of strong interspecies correlations while retaining analytical transparency. The BF interaction is modeled by a properly regularized attractive contact potential, enabling the exploration of both the attractive and repulsive energy branches.
We determine the minimal BB repulsion required to ensure mechanical stability of the mixture by evaluating the inverse compressibility matrix over the full range of BF coupling strengths, within the domain of validity of the LOCV approximation. The interaction contribution to the energy is benchmarked against available experimental data and Quantum Monte Carlo results in the single-impurity limit, showing good agreement.
Our analysis reveals how the critical BB coupling depends on interaction strength, density imbalance, and mass ratio. In particular, we find that mixtures with equal boson and fermion masses exhibit enhanced stability, requiring the smallest BB repulsion to prevent mechanical instability. In this case, a relatively small BB interaction is sufficient to stabilize attractive mixtures for all values of the BF interaction. These results provide a theoretical framework for assessing stability conditions in experimentally realizable two-dimensional Bose–Fermi mixtures with tunable interactions.
\end{abstract}
\maketitle

\section{Introduction}

Bose-Fermi (BF) mixtures constitute a paradigmatic platform for studying correlated quantum matter composed of particles with different quantum statistics. Since the early investigations of dilute $^3$He-$^4$He mixtures\til\cite{BBP-1967,Bardeen-1966,Wheatly-1968,Edwards-1971,Cohen-1977,krotscheck-1993,Boronat-1989,Boronat-1994}, such systems have provided insight into mediated interactions \cite{Wheatly-1968}, impurity physics \cite{Krotscheck-1988},  phase separation \cite{Reppy-1967,Edwards-1961},  $p$-wave pairing \cite{Edwards-1971,Kagan-2013}, dual superfluids \cite{Kagan-2013}, and effects beyond the mere remit of condensed matter physics\til\cite{Iachello-1991,Maeda-2009,Stewart-2017,Tajima-2024}. 

In the context of ultracold atomic gases\til\cite{Bloch-2008}, the advent of Feshbach resonances\til\cite{Grimm-2010} has enabled unprecedented control over interspecies interactions, allowing the realization of BF mixtures with tunable coupling strengths in a variety of atomic combinations\til\cite{Zaccanti-2024}.
Several studies have appeared in both the experimental\til\cite{Jin-2008,Zwierlein-2011,Zwierlein-2012,Zwierlein-2012a,Ketterle-2012,Jin-2013,Jin-2013a,Porto-2015,Salomon-2014,Salomon-2015,Jin-2016,Pan-2017,Gupta-2017,DeSalvo-2017,Ferlaino-2018,Grimm-2018,DeSalVo-2019,Valtolina-2019,Zwierlein-2020,Khoon-2020,Ozeri-2020,Fritsche-2021,Schindewolf-2022,Bloch-2022,Duda-2023,Patel-2023,Baroni-2024,Yan-2024,Chuang-2024} and theoretical \cite{Sachdev-2005,Stoof-2005,Schuck-2005,Avdeenkov-2006,Pollet-2006,Rothel-2007,Barillier-2008,Pollet-2008,Bortolotti-2008,Parish-2008,Watanabe-2008,Fratini-2010,Zhai-2011,Song-2011,Ludwig-2011,Fratini-2012,Anders-2012,Bertaina-2013,Fratini-2013,Sogo-2013,Guidini-2014,Guidini-2015,Kharga-2017,Ohashi-2019,Manabe-2020,Ohashi-2021,Schmidt-2022,Tajima-2024,Bruun-2024,Tajima-2024b,Pisani-2025,Gualerzi-2025,Foster-2026} literature on BF mixture with tunable interactions.

A central issue in the realization of quantum mixtures is their mechanical stability\til\cite{Modugno-2002,Lous-2018}. Unlike purely fermionic systems, Bose–Fermi mixtures are not intrinsically stabilized by Pauli pressure alone. 
Determining the minimal boson-boson (BB) interaction required to stabilize the mixture is therefore a fundamental problem, especially when the interspecies interaction is tuned across a Feshbach resonance.

Experimental\til\cite{Modugno-2002,Lous-2018} and theoretical\til\cite{Molmer-1998,Viverit-2000,Yi-2001,Feldmeier-2002,Viverit-2002,Capuzzi-2003,Chui-2004,Yabu-2014,DAlberto-2024} efforts in this direction have concentrated mainly on  repulsive or weakly-attractive interspecies interaction. 
In three dimensions, the stability of resonant BF mixtures has been studied using mean-field theory for narrow resonances \cite{Parish-2008} possibly with related Gaussian corrections \cite{Sachdev-2005}, variational methods \cite{Zhai-2011,Foster-2026}, and $T$-matrix approximations \cite{Manabe-2020,Ohashi-2021,Gualerzi-2025}. These works have clarified the interplay between pairing, molecule formation, and phase stability across the resonance. 

However, the situation in two dimensions is qualitatively different. Two-dimensional scattering is characterized by a logarithmic energy dependence and the existence of a two-body bound state for arbitrarily weak attraction\til\cite{Averbuch-1986,Hammer-2010,Bloom-1975}. 
Moreover, tight confinement enables confinement-induced resonances \cite{Olshanii-1998,Haller-2010}, providing an additional experimental knob for controlling interactions in quasi-2D geometries. This situation is particularly suitable for a BF mixture with a tunable BF interaction since an additional and independent BB interaction is required to guarantee its mechanical stability \cite{Zhai-2011,Ohashi-2021,Gualerzi-2025}.

Despite increasing experimental and theoretical interest in low-dimensional mixtures, a systematic investigation of the mechanical stability of a two-dimensional BF mixture with a tunable interspecies interaction is still lacking. Existing theoretical studies of 2D BF systems have focused primarily on pairing, polarons, or collective properties \cite{Mathey-2006,Subasi-2010,Noda-2011,Schmidt-2022,Pisani-2025} and the problem of stability has been addressed only in restricted weak-coupling regimes \cite{DAlberto-2024} or for dilute $^3$He-$^4$He mixtures~\cite{Krotscheck-2000,Krotscheck-2001}. In particular, a nonperturbative analysis of stability across the full interaction crossover in two dimensions is currently missing.

In this work, we fill this gap by analyzing the mechanical stability of a homogeneous two-dimensional Bose–Fermi mixture at zero temperature with a tunable BF interaction and a weak BB repulsion.   We focus on mixtures with a majority of fermions, in line with previous studies for resonant Bose-Fermi mixtures in 3D \cite{Zhai-2011,Guidini-2015,Pisani-2025}.
To treat the interspecies interaction nonperturbatively, we employ the lowest-order constrained variational (LOCV) approach\til\cite{Pandharipande-1971,Pandharipande2-1971,Pandharipande-1973,Pethick-2002,Heiselberg-2011,Zhai-2011,Taylor-2011,Grochowski-2020, Gawryluk-2024}, which incorporates short-range correlations at the two-body level  and naturally connects to the polaron problem in the single-impurity limit. 
The variational nature of the LOCV approach allows us to describe both the attractive (lower) and repulsive (upper) branches of an attractive contact interaction modeling the effective 2D scattering length resulting after confinement to two dimensions from a Feshbach or confinement-induced resonance. These two branches effectively correspond to attractive or repulsive BF interactions, respectively \cite{Massignan-2014}. 

By solving the LOCV equations across the full range of coupling strengths and evaluating the inverse compressibility matrix, we determine the minimal boson–boson repulsion required to stabilize the homogeneous mixture against thermodynamic instability. We analyze the dependence of this critical BB interaction on the interspecies coupling, density ratio, and mass imbalance, and discuss the regime of validity of the approach, particularly in the molecular limit where higher-order correlations become increasingly important.

The paper is organized as follows. 
In Sec.\til\ref{sec:LOCVtheory} we present the LOCV formalism for the two-dimensional Bose–Fermi mixture. 
In Sec.\til\ref{sec:locvequsol} we solve the LOCV equation and obtain the  BF pair correlation function for the attractive and repulsive branches. 
In Sec.\til\ref{sec:attrepbran} we compute the corresponding energy branches and benchmark our results against available experimental and Monte Carlo data.
In Sec.\til\ref{sec:mechstab} we analyze the mechanical stability of the mixture and determine the critical BB interaction required for stability, discussing the dependence on varying concentration and mass ratio.
Finally, in Sec.\til\ref{sec:concl} we draw our conclusions and outline possible extensions.

\section{The Locv Approximation}
\label{sec:LOCVtheory}

We consider a homogeneous atomic mixture of $\NB$  bosons with mass $\mB$ and $\NF$  fermions with mass $\mF$ $(\NB\leq \NF)$, within a square of area  $A$ at zero temperature. 
We first consider the Hamiltonian in the presence of interaction between bosons (B) and fermions (F) only:
\begin{equation}
\label{equ:HBF}
   H_{\rm BF}=\sum_{i=1}^{N_{\rm B}} -\frac{\nabla_i^2}{2m_{\rm B}} +
    \sum_{j=1}^{N_{\rm F}} -\frac{\nabla_j^2}{2m_{\rm F}}  + \sum_{i=1}^{N_{\rm B}} \sum_{j=1}^{N_{\rm F}} U_\mathrm{BF}(\vec{r}_i-\vec{R}_j) 
\end{equation}
(we set $\hbar=1$).
We assume that the interaction potential \( U_\mathrm{BF}(r) \) is short-ranged, central, and spherically symmetric.
To describe the ground-state of the above Hamiltonian, we make use of a variational wave-function  of the Jastrow-Slater (JS) form\til\cite{Jastrow-1955} 
\begin{equation}
    \label{PSI}
\Psi(\vec{\mathbf{r}},\vec{\mathbf{R}})=\prod_{i,j}f(|\vec{r}_i-\vec{R}_j|)\Phi_\mathrm{B}(\vec{\mathbf{r}})\Phi_\mathrm{F}(\vec{\mathbf{R}}),
\end{equation}
where \( \vec{\mathbf{r}} = (\vec{r}_1, \ldots, \vec{r}_{N_{\rm B}}) \) are the bosonic coordinates and \( \vec{\mathbf{R}} = (\vec{R}_1, \ldots, \vec{R}_{N_{\rm F}}) \) the fermionic ones. 
Here
\begin{equation}
    \Phi_\mathrm{B}(\vec{r}_1,\ldots,\vec{r}_{\NB})=\bigg(\frac{1}{A}\bigg)^{\frac{\NB}{2}}
    \label{phiB}
\end{equation}
is the wave function of $N_{\rm B}$ bosons condensed in the zero-momentum state, while 
\begin{equation}
    \Phi_\mathrm{F} (\vec{R}_1,\ldots,\vec{R}_{\NF}) = \bigg(\frac{1}{A}\bigg)^{\frac{\NF}{2}} \lVert e^{i\vec{k}_n\cdot \vec{R}_m} \rVert,
\end{equation}
with $n,m=1,\ldots,\NF$.
is a Slater determinant $\lVert...\rVert$ of the lowest $N_{\rm F}$ normalized plane-wave states within the square of area $A$. The function $f(|\vec{r}_i-\vec{R}_j|)$ introduces short-range correlations between the bosonic and fermionic components.

Within the Jastrow variational approach\til\cite{Jastrow-1955}, the total energy density $\mathcal{E}_\mathrm{BF}$ is given by (\ref{equ:HBF})
\begin{align}
 \label{equ:fullexpval}
          \mathcal{E}_\mathrm{BF}&\equiv\frac{1}{A} \frac{\bra{\Psi} H_\mathrm{BF} \ket{\Psi}}{\braket{\Psi}{\Psi}} ,
\end{align}
with $\Psi$ determined by the ansatz (\ref{PSI}). The above expectation value  can be recast as an expansion in terms of the number of correlated particles (in full analogy with the classical cluster development of Ursell and Mayer\til\cite{deBoer-1949}). The LOCV scheme retains only the lowest non-trivial order of the above expansion. That is, only direct two-particle correlations are retained.
With this approximation, one can show that the total energy density of the Bose-Fermi mixture in the thermodynamic limit $(\NF,\NB,A \rightarrow \infty)$ acquires the form\til\cite{Chang-2006,Zhai-2011, Cordioli-2025}
\begin{align}
 \label{equ:2DBFexpval}
          \mathcal{E}_\mathrm{BF} &\simeq\hspace{1mm}\nF E_\mathrm{FG} + \nonumber \\
          & \nB \nF\int d^2r f(r)\left[-\frac{\nabla^2 }{2m_r}+ U_\mathrm{BF}(r)\right]f(r),
\end{align}
with $\nF=\NF/A$, $\nB=\NB/A$ and $E_{\rm FG}$ the total energy of  non-interacting spinless Fermi gas. In two dimensions, $E_{\rm FG}=\eF/2$, where $ \eF = k_\mathrm{F}^2 / (2\mF)$ is the Fermi energy and $k_{\rm F}=(4\pi \nF)^{1/2}$ is the Fermi wave vector.

By minimizing the energy functional in Eq.\til(\ref{equ:2DBFexpval}) with respect to $f(r)$, one obtains the Euler-Lagrange equation for the stationary minimum of the variational problem at hand. 
The existence of this minimum is not guaranteed a priori if the energy expectation value \til(\ref{equ:fullexpval}) is evaluated by a truncated cluster expansion\til\cite{Emery-1958}. As argued in Ref.\til\cite{Emery-1958,Clark-1972,Krotscheck-1977}, a subsidiary condition on the form of $f(r)$ must be introduced to guarantee the existence of a physically sound solution of the Euler-Lagrange  equation. 

Since $f(r)$ is expected to reach the uncorrelated limit $f(r)\rightarrow 1$ for large $r$, an effective ``small" parameter for the cluster expansion is provided by the expression\til\cite{Krotscheck-1977}
\begin{equation}
\label{equ:sp}
    \varepsilon\equiv \nF \int d^2r \, \left[ f^2(r) -1 \right].
\end{equation}
To justify a truncation of the cluster series
and ensure the convergence of its  summation, the parameter\til(\ref{equ:sp}) must remain finite, that is, the correlation function $f(r)$ should approach one sufficiently fast as $r$ increases.

In the LOCV scheme, this ``healing constraint" on $f(r)$ is enforced by requiring that, on average, a circle of radius $d$ (with $d$ the \textit{healing distance}) around any given boson may contain only one correlated fermion.
If $g(r)$ is the BF pair distribution function, the probability of finding a fermion at distance $r$ from a boson is $\nF\, g(r) d^2r$. As $g(r) \approx f^2(r)$ at the lowest cluster order\til\cite{Jastrow-1955},
the above constraint translates into the following expression
\begin{equation}
\label{locvconstraint}
    \nF\int_{r<d} d^2r f(r)^2 = 1.
\end{equation}
The energy functional\til(\ref{equ:2DBFexpval}) is thus
minimized subject to the constraint (\ref{locvconstraint}) by the introduction of a Lagrange multiplier ($\lambda'\equiv\lambda \nB$ for convenience). In this way, one obtains the LOCV equation 
\begin{equation}
\label{equ:2DLOCVeq}
        \left[-\frac{\nabla_r^2 }{2m_r}+ U_{\rm BF}(r)\right]f(r)=\lambda f(r),
\end{equation}
with $m_r=m_{\rm B} m_{\rm F}/(m_{\rm B}+m_{\rm F})$ the reduced mass. This is effectively the Schr\"odinger equation for the relative motion of the two-body problem in vacuum \cite{Boronat-2005,Whitehead-2016}, where $\lambda$ can be interpreted as the effective interaction energy per BF pair. Note that in Eqs.~\eqref{equ:sp} and \eqref{locvconstraint}
we have explicitly assumed $n_{\rm F}\geq n_{\rm B}$.
At distances greater than the healing distance $d$, the two-particle correlations are assumed to be absent and the correlation function must smoothly match the uncorrelated limit $f(r)\rightarrow 1$. These two properties are enforced by the two boundary conditions at $ r = d $, respectively
\begin{subequations} 
    \begin{align}
        f(r= d) &= 1, \label{boundaryconda} \\
        \frac{d f}{d r}\bigg|_{r=d} &= 0.                    \label{boundarycondb}
     \end{align}
\end{subequations}

Therefore the many-body problem of evaluating the total energy\til(\ref{equ:HBF}) is reduced to the effective one-body problem
\til(\ref{equ:2DLOCVeq}) with  boundary conditions (\ref{boundaryconda})  and (\ref{boundarycondb}).
In the next section we will obtain the solution of the LOCV Eq.\til(\ref{equ:2DLOCVeq})  for a specific choice of the pair potential $U_\mathrm{BF}(r)$.
We will be interested in the ground state and first excited state solutions of the LOCV equation\til\eqref{equ:2DLOCVeq} for an attractive contact potential and we will find that, in the limit $\nB \rightarrow 0$, these two solutions reduce to the lower (attractive) and  upper (repulsive) energy branches of the associated polaron problem, respectively.

We finally remark that the solution of Eq.~\eqref{equ:2DLOCVeq} with the associated boundary conditions does not depend on whether one is dealing with BF or FF mixtures (while the total energy depends on it). For this reason, the results for $f(r)$ or $\lambda$ that we will discuss below, can be compared with corresponding results for FF mixtures, when available.

\section{Solution of the LOCV equation}
\label{sec:locvequsol}

The BF mixture under study is assumed to be dilute such that the
range of all interactions is smaller than the average inter-particle distance. 
This justifies the adoption of an effective pseudo-potential of the point-contact form, which can be regularized by the introduction of the $s$-wave scattering length $a_\mathrm{BF}$ through appropriate Bethe-Peierls conditions on the wave-function \cite{Bethe-1935}.
In two dimensions, the scattering amplitude has a logarithmic dependence at low energy \til\cite{Averbuch-1986,Hammer-2010}, thus suggesting the use of the parameter $\eta = -\ln(\kF a_\mathrm{BF})$ as the dimensionless strength of the BF interaction\til\cite{Bloom-1975, Bertaina-2011}.

For an attractive  potential in 2D, a bound state always exists and the scattering length $a_\mathrm{BF}$ essentially corresponds to its radius. 
The weakly attractive regime corresponds to a bound-state radius much larger than the average distance of a fermion from a given boson ($\simeq k_{\rm F}^{-1}$ for mixtures with $n_{\rm B} \le n_{\rm F}$) and is thus characterized by large and negative values of $\eta$. In the strongly-attractive regime, the bound state radius is instead much smaller than such a distance and thus  corresponds to large and positive values of $\eta$.
Here,
we adopt the definition of the scattering length by which the bound state binding energy is given by 
\begin{equation}
\label{2DbindingE}
    \epsilon_0 = \frac{1}{2 m_r \left(a^{}_\mathrm{BF} e^{\gamma}/2\right)^2 } ,
\end{equation}
with $\gamma \simeq 0.577$ the Euler-Mascheroni constant \til\cite{Bertaina-2011}. An alternative definition sometimes used in the literature incorporates in the scattering length the factor $e^\gamma/2$ appearing in Eq.~\eqref{2DbindingE}.

The aim of this section is to solve Eq.\til(\ref{equ:2DLOCVeq}) across the full crossover between these two coupling  regimes, which in practical terms is exhausted within the range $-2 \lesssim \eta \lesssim +2$.
In particular, we are interested in the ground state and first excited state solutions, also known as the attractive and repulsive branches of the associated polaron $(\nB \rightarrow 0)$ problem, respectively\til\cite{Massignan-2014}.
Along the attractive branch, atoms start to experience a weak attraction at $\eta \lesssim -2$ leading to the formation of a tight molecular state in the limit $\eta \gtrsim 2$. The corresponding correlation function $f(r)$ can be seen as the two-body bound state wave function embedded into a many-body environment.
Along this branch, the total energy of the system is always lower than that of the non-interacting mixture\til\cite{Chang-2004,Astra-2004}.
Along the repulsive branch, the effective interaction is instead repulsive and its strength increases as $\eta$ spans the opposite range from +2 to -2. The corresponding correlation function $f(r)$ can be seen as the embedding of the two-body scattering problem at threshold into a many-body environment.  Along this branch, the total energy of the system is always greater than that of the non-interacting mixture\til\cite{Taylor-2011,Randeria-2011,Bertaina-2021}.

For a contact potential, the Schr\"odinger equation\til(\ref{equ:2DLOCVeq}) can be recast into that for a free particle subject to a specific boundary condition at the origin, namely the Bethe-Peierls boundary condition\til\cite{Bethe-1935}. In two dimensions, it reads (see \cite{Werner-2012} and references therein)
\begin{equation}
    \left[r\frac{d}{dr}-\frac{1}{\ln(r/a_\mathrm{BF})}\right]f(r) = \mathcal{O}(r) \phantom{aaaaa} (r\to 0)
    \label{bpcond}
\end{equation}
or, equivalently,
\begin{equation}
f(r) = C \; \ln(r/a_\mathrm{BF}) + \mathcal{O}(r) \phantom{aaaaa} (r\to 0), 
    \label{bpcond2}
\end{equation}
where $C$ is a constant.
Applying conditions (\ref{bpcond}), (\ref{boundaryconda}) and looking for the lowest positive energy solution of Eq.\til(\ref{equ:2DLOCVeq}), one obtains the correlation  function of  the repulsive branch
$\left(\lambda>0, \; k \equiv \sqrt{ 2m_r \lambda } \right)$  \cite{Gawryluk-2024,Whitehead-2016}
\begin{equation}
\label{2Drepulsivesolution}
    f^+(r)=\frac{\ln(k a_{\rm BF}e^\gamma/2)J_0(k r)-\frac{\pi}{2}Y_0(k r)}{\ln(k a_{\rm BF}e^\gamma/2)J_0(k d^+)-\frac{\pi}{2}Y_0(k d^+)},
\end{equation}
with $J_0(x)$ and $Y_0(x)$ the Bessel functions of zeroth order.

In an analogous way, but now looking for the lowest negative energy solution  of Eq.~(\ref{equ:2DLOCVeq}), one obtains the following expression
for the correlation  function of the attractive branch
$\left(\lambda<0,\; \kappa \equiv   \sqrt{- 2m_r \lambda }\right)$ : 
\begin{equation}
\label{2Dattractivesolution}
    f^-(r)=\frac{\ln(\kappa a_{\rm BF} e^\gamma/2)\,I_0(\kappa r)+K_0(\kappa r)}{{\ln(\kappa a_{\rm BF}e^\gamma/2)\,I_0(\kappa d^-)+K_0(\kappa d^-)}},
\end{equation}
with $I_0(x)$ and $K_0(x)$ the modified Bessel functions of zeroth order. Note that to obtain Eq.~\eqref{2Dattractivesolution} we have replaced $k \to i \kappa$ in Eq.~\eqref{2Drepulsivesolution} \cite{Gawryluk-2024,Whitehead-2016} and used the relations $J_0(ix)=I_0(x)$  and $Y_0(ix)=iI_0(x)-2 K_0(x)/\pi$ with $x$ real and positive.

The quantities  $\kappa$ and $d^-$ or $k$ and $d^+$ are then obtained by simultaneously solving  Eqs.~(\ref{locvconstraint}) and  (\ref{boundarycondb}) (with $f(r)$ given by Eq.~\eqref{2Dattractivesolution} or Eq.~\eqref{2Drepulsivesolution}, respectively) for a given  choice of the coupling parameter $\eta$\til\cite{Cordioli-2025}.

\begin{figure}[t]
\centering
\includegraphics[width = 0.48\textwidth]{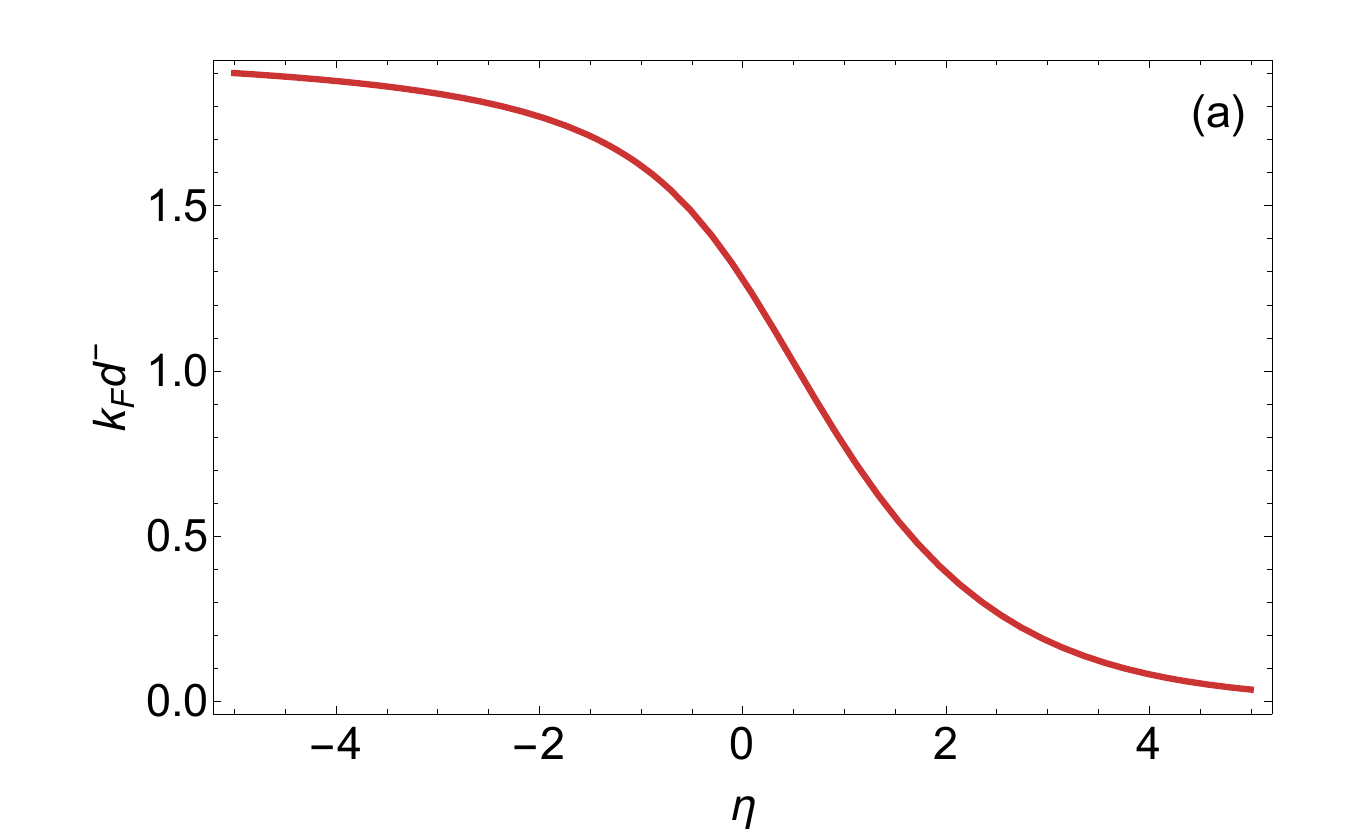}
\includegraphics[width = 0.48\textwidth]{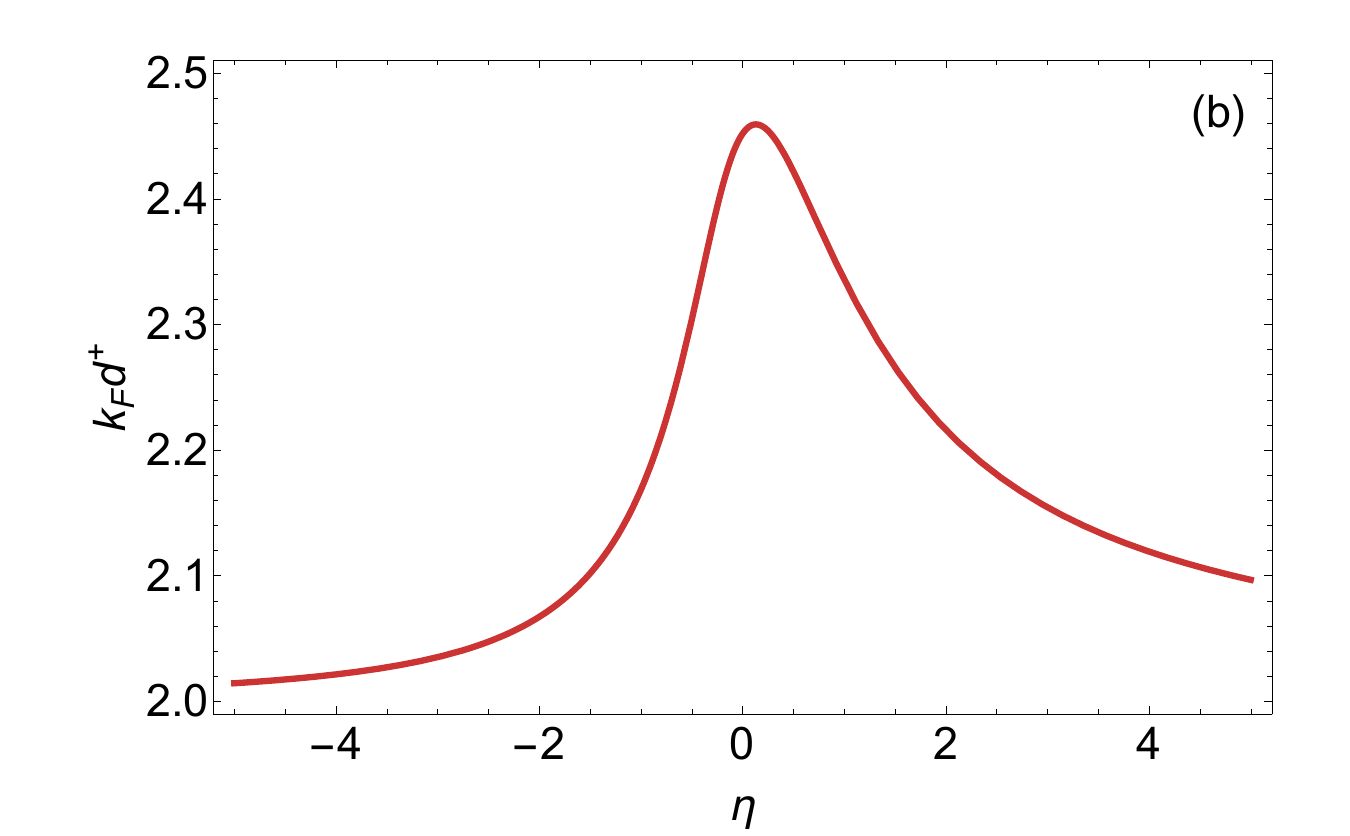}
\caption{Dimensionless healing distance (a) $\kF d^-$ and (b)  $\kF d^+$   
versus the coupling strength  \( \eta = -\ln(\kF a_\mathrm{BF}) \)  for the attractive and repulsive branches, respectively.}
 \label{fig:Healingdistance}
\end{figure}

In Fig.\til\ref{fig:Healingdistance} the dimensionless healing distances $\kF d^\pm$ versus the coupling parameter $\eta$ are shown for the attractive (a) and repulsive (b) branches.
We first notice that 
$k_{\rm F}d^- < 2$ while $k_{\rm F}d^+ >2$ for all coupling strengths. This can be understood by observing that, in the absence of interaction,  the average  distance of a fermion from a given boson $r_n$ is given by $r_n= 1/(\pi \nF)^{1/2} = 2/k_{\rm F}$.  The healing distance $d$ is thus smaller than $r_n$ for an attractive interaction and larger than $r_n$ for a repulsive one.
Clearly, this uncorrelated value is reached in the weak-coupling limits of the two branches, corresponding to $\eta\to-\infty$ for $k_{\rm F} d^-$ and $\eta\to+\infty$ for $k_{\rm F} d^+$. 
In the strong-coupling limit $\eta \rightarrow +\infty$ of the attractive branch, which corresponds to the formation of tightly bound BF molecules of radius $a_\mathrm{BF}$, one has $d^-\sim a_{\rm BF}$ and thus $\kF d^- \rightarrow 0$. 
For the attractive branch, the overall behavior of $k_{\rm F} d^-$ as a function of $\eta$ is then just a monotonic evolution between these two limits.

For the repulsive branch, instead, $k_{\rm F}d^+$ first increases as the effective repulsion, proportional to $k_{\rm F} a_{\rm BF}=e^{-\eta}$, increases (i.e, when $\eta$ decreases from $\eta \to +\infty$ to smaller values), to reach a maximum for $\eta\simeq 0$. When the effective repulsion increases further, $\kF d^+$ starts to decrease, reaching the uncorrelated value 2 in the strongly repulsive limit $\eta \to -\infty$. 

We note in this respect that $\eta < 0$ 
corresponds to $k_{\rm F} a^{}_{\rm BF} > 1$. For a genuine repulsive potential, for which the range $r_0$ of interaction is of the same order (or larger) than $a^{}_{\rm BF}$, this means $k_{\rm F} r_0 > 1$. Therefore, the system is outside the universality regime of the interaction, with the results depending on the detailed form of the potential and not only on the scattering length $a_{\rm BF}$. In addition, when $\eta$ approaches $-\infty$ the bound-state energy of the attractive contact interaction approaches the scattering threshold, making the repulsive branch increasingly unstable towards decay to the attractive branch. Therefore, the region $\eta < 0$, which we report here for completeness, should be taken with care. Still, we observe that for a genuinely repulsive interaction the limit $k_{\rm F}a_{\rm BF} \to \infty$ is a high-density limit in which the Pauli pressure dominates over the BF interaction. The distribution of the fermions should therefore not be affected by the presence of bosons, and the correlation distance $d^+$ should recover the uncorrelated value $r_n$, as indeed observed in Fig.~\ref{fig:Healingdistance}.

\begin{figure}[t]
\includegraphics[width = 0.48\textwidth]{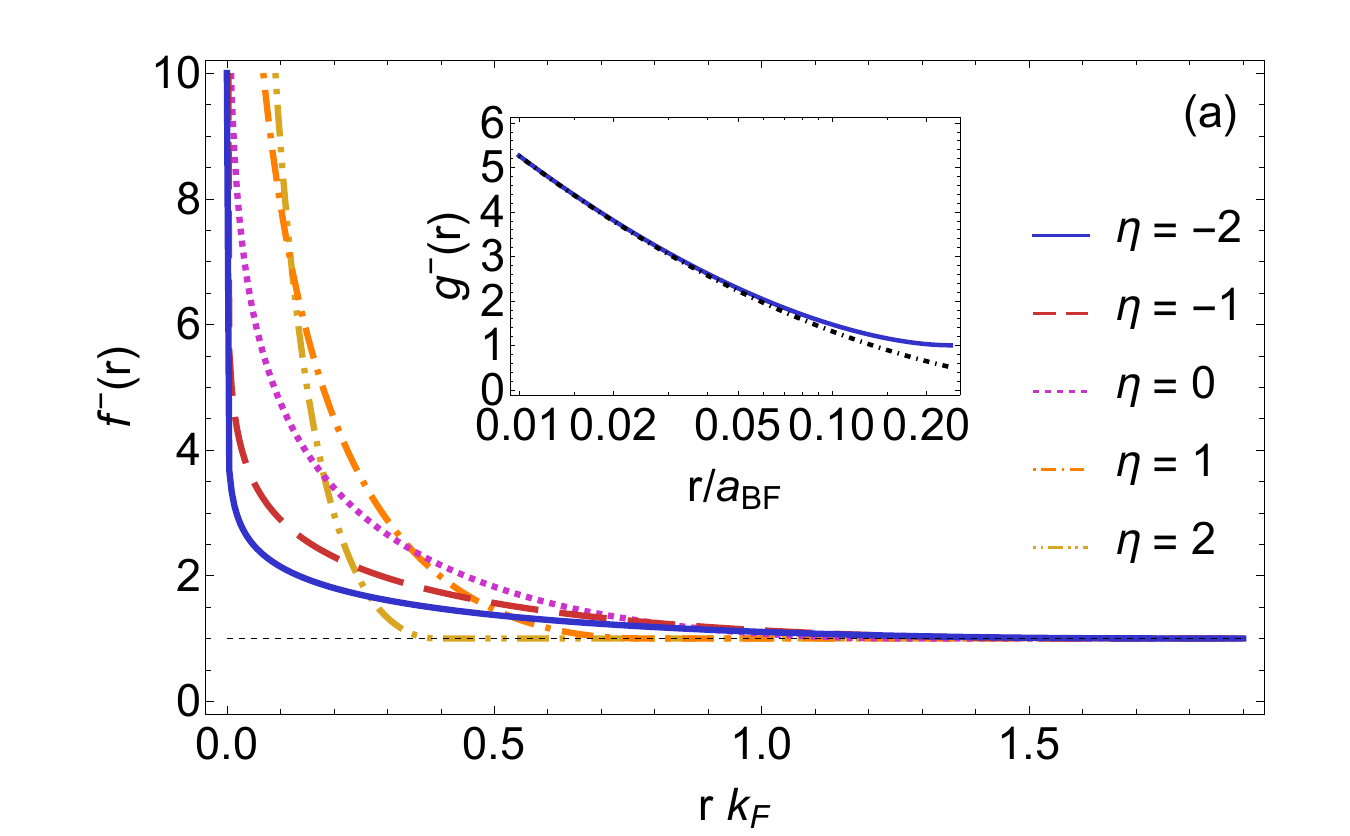}
\includegraphics[width = 0.48\textwidth]{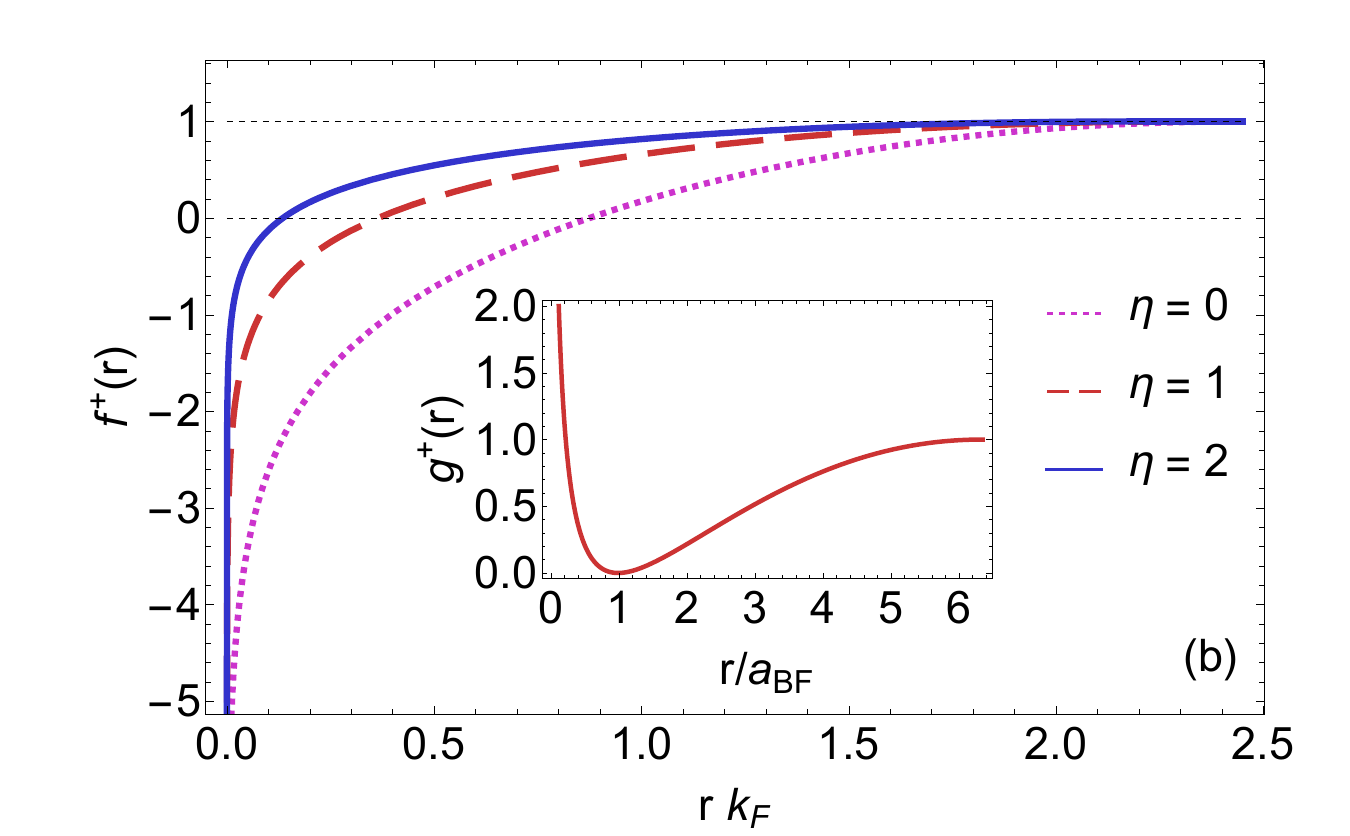}
 \caption{ (a) Correlation function $f^-(r)$ for the attractive branch, for several interaction strengths ranging from $\eta = -2$ (weak attraction) to $\eta = 2$ (strong attraction). Inset: pair correlation function $g^-(r)=f^-(r)^2$ for $\eta=-2$ (solid line) together with its small $r$ asymptotic behavior $\propto \ln^2(r/a_\mathrm{BF})$ (dashed-dotted line). (b) Correlation function $f^+(r)$ for the repulsive branch, for three interaction strengths ranging from $\eta = 0$ (strong repulsion) to $\eta = 2$ (weak repulsion). Inset: pair correlation function $g^+(r)=f^+(r)^2$ for $\eta=1$.}
    \label{fig:fcorr}
\end{figure}

The corresponding correlation functions $f^-(r)$ and $f^+(r)$ are shown in Fig.\til\ref{fig:fcorr} for representative values of the interaction strength $\eta$. In Fig.\til\ref{fig:fcorr}(a) $f^-(r)$ is found to be consistently larger than the uncorrelated asymptotic value (dotted line) and highly peaked at short distances, as expected for an attractive type of interaction. 
We also report the pair distribution function $g^-(r)$ ($=f^-(r)^2$ apart from constants) in the corresponding inset: we note that the diverging behavior at small distances is due to the choice of a zero-range interaction and it
is not observed for finite-range interactions\til\cite{Chang-2004,Chang-2005}, where a strong but finite enhancement of pair correlations is found at short distances. 

In Fig.\til\ref{fig:fcorr}(b), $f^+(r)$ is consistently smaller than the uncorrelated value (upper dotted line) and displays a node whose position shifts to larger distances as the repulsion strength increases ($\eta \rightarrow 0^+$). In particular, the node is located at $r\simeq a_{\rm BF}$. Therefore, coming from large $r$ the pair correlation function $g^+(r)=f^+(r)^2$ is progressively suppressed at smaller $r$ until $r\simeq a_{\rm BF}$ (see inset), reflecting the expected behavior for a repulsive potential\til\cite{Randeria-2011,Pilati-2010,DAlberto-2024}.  For $r\lesssim a_{\rm BF}$, however, one is in the non-universal region of the correlation function, and the underlying attractive nature of the contact interaction emerges, producing strong attractive correlations at short distances. 
\section{Attractive and repulsive Energy-Branches}
\label{sec:attrepbran}

\begin{figure}[t]
\includegraphics[width = 0.48\textwidth]{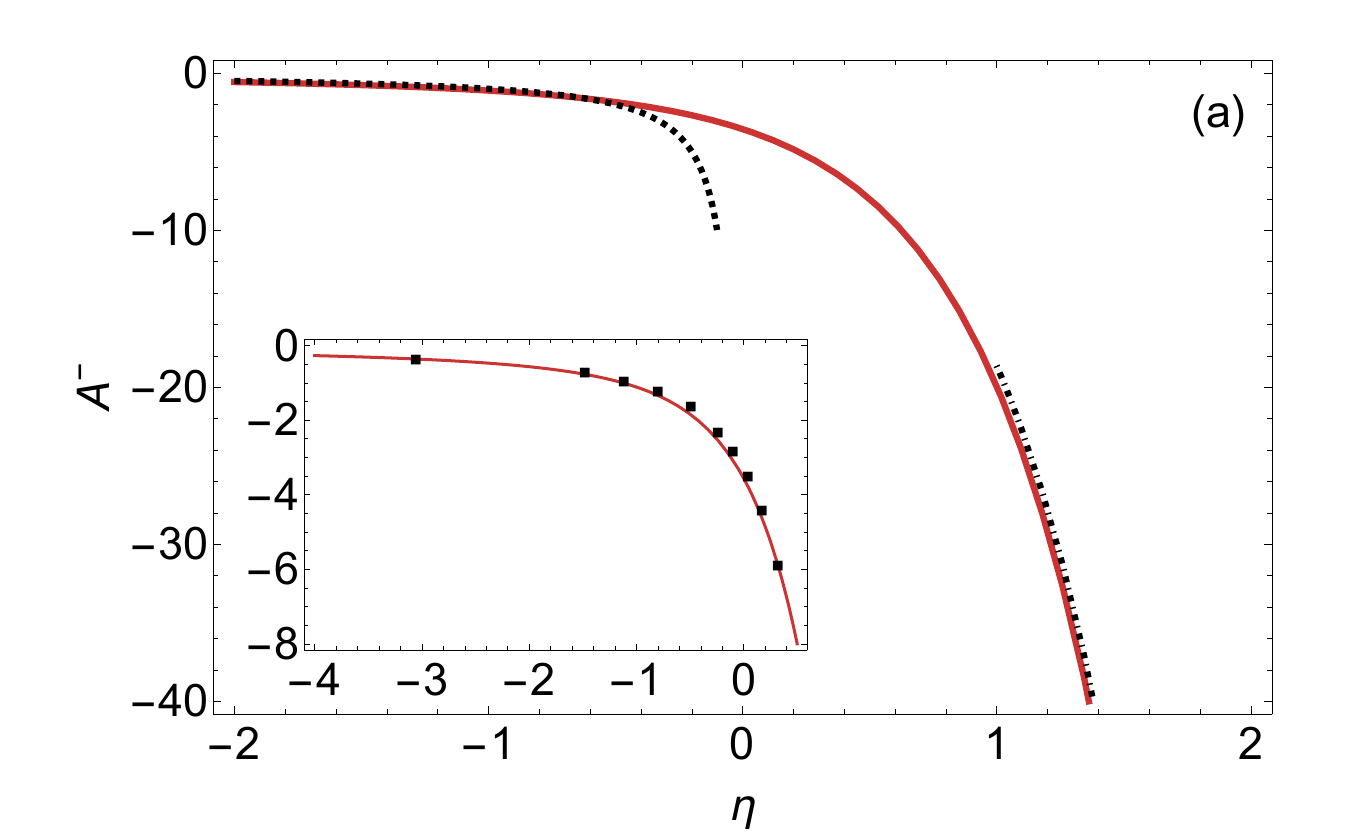}
\includegraphics[width = 0.48\textwidth]{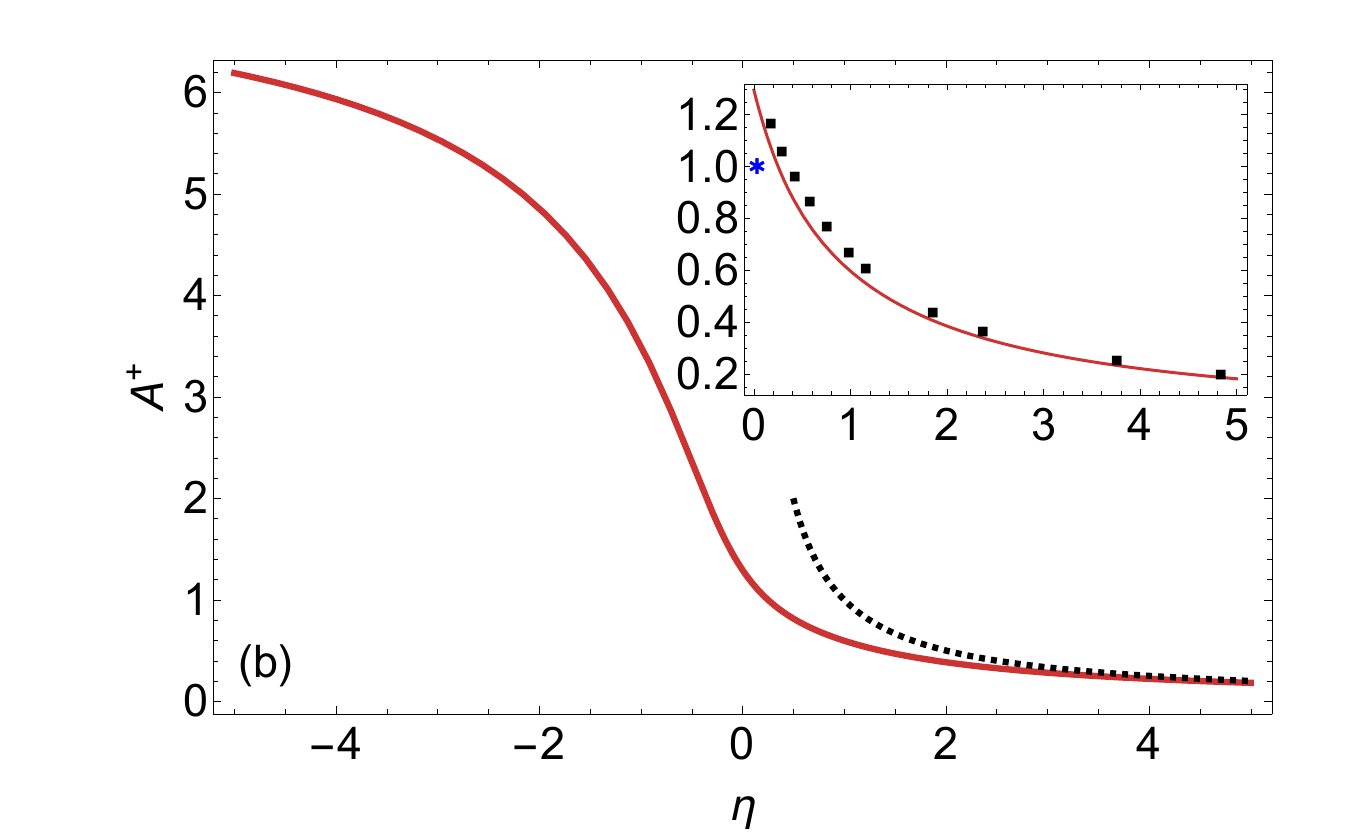}
\caption{Dimensionless quantities $A^-$ and $A^+$ determining the BF interaction energy as a function of the interaction strength  \( \eta = -\ln(\kF a_\mathrm{BF}) \)  for the attractive and repulsive branches, respectively.
Insets: (a) comparison between $A^-$ (line) and experimental data for the attractive polaron energy normalized by $\epsilon_{\rm F}$ \cite{Koehl-2012} (squares);
(b) comparison between  $A^+$ (line) and QMC results for the polaron energy normalized by $\epsilon_{\rm F}$ for hard-disk BF repulsive interaction~\cite{Bertaina-2021} (squares); the value obtained in \cite{Parish-2012} within a $T$-matrix approximation is also reported (asterisk). Dotted lines in both panels: weak-coupling approximation\til(\ref{equ:mubbench}); dash-dotted line in (a): strong-coupling approximation\til(\ref{equ:scbench}).} 
\label{fig:Apm}
\end{figure}

With the solutions of Eq.\til($\ref{equ:2DLOCVeq}$) the energy density  (\ref{equ:2DBFexpval}) now reads 
\begin{equation}
\mathcal{E}^{\pm}_\mathrm{BF} =\nF E_\mathrm{FG} + \nB \lambda^{\pm},
\label{equ:enebf}
\end{equation}
where $\lambda^-$ and $\lambda^+$ represent the interaction energies per BF pair and are given by 
\begin{align}
    \lambda^- &\equiv -\frac{\kappa^2}{2m_r} = \frac{1 + b}{2b} A^-\, \eF, \qquad A^- = -2\frac{\kappa^2}{k_{\rm F}^2}, \\
    \lambda^+ &\equiv \;\; \frac{k^2}{2m_r} = \frac{1 + b}{2b} A^+\, \eF, \qquad A^+ = 2\frac{k^2}{k_{\rm F}^2},
\end{align}
with $ b \equiv m_{\rm B}/m_{\rm F}$ the boson-fermion mass ratio.

The dimensionless quantities \( A^-\) and $A^+$ determine the contribution of the BF interaction to the energy density \eqref{equ:enebf} for the attractive and repulsive branches, respectively. Within the LOCV approximation $A^\pm$ depends only on the dimensionless interaction strength $\eta$. In the single-impurity limit $(\nB \rightarrow 0)$  and for equal masses $A^\pm$ corresponds to the energy of the polaron normalized by the Fermi energy $\epsilon_{\rm F}$. More generally, once multiplied by the coefficient $(1+b)/2b$, $A^\pm$ corresponds to the contribution of the BF interaction to the boson chemical potential $\mu_{\rm B}$ for the two branches (normalized by $\eF$).
In the weak coupling limits of the attractive ($\eta \lesssim -1$) and repulsive ($\eta \gtrsim 1$) branches, the quantities $A^\pm$ are  expected to recover the leading order perturbative expression of the boson chemical potential due to the BF interaction as computed in Ref.\til\cite{DAlberto-2024} 
\begin{equation}
\label{equ:mubbench}
A^\pm= \frac{1}{\eta}  + O(1/\eta^2) .
\end{equation}

In the strong coupling limit of the attractive branch ($\eta \gtrsim 1$) all bosons are expected to bind with fermions into molecules with binding energy $\epsilon_0$. In this limit $\mu_{\rm B}\simeq -\epsilon_0$ \cite{Pisani-2025}, which implies  
\begin{equation}
\label{equ:scbench}
A^-\simeq -\frac{2b}{1+b}\frac{\epsilon_0}{\eF} =-8 e^{2\eta-2\gamma},
\end{equation}
to leading order. 

Both quantities $A^\pm$ are shown in Fig.~\ref{fig:Apm} as a function of $\eta$, along with their weak coupling\til(\ref{equ:mubbench}) (dotted lines) and strong coupling\til(\ref{equ:scbench}) (dash-dotted line in panel (a)) benchmarks. In particular, we have numerically verified that in the strong coupling limit $A^-$ approaches \eqref{equ:scbench} with a subleading correction linearly proportional to $\eta$. 
\begin{figure}[t]
   \centering
    \includegraphics[width=0.48\textwidth]{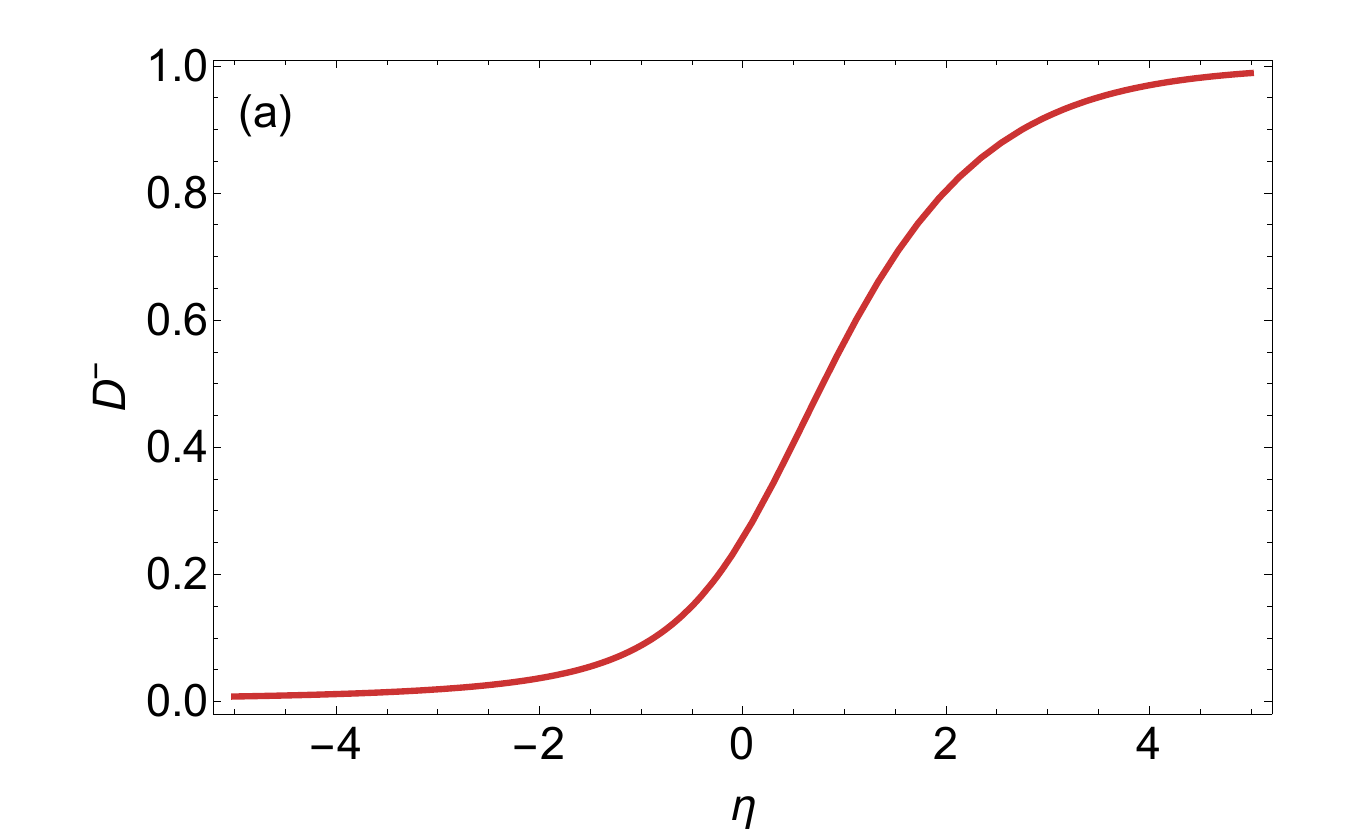}
    \includegraphics[width=0.48\textwidth]{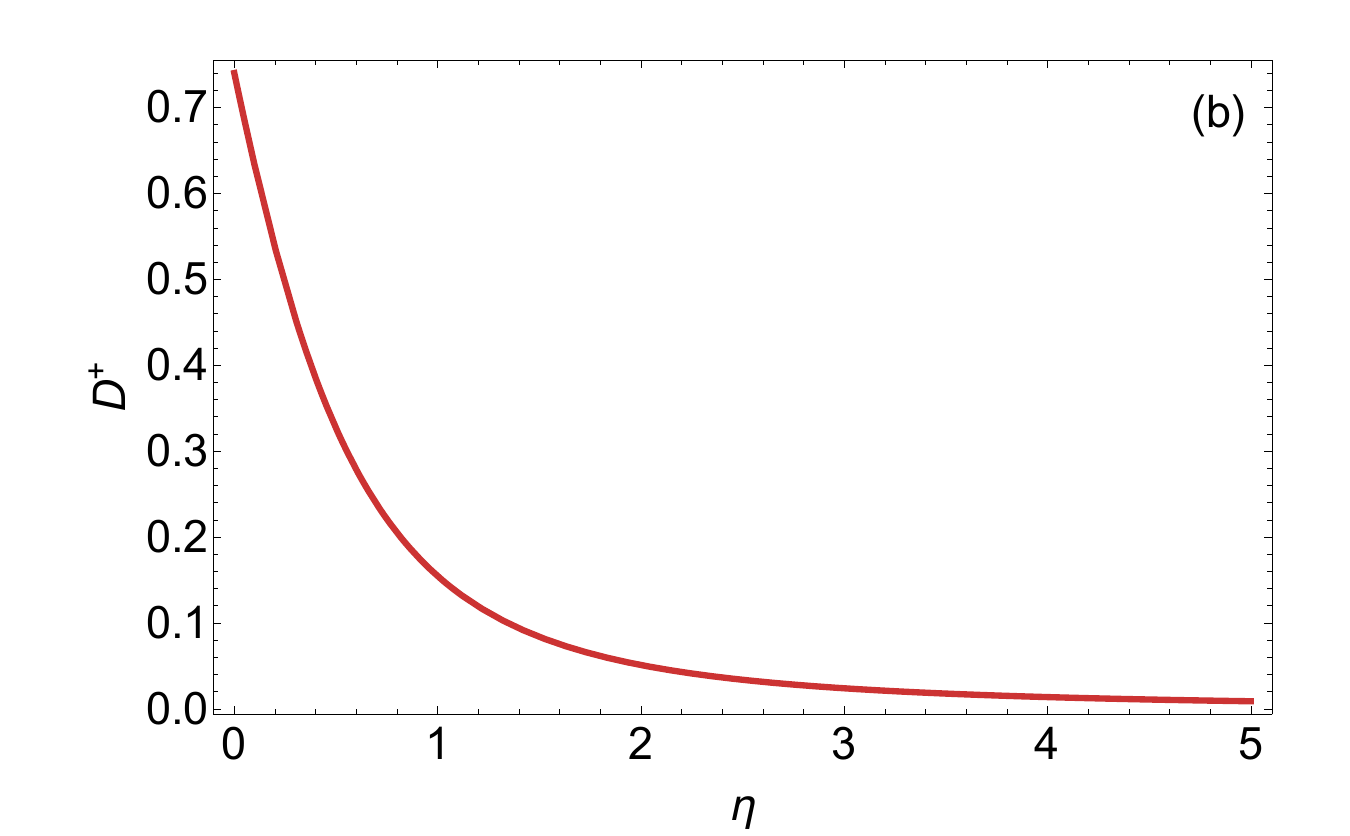}
    \caption{Relative quantum depletion coefficients $D^-$ (a) and $D^+$ (b) as a function of the interaction strength  \( \eta = -\ln(\kF a_\mathrm{BF}) \)  for the attractive and repulsive branches, respectively.}
    \label{fig:2Ddep}
\end{figure}
The 2D attractive and repulsive polarons have been the subject of both experimental and theoretical investigations. 
Specifically, the attractive branch was experimentally studied
in a two-dimensional Fermi-Fermi mixture 
by measuring its spectral function by
momentum-resolved photoemission spectroscopy in Ref.\til\cite{Koehl-2012}. 
The repulsive polaron was computed with fixed-node diffusion Monte Carlo simulations in Ref.\til\cite{Bertaina-2021} adopting hard- and soft-disk potentials. 

In the inset of Fig.~\ref{fig:Apm}(a) we compare $A^-$ with the experimental data for the polaron energy of the attractive branch from Ref.\til\cite{Koehl-2012}, showing a very good agreement between the two quantities.

In the inset of Fig.~\ref{fig:Apm}(b) $A^+$ is compared with Monte Carlo data for the polaron energy of the repulsive branch from Ref.\til\cite{Bertaina-2021}, as well as with the value $A^+=1$ at $\epsilon_0/\eF =2.7$ (corresponding to $\eta=0.034$) obtained in \cite{Parish-2012} within a $T$-matrix approximation. Our results  for the repulsive polaron energy agree very well with the QMC results for the repulsive polaron, including the strongly repulsive region close to $\eta=0$. This result is noteworthy, considering also that the QMC simulations were performed for a purely repulsive potential, while our results are for the repulsive branch of an attractive contact interaction. This agreement, besides validating the LOCV approximation, corroborates our previous statement (and the finding of Ref.\til\cite{Bertaina-2021} obtained by comparing different repulsive potentials) that the region $\eta > 0$ can be considered universal, i.e., independent of the details of the two-body interaction.
 
As a final remark on the polaron limit, we wish to emphasize that the LOCV approximation does not coincide with the widely used Chevy ansatz \cite{Chevy-2006,Massignan-2014}, as can be checked by comparing the corresponding integral/differential equations. Rather, the LOCV approximation provides a distinct variational approach based on a real-space two-body Bose--Fermi correlation function $f(r)$ subject to the healing constraint. The physical content of the LOCV approximation is the inclusion of short-range two-body Bose--Fermi correlations in a nonperturbative way, while neglecting explicit higher-order correlations such as multiple particle-hole excitations, irreducible three-body correlations, and correlations among bosons. 
The good agreement shown in Fig.~\ref{fig:Apm} with experimental and Quantum Monte Carlo results indicates that LOCV provides a reasonable estimate of the polaron energy in the considered regimes.
However, higher-order correlations may become important at strong Bose--Fermi coupling and large boson density. In particular, effects such as the formation of many-particle clusters, discussed for the Bose polaron limit in Refs.~\cite{Christianen-2022,Christianen-2024} are clearly out of reach of the present LOCV wave function.

\section{Mechanical Stability}
\label{sec:mechstab}
The Bose-Fermi mixture described by the Hamiltonian~(\ref{equ:HBF}) is, in general, mechanically unstable. For a weak BF interaction,  perturbative calculations show that the compressibility matrix of a BF mixture in the absence of BB repulsion is negative definite for both repulsive and attractive BF interactions \cite{DAlberto-2024}. Like for 3D dilute BF mixtures \cite{Yu-2012}, for a small boson concentration this instability can be interpreted as due to the attractive nature of indirect BB interactions mediated by the fermions. 
To prevent this instability, it is necessary to include an additional BB interaction term \( H_\mathrm{BB} \)  that provides sufficient repulsion among the bosons and contrasts their indirect attraction.

For a dilute weakly-interacting Bose gas, the BB interaction  can be modeled as a contact potential
\begin{equation}
\label{equ:HBB}
H_\mathrm{BB}=\frac{1}{2}g_\mathrm{BB}\sum_{i \neq j} \delta(\vec{r}_i-\vec{r}_j).
\end{equation}

In two dimensions, the coupling constant $g_\mathrm{BB}$ can be expressed in terms of the gas parameter $\nB a_\mathrm{BB}^2$ as $g_{\rm BB}=(4\pi/m_{\rm B}) \zeta$, with $\zeta\equiv -1/\ln(\nB a_\mathrm{BB}^2)$.
In this way, in the absence of BF interactions, the expectation value of $H_{\rm BB}$ over the variational ansatz \eqref{PSI} reproduces the ground-state energy of a dilute 2D Bose gas to leading order in the dimensionless BB coupling strength $\zeta$\til\cite{Schick-1971}. 

We notice in this respect that the use of the uncorrelated form \eqref{phiB} for $\Phi_\mathrm{B}(\vec{\mathbf{r}})$ in the variational wave-function \eqref{PSI} assumes a weak BB repulsion (such that BB Jastrow factors can be ignored). 
A comparison of the condensate fraction of a 2D Bose gas computed with QMC simulations \til\cite{Boronat-2005} with that obtained by the Bogoliubov weak-coupling theory \til\cite{Schick-1971}, shows that the two results start to disagree as $\nB a_\mathrm{BB}^2 \gtrsim 10^{-2}$ (whereby the condensate fraction falls below about 80$\%$)\til\cite{Boronat-2005}. We can thus take the condition  $\nB a_\mathrm{BB}^2\lesssim 10^{-2}$, corresponding to $\zeta \lesssim 0.2$, as an approximate upper boundary above which the BB repulsion cannot be considered weak and the use of an uncorrelated  $\Phi_\mathrm{B}(\vec{\mathbf{r}})$ is no longer justified.

After a lengthy calculation \cite{Cordioli-2025}, the expectation value of \( H_\mathrm{BB} \) on the Jastrow-Slater wave-function\til(\ref{PSI}) acquires the form \cite{Zhai-2011,Yu-2012err}
\begin{align}
    \mathcal{E}_\mathrm{BB}&=\frac{1}{A}\frac{\bra{\Psi}H_\mathrm{BB}\ket{\Psi}}{\braket{\Psi}{\Psi}} \nonumber \\
    &=\frac{1}{2}g_\mathrm{BB}\nB^2\left(1+4\nF\int d^2r \; [f(r)-1]^2\right).
\end{align}

Adding the contribution of the direct BB interaction, the total energy density of the mixture becomes 
\begin{equation}
    \label{equ:EBBBF}
\mathcal{E}_\mathrm{BF}+\mathcal{E}_\mathrm{BB}=\frac{1}{2}\eF \nF+\frac{1+b}{2b}A \, \eF n_{\rm B}+\frac{1}{2}g_{\rm BB}n_{\rm B}^2\left[1+4D\right],
\end{equation}
where the quantity
\begin{equation}
\label{condensatefracRistig}
    D\equiv\nF\int d^2r \hspace{1mm}[f(r)-1]^2
\end{equation}
can be interpreted approximately as a relative condensate depletion ($(n_{\rm B}-n_0)/n_{\rm B}$) due to the BF interaction \cite{Zhai-2011,Pethick-2002,Ristig-1975}.

Figure \ref{fig:2Ddep} reports this quantity for both the attractive  ($D^-)$ and repulsive ($D^+$) branches. 
For the attractive branch we find that 
the behavior of $D^-$ is analogous
to its 3D counterpart in \cite{Zhai-2011} and that it converges asymptotically to 1 showing no transition to the normal state, as also found in the 3D system  \cite{Zhai-2011}.

Differentiating the energy density (\ref{equ:EBBBF}) with respect to \( \nF \) and \( \nB \), one obtains the chemical potentials of the two species
\begin{align}
    \frac{\muF}{\eF} &=1+\frac{1+b}{2b}\big[A(\eta)-\frac{1}{2}A'(\eta)\big]x-\frac{2}{b} D'(\eta)\, \zeta \, x^2  , \label{equ:muf} \\ 
    \frac{\muB}{\eF} &=\frac{1+b}{2b}A(\eta)+\frac{1}{b}\big[1+4D(\eta)\big] \,(2\zeta+\zeta^2) \,x, \label{equ:mub}
\end{align}
where $x\equiv\frac{\nB}{\nF}$ is the boson concentration and we have made explicit the dependence of the two quantities $A$ and $D$ on the BF 
coupling strength $\eta$. 
A discussion of the behavior of the chemical potentials $\mu_{\rm B}$ and $\mu_{\rm F}$ is presented in the appendix\til\ref{app:chempot}.

The mechanical stability of the mixture requires that the stability matrix (or inverse compressibility)
\begin{gather}
M  =
\begin{pmatrix}
\frac{\partial\mu_{\rm F}}{\partial \nF}&\frac{\partial\mu_{\rm F}}{\partial n_{\rm B}}\vspace{1mm}\\
\frac{\partial\mu_{\rm B}}{\partial \nF}&\frac{\partial\mu_{\rm B}}{\partial n_{\rm B}}
\end{pmatrix}
\end{gather}
be positive definite. This corresponds to the conditions 
\begin{align}
\label{equ:smcond1}
&\Tr M = \frac{\partial\muF}{\partial \nF}+ \frac{\partial\muB}{\partial \nB}>0\\
\label{equ:smcond2}
&\det M = \frac{\partial\muF}{\partial \nF}\frac{\partial\muB}{\partial \nB}- \frac{\partial\muF}{\partial \nB}\frac{\partial\muB}{\partial \nF}>0
\end{align}
to be satisfied simultaneously \til\cite{Viverit-2000}.

Given that $M$ is symmetric by construction,
condition\til(\ref{equ:smcond2}) implies that
its diagonal elements must have the same sign. On the other hand, the quantity $\partial\muF/\partial \nF$ is always positive due to Fermi pressure, so, once condition \eqref{equ:smcond2} is satisfied, it follows that $\partial\muB/\partial \nB$ is guaranteed to be positive and condition\til(\ref{equ:smcond1}) is automatically satisfied.
Therefore, condition\til(\ref{equ:smcond2}) alone determines in practice the  stability of the mixture.

It is interesting to notice that in the polaron limit $\nB\to 0$, with simple manipulations, one can write $\det M =(\partial\eF/\partial\nF) F_\mu$. Here, $F_\mu$ is the polaron-polaron mediated interaction at constant medium chemical potential discussed in recent works \cite{Levinsen-2025,Levinsen-2026}. We thus see that in this limit the stability of the mixture is equivalent to the requirement that the polaron-polaron mediated interaction at constant medium chemical potential be repulsive. We further notice that the factor $4D$
appearing in Eq.~\eqref{equ:EBBBF} produces an enhancement of the boson-boson repulsion, in line with the medium enhanced polaron-polaron repulsion $F_n$ at constant medium density pointed out in \cite{Levinsen-2025,Levinsen-2026}. 

Returning to the stability condition \eqref{equ:smcond2},
one obtains for the determinant of the stability matrix within the LOCV approximation
\begin{equation}
    \det M=\left(\frac{\eF}{\nF}\right)^2\left[c_0+c_1\zeta+c_2\zeta^2+c_3\zeta^3+c_4\zeta^4\right] ,
    \label{detM}
\end{equation}
with coefficients  
\begin{align}
    c_0&=-\left(\frac{1+b}{2b}\right)^2 A_1^2\\
    c_1&=\frac{2}{b}\bigg\{D_0+ \frac{1+b}{2b}\bigg[4D' A_1 -A_2 D_0\bigg]x\bigg\} \\ 
    c_2&=\frac{1}{b}\bigg\{3 D_0+\frac{1+b}{2b}\bigg[4D' A_1 -3 A_2 D_0 \bigg]x \nonumber \\ 
    & \hspace{2cm} +\frac{4}{b}\left[ D_2  D_0 -4D'^2\right]x^2 \bigg\} \\
    c_3&=\frac{2}{b}\bigg\{ D_0 -\frac{1+b}{2b} A_2 D_0 x+\frac{1}{b}\left[3 D_2 D_0 -8D'\right]x^2\bigg\} \\    
    c_4&=\left(\frac{2}{b}\right)^2\left[ D_2  D_0 -D'^2\right]x^2,
\end{align}
having defined
$
A_1 \equiv A-\frac{1}{2}A',\;
A_2 \equiv \frac{1}{2}A'-\frac{1}{4}A'',\;
D_0 \equiv 1+4D$ and 
$D_2 \equiv D'+\frac{1}{2}D''$.

It is worth noting that the coefficient $c_0$ is always negative, thus showing that a non-zero BB repulsion is required for stability. 
For a fixed BF coupling $\eta$, the solution of the equation $\det M=0$
provides the critical BB coupling
$\zeta_c$ below which mechanical stability is lost.

\begin{figure}[t]
\centering
\includegraphics[width = 0.48\textwidth]{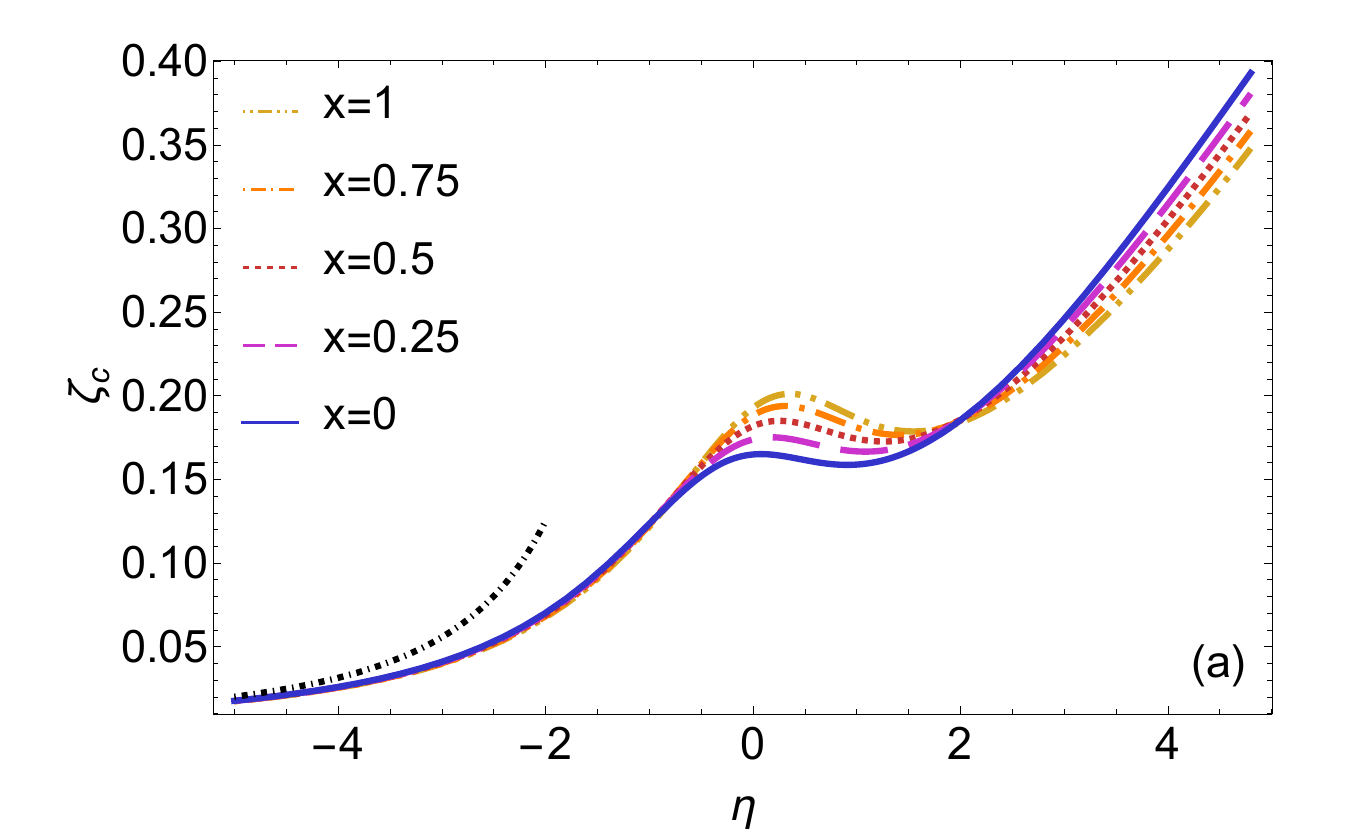}
\includegraphics[width = 0.48\textwidth]{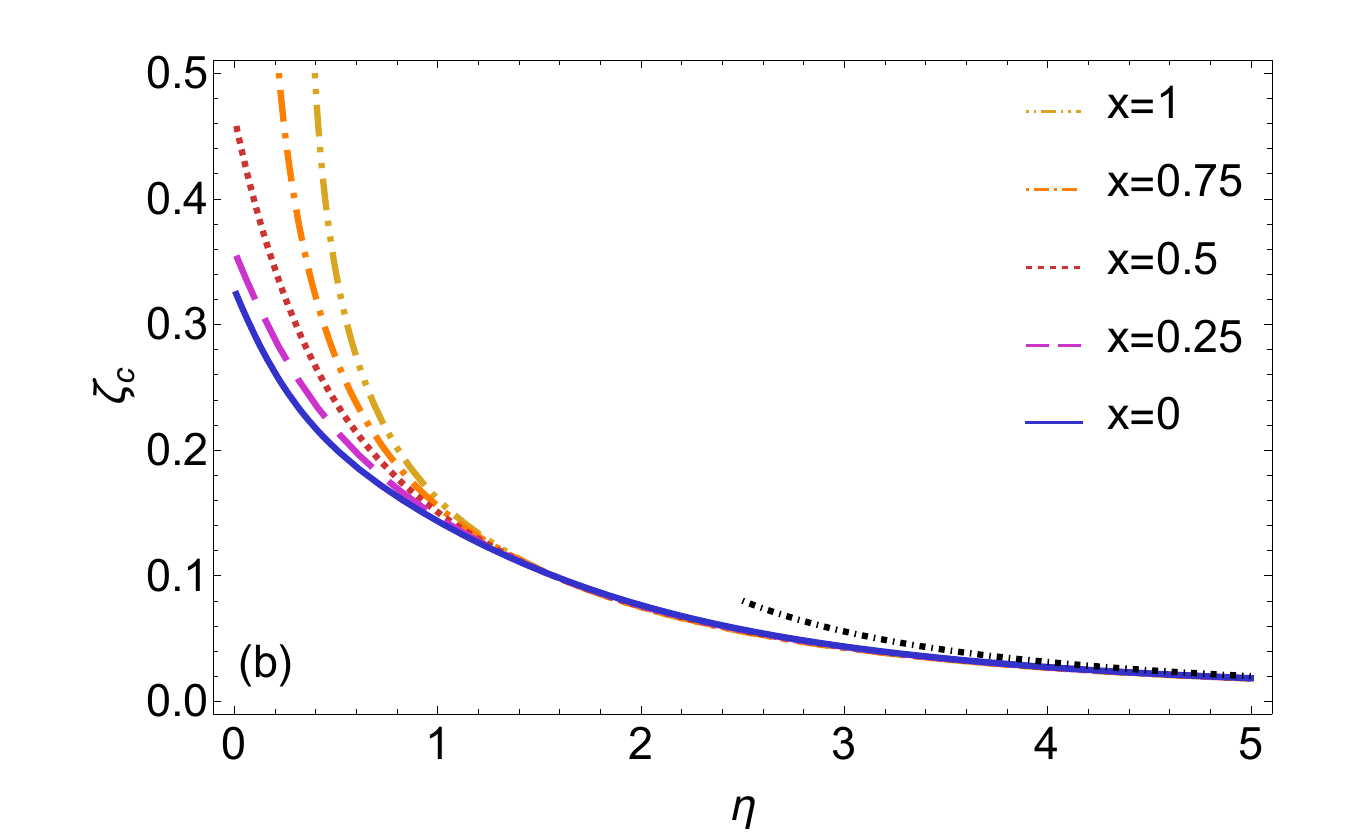}
\caption{ Critical value of the BB repulsion strength $\zeta_c$ as a function of the BF interaction strength \( \eta = -\ln(\kF a_\mathrm{BF}) \) for $m_{\rm B}=m_{\rm F}$ and different values of the boson concentration  $x$  for the attractive (a) and repulsive (b) branches.
The system is stable for $\zeta > \zeta_c$.}
\label{fig:2Dstab}
\end{figure}

In Fig.~\ref{fig:2Dstab}(a) this critical curve is reported for the attractive branch,  at different boson concentration $x$ for $m_{\rm B}= m_{\rm F}$.
In the weak coupling regime, this curve recovers the  perturbative result \til\cite{DAlberto-2024} (dotted line)
\begin{equation}
\label{stab2Dbenchmark}
    \zeta_c \simeq \frac{1}{2\eta^2}.
\end{equation}
Moving towards intermediate values of the BF coupling strength $\eta$,  the  BB interaction required for stability steadily increases.
At small boson concentration, this increase can be interpreted as to compensate for the increasing indirect attraction among the bosons. 

Beyond intermediate couplings ($\eta \gtrsim 0$), the critical BB coupling strength $\zeta_c$ first decreases, to then increase again with $\eta$ for even larger values ($\eta\gtrsim 1$). The initial decrease of $\zeta_c$ is similar to what observed in 3D with the LOCV approximation \cite{Zhai-2011} and by diagrammatic methods \cite{Gualerzi-2025}. This is the expected physical behavior, since the original BF mixture progressively becomes a Fermi-Fermi mixture of fermionic BF molecules and unpaired fermions, with a stabilizing effect of Fermi pressure. 
The subsequent increase of $\zeta_c$ at large $\eta$ is observed only in 2D. We attribute it to a shortcoming of the present approach.
It has been observed already in 3D that the Jastrow-Slater wave function $\eqref{PSI}$ becomes progressively less reliable in the strong-coupling limit when fermionic BF molecules form. 
In this regime, correlations between BF molecules and unpaired fermions or among molecules become important. These direct three-body and four-body correlations are missed by the JS wave-function $\eqref{PSI}$, which directly describes only two-body BF correlations (with three-body correlations arising only indirectly). In 3D, this shortcoming leads to a critical repulsion $\zeta_c$ that progressively decreases but never vanishes at large $\eta$, contrary to what expected on physical grounds for the fermionic  weakly-interacting FF mixture that is effectively obtained in this limit. In 2D its consequences are even more drastic and lead to a $\zeta_c$ that increases indefinitely at large $\eta$. On these grounds, for the attractive branch, the regime of validity of our results for $\zeta_c$ is effectively limited to $\eta \lesssim  1\textnormal{--}2$, after which the large $\eta$ growth of $\zeta_c$ sets in.   
This growth is however just an artifact of the LOCV approximation, and  $\zeta_c$ is expected to continue to decrease at large $\eta$.  Therefore, we  can safely conclude from Fig.~\ref{fig:2Dstab}(a) that, for $m_{\rm B}=m_{\rm F}$, a BB repulsion $\zeta$ as small as $\sim 0.2$ is sufficient to stabilize attractive BF mixtures with $x\le 1$ for all values of the BF interaction $\eta$.

Concerning the nature of the instability occurring below $\zeta_{\rm c}$, we have numerically verified that for the attractive branch the signs of the off-diagonal matrix elements are always negative. We thus conclude that the unstable eigenmode of the stability matrix  corresponds to in-phase density fluctuations of the two components, leading to collapse \cite{Viverit-2000} (which could possibly be arrested by droplets formation \cite{Rakshit3D-2019,Rakshit2D-2019,Foster-2026}).

\begin{figure}[t]
\includegraphics[width = 0.48\textwidth]{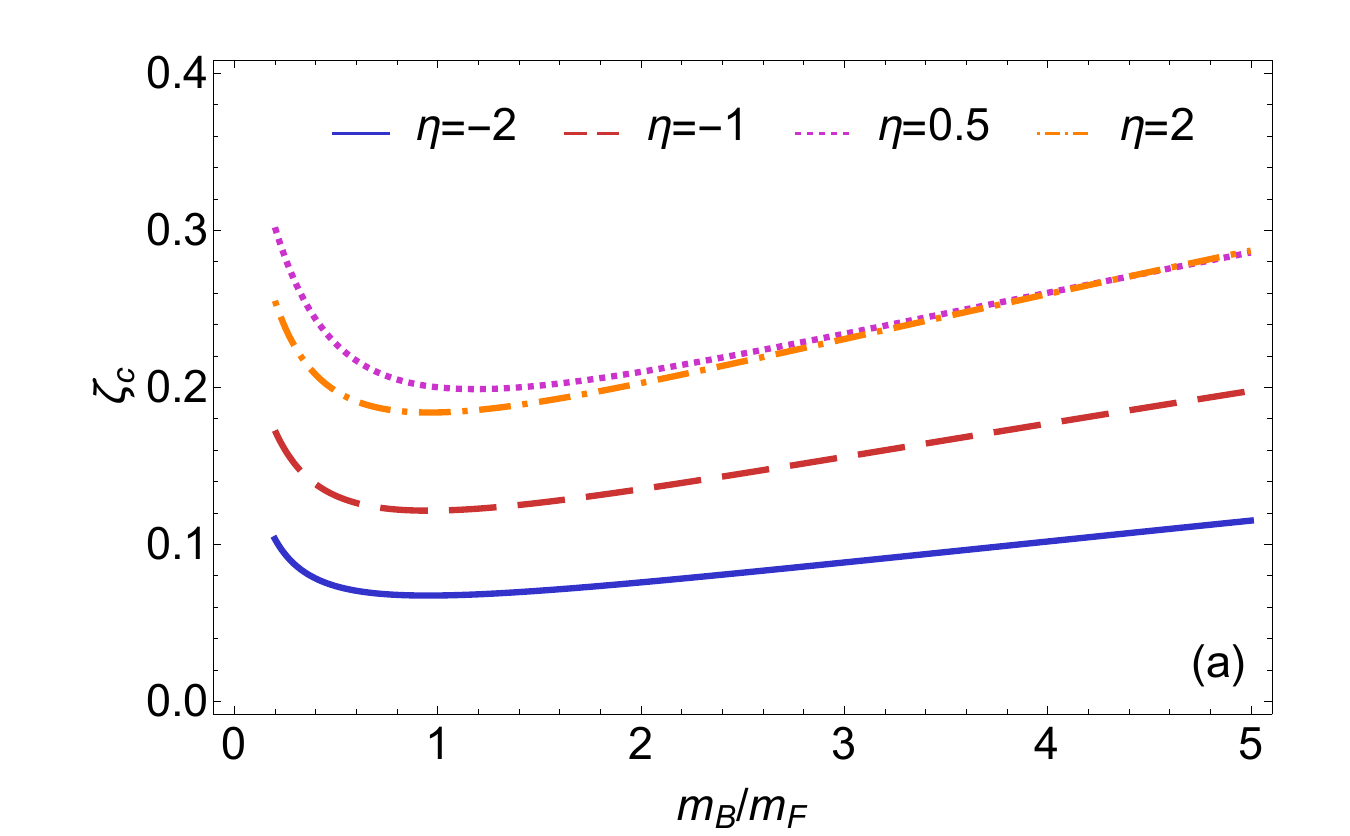}
\includegraphics[width = 0.48\textwidth]{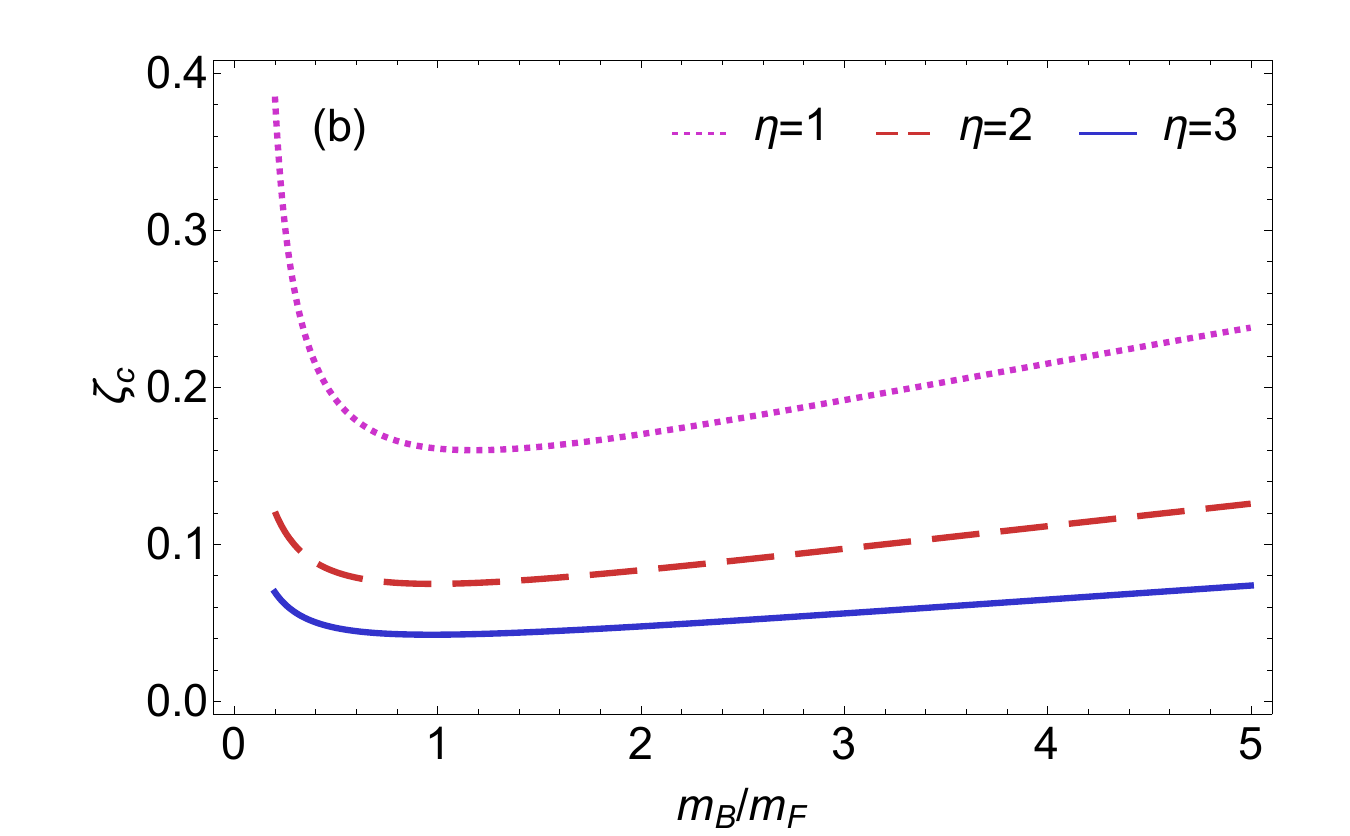}
\caption{Critical BB coupling $\zeta_\mathrm{c}$ as a function of the mass ratio $m_{\rm B}/m_{\rm F}$ for $n_{\rm B}=n_{\rm F}$ and different values of the BF coupling strength $\eta$ for the attractive (a) and repulsive (b) branches. The system is stable for $\zeta > \zeta_c$.}
\label{fig:chicmr}
\end{figure}

For the repulsive branch, instead, we present in  Fig.~\ref{fig:2Dstab}(b) the critical BB repulsion $\zeta_c$  for $\eta > 0$ since, as previously argued, only above this coupling the results for the repulsive branch can be considered universal. 
In addition, in our case in which a repulsive potential is mimicked by the repulsive branch  of an attractive contact potential, the instability towards decay into the attractive branch becomes increasingly important when $\eta$ becomes large and negative, as pointed out above. Therefore, it would be meaningless discussing the mechanical stability of the repulsive branch in a regime in which, for an attractive contact potential, this branch would be dynamically unstable towards decay into the attractive one.

We observe in Fig.~\ref{fig:2Dstab}(b) that the amount of BB repulsion needed for stability increases with increasing BF repulsion $(\eta \rightarrow 0^+)$. When the regime of strong repulsion $\eta < 1$ is entered,  a marked dependence of $\zeta_c$ on the concentration sets in, while
in the weakly repulsive regime $(\eta \rightarrow +\infty)$  the critical BB repulsion
$\zeta_c$ becomes independent of the concentration, consistently with the perturbative result\til(\ref{stab2Dbenchmark}) (dotted line)\til\cite{DAlberto-2024}. 

Concerning the nature of the mechanical instability, for the repulsive branch we have numerically verified that the signs of the off-diagonal matrix elements are always positive. The unstable mode of the stability matrix is thus
out-of-phase, indicating an instability towards phase separation \cite{Viverit-2000}.

Finally, we wish to discuss the dependence of $\zeta_c$ on the mass ratio $m_{\rm B}/m_{\rm F}$. Given the weak dependence of $\zeta_c$ on the concentration $x$ for both branches in the respective physical regions ($\eta \lesssim 2$ for the attractive branch and $\eta\gtrsim 1$ for the repulsive one), in Fig.\til\ref{fig:chicmr} we focus on the case of equal densities $n_{\rm B} = n_{\rm F}$ and present $\zeta_c$ as a function of the mass ratio $m_{\rm B}/m_{\rm F}$ for different coupling strengths for both branches.
We see that the mixtures with equal masses are the most stable ones.
This result can be understood by neglecting in Eq.~\eqref{detM} all terms of order higher than one in $\zeta$ (since $\zeta$ is a small number in the relevant coupling region)  and setting $x=0$ (since the results depends only weakly on $x$). In this way, one obtains the approximation 
\begin{equation}
\zeta_c\simeq\frac{(1+b)^2}{8 b}
\frac{[A(\eta)-\frac{1}{2}A'(\eta)]^2}{1+4D(\eta)}
\end{equation}
which shows that the minimum value of $\zeta_c$ is obtained for $b=1$, independently of the coupling strength $\eta$. A similar equation in the limit $x\to 0$ was obtained in 3D in \cite{Zhai-2011}.

\section{Conclusions}
\label{sec:concl}

In this work we have investigated the mechanical stability of two-dimensional homogeneous Bose-Fermi mixtures at zero temperature  in the presence of a tunable BF interaction and a weak BB repulsion. We have modeled the tunable BF interaction with a properly regularized attractive contact potential. 
The use of the lowest order constrained variational approach allows the investigation of both attractive and repulsive BF mixtures by selecting the so-called attractive and repulsive branches when solving the LOCV equations for the underlying attractive contact interaction.
In the single-impurity limit ($\nB \to 0$), the results for the polaron energy compare well with experimental data \cite{Koehl-2012} for attractive BF mixtures and Quantum Monte Carlo data for repulsive BF mixtures \cite{DAlberto-2024}. At finite bosonic densities, we have successfully benchmarked our results with perturbative calculations \cite{DAlberto-2024} (in the regimes where perturbation theory is valid).

By analyzing the inverse compressibility matrix, we have determined the critical BB repulsion required for stability as a function of the tunable BF interaction for different density and mass ratios. For equal masses, we have found that a BB repulsion $\zeta \simeq 0.2$ (corresponding to $n_{\rm B} a_{\rm BB}^2 \simeq  10^{\rm -2}$) is sufficient to provide stability of attractive BF mixtures from weak to strong values of the BF interaction strength $\eta$. 

Finally, we have found that for both attractive and repulsive mixtures the critical BB repulsion $\zeta_c$ for stability increases when the mass ratio $m_{\rm B}/m_{\rm F}$ is varied from one. 

Our results provide a quantitative stability criterion for experimentally realizable two-dimensional Bose–Fermi mixtures with tunable interactions, and identify the parameter window where homogeneous mixtures can be realized without collapse or demixing. These findings are relevant for ongoing experiments in quasi-two-dimensional geometries and may serve as a guide for the exploration of mediated interactions, polaron physics at finite impurity concentration, and possible collective or pairing phenomena in mixed quantum gases.

As a perspective, it would be interesting to improve the present approach in the molecular regime in which the inclusion of three- and four-body correlations becomes increasingly important.   This could be done, in principle, by extending the JS single-determinant wave-function with a Pfaffian wave-function\til\cite{Bajdich-2006,Carlson-2003,Bertaina-2013} whereby the correct nodal structure stemming from the simultaneous presence of single-particle and pairing states can be accommodated.

Data for reproducing the figures are available online \cite{DatasetZenodo}.

\begin{acknowledgments}
L.P. and P.P. acknowledge financial support from the Italian Ministry of University and Research (MUR) under project PRIN2022, Contract No. 2022523NA7. P.P. also
acknowledges financial support from the European Union - Next Generation EU through MUR
projects PE0000023-NQSTI (Italy).
\end{acknowledgments}
\begin{figure*}[ht]
\includegraphics[width = 0.45\textwidth]{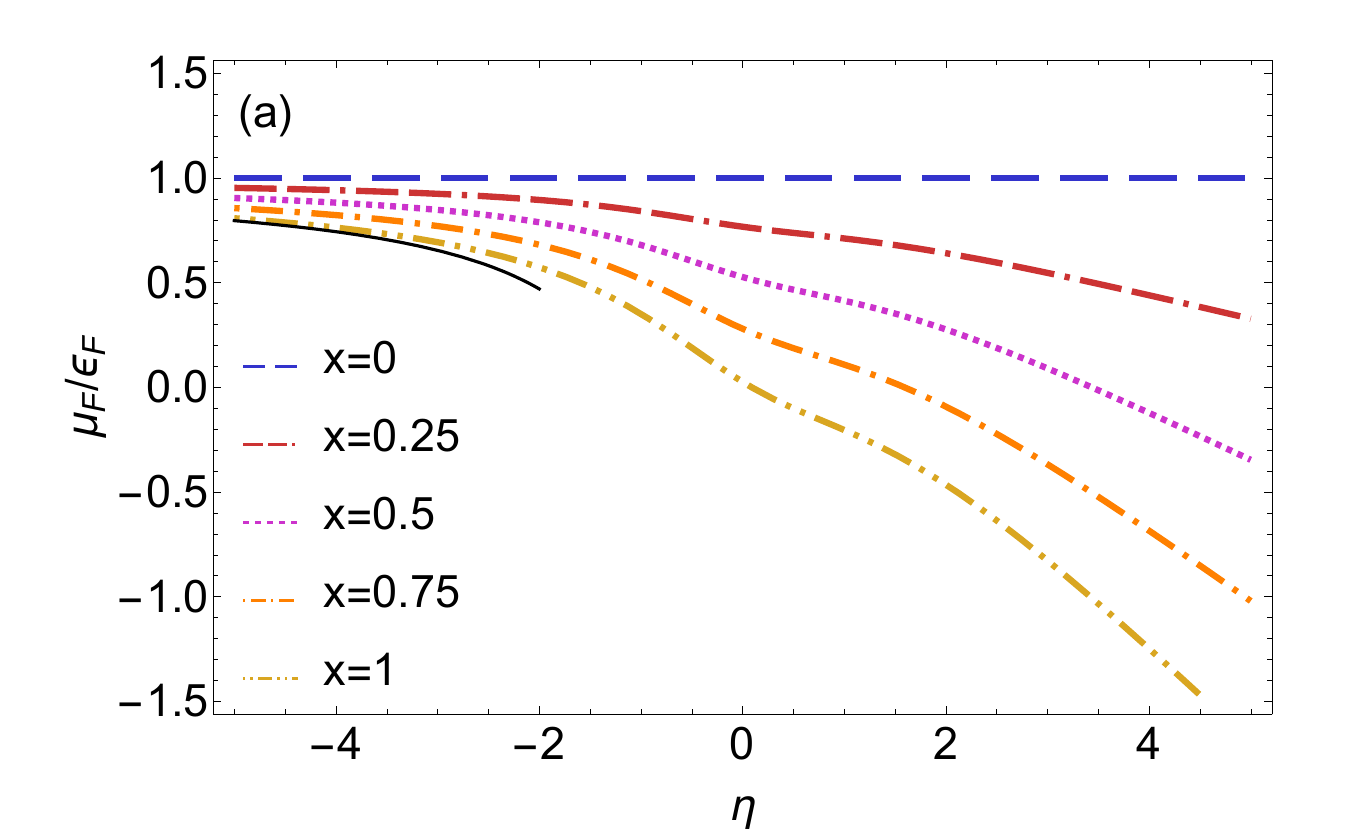}
\includegraphics[width = 0.45\textwidth]{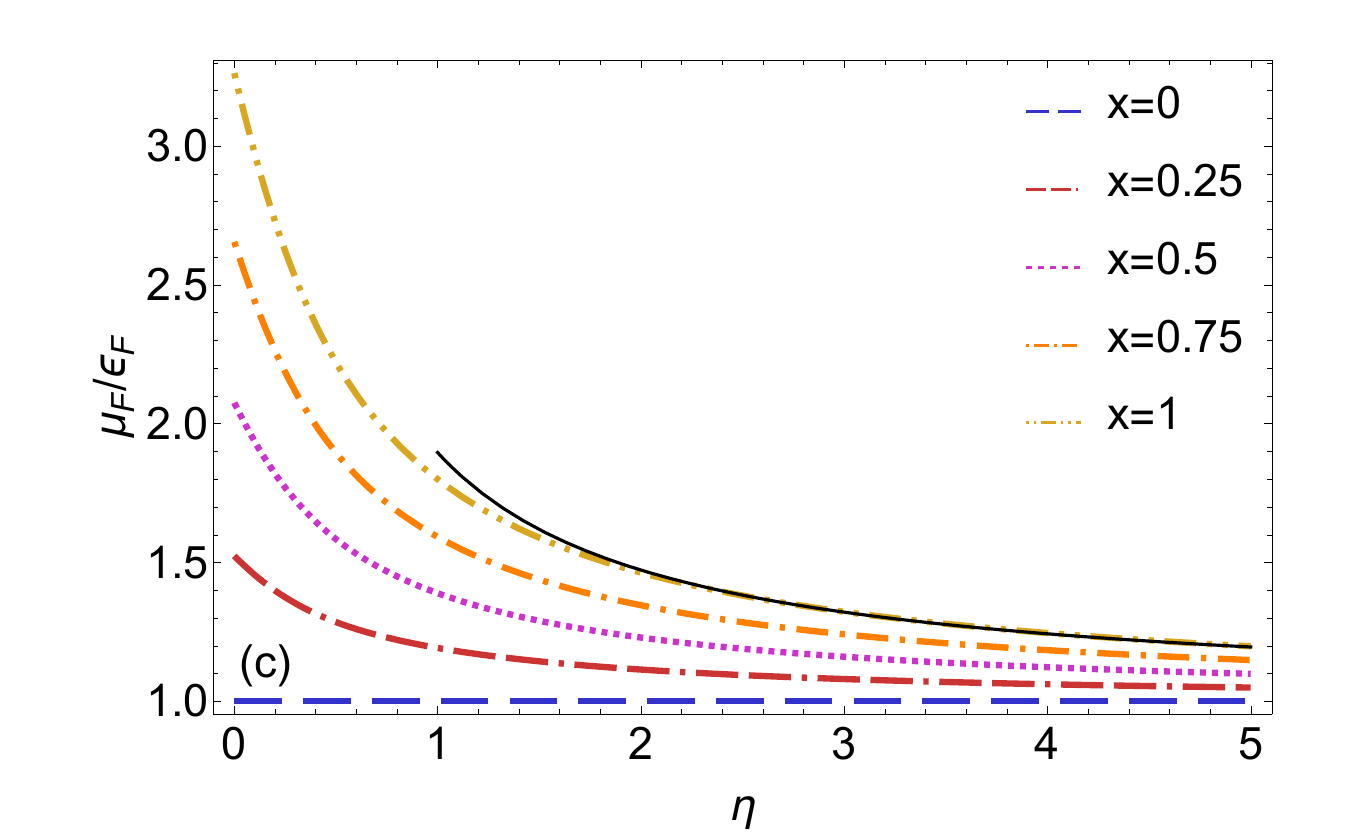}
\includegraphics[width = 0.45\textwidth]{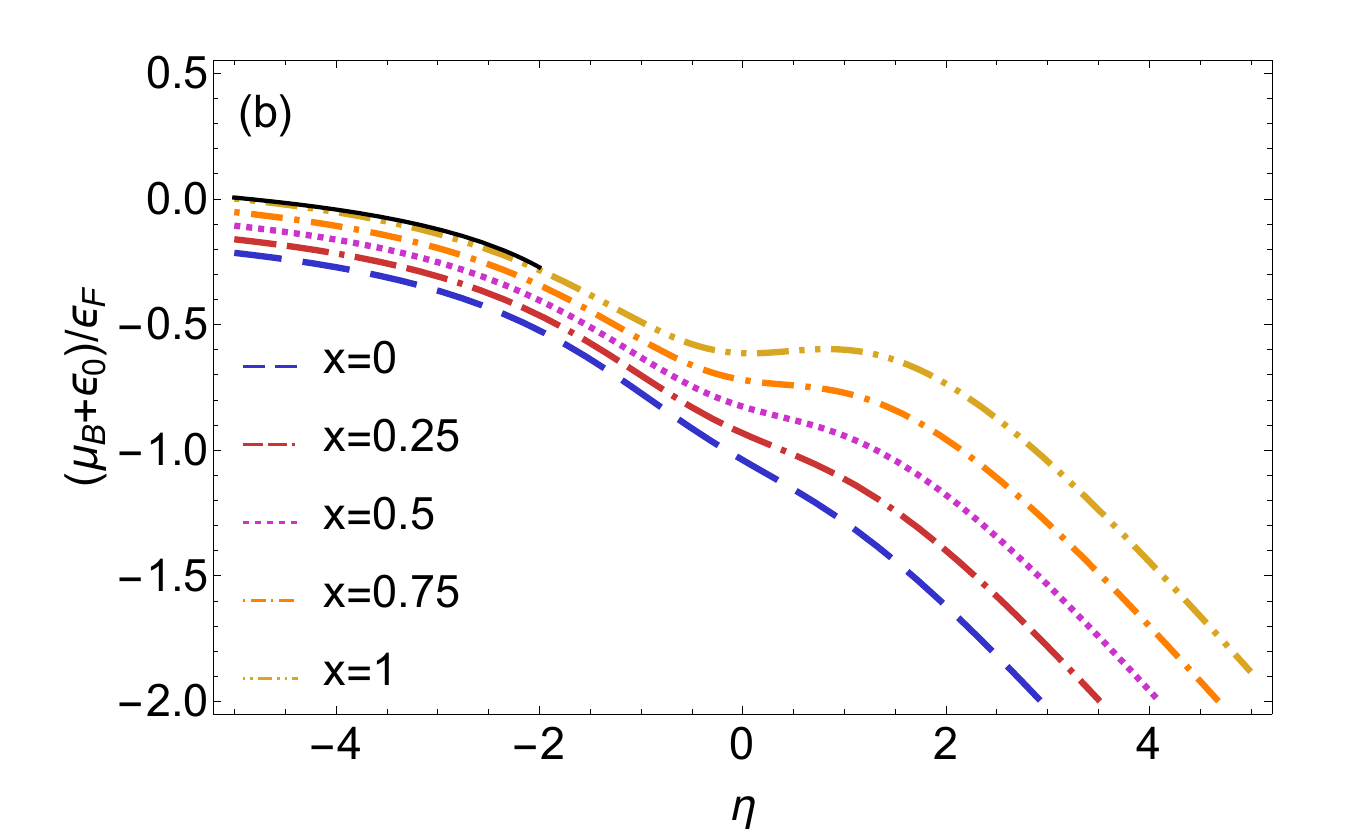}
\includegraphics[width = 0.45\textwidth]{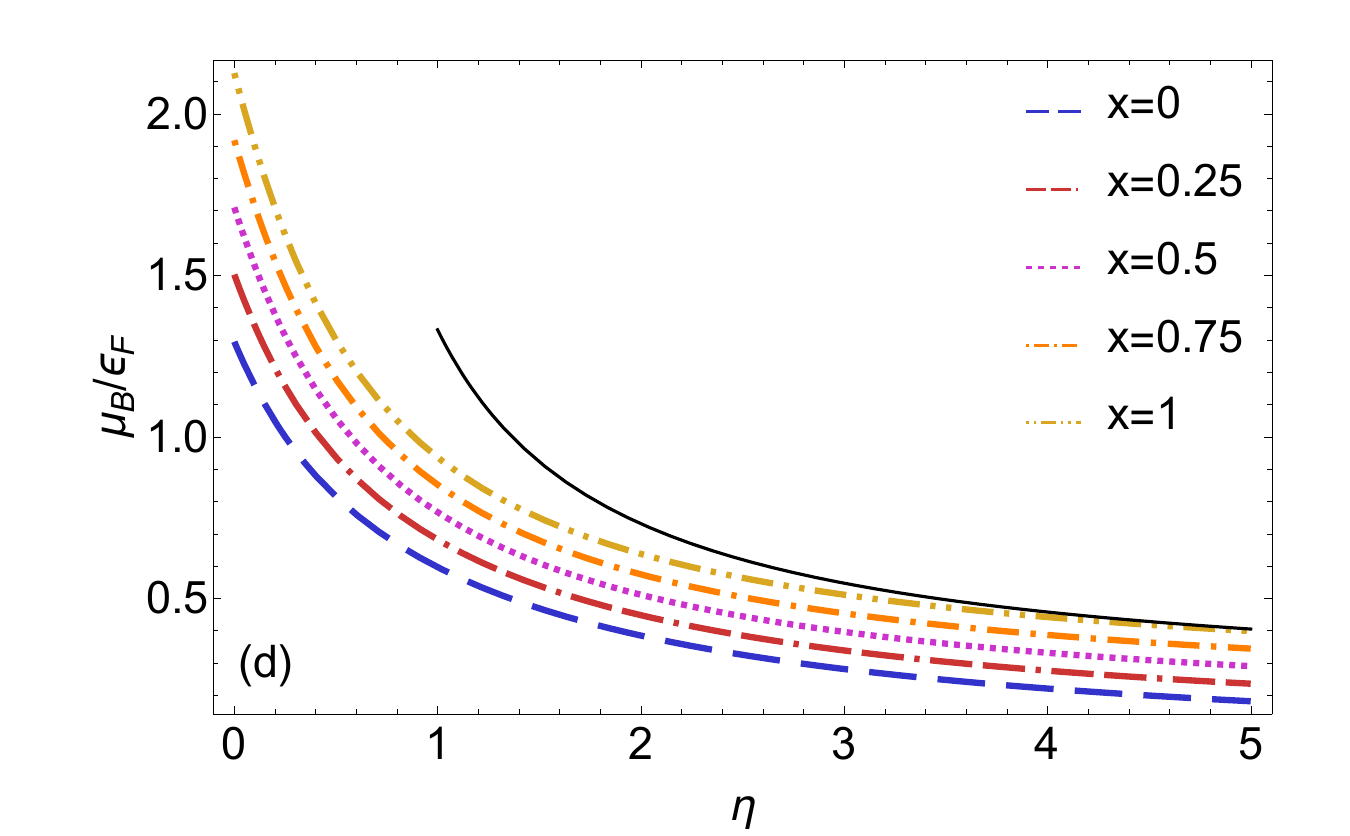}
\caption{(a) Fermionic chemical potential $\muF$ and (b) shifted  bosonic chemical potential $(\muB+\be)$ (normalized by $\eF$) for the attractive branch as a function of the coupling strength $\eta$ for several concentrations $x$. (c) Fermionic chemical potential $\muF$ and (d) bosonic chemical potential $\muB$ (normalized by $\eF$) for the repulsive branch as a function of the coupling strength $\eta$ for several concentrations $x$.
In all panels we have used $m_{\rm B}=m_{\rm F}$ and $\zeta=0.1$.}
\label{fig:chempot}
\end{figure*}

\appendix
\section{Chemical Potentials and molecular limit}
\label{app:chempot}

In this appendix we analyze the behavior of the fermionic and bosonic chemical potentials $\muF$ and $\muB$ obtained in sec.\til\ref{sec:mechstab} and given by the Eqs.\til(\ref{equ:muf}) and\til(\ref{equ:mub}), respectively.
These two quantities are shown in Fig.\til(\ref{fig:chempot}) for both the attractive (panels (a),(b)) and repulsive (panels (c),(d)) branches and for a representative BB repulsion strength $\zeta=0.1$. For the chemical potential  $\muB$ we subtract the leading contribution $-\epsilon_0$ of the molecular binding energy thus removing its diverging behavior in the strong coupling limit.

We start by considering the attractive branch. In the single impurity limit $(x = 0)$ we see that $\muF$ recovers the Fermi energy of a non-interacting gas whereas $\muB$ represents the binding energy of the polaron, $A^- \eF$. 
At finite concentration $x$ and in the weak coupling regime of $\eta$, our results  recover the perturbative expressions
\begin{align}
    \frac{\muF}{\eF} &= 1+\frac{x}{\eta}  \label{equ:muF2Dbench} \\ 
    \frac{\mu_{\rm B}}{\eF} &= 2\,x\, \zeta+\frac{1}{\eta} ,
     \label{equ:muB2Dbench} 
\end{align}
valid to leading order in the small parameters $1/\eta$ and $\zeta \til$\cite{DAlberto-2024}.
The asymptotic expressions\til(\ref{equ:muF2Dbench}) and\til(\ref{equ:muB2Dbench}) (with $\epsilon_0/\eF$ negligible in their regime of validity) are reported for $x=1$ as black  solid lines in  panels (a) and (b) of Fig.\til\ref{fig:chempot}.

As the interaction $\eta$ increases, we see that both $\muF$ and $\muB+\epsilon_0$ monotonically decrease tending towards large negative values. 
We recall however that, as already mentioned above, at large values of $\eta$ the present approach is no longer valid for the attractive branch.

For the repulsive branch, we observe in Fig.\til\ref{fig:chempot}(c) and (d) a monotonic increase of both chemical potentials as the intensity of the repulsion is increased, as expected on physical grounds.
In the weak coupling region $1/\eta \to 0^+$ our results recover the perturbative benchmarks\til\eqref{equ:muF2Dbench} and\til\eqref{equ:muB2Dbench} (black solid lines). 

\section{Auxiliary conditions to the LOCV Equation}
\label{app:meijer}

In this appendix we provide the equations obtained by implementing the conditions~(\ref{locvconstraint}) and  (\ref{boundarycondb}) in Eqs.\til\eqref{2Dattractivesolution} and\til\eqref{2Drepulsivesolution} for the correlation function $f(r)$ in the attractive and repulsive  regimes, respectively.
Condition (\ref{boundarycondb})  provides the relation between the quantity $k d$ (or $\kappa d$) and the scattering length $a_\mathrm{BF}$ and involves 
modified Bessel functions of order one  for the attractive branch
\begin{equation}
    \ln(\kappa a_{\rm BF}e^\gamma/2)\, I_1( \kappa d)=K_1( \kappa d)
\end{equation}
and  Bessel functions of order one for the repulsive branch 
\begin{equation}
   \ln(k a_{\rm BF}e^\gamma/2) \, J_1(k d)=\frac{\pi}{2}Y_1(k d).
\end{equation}

Upon analytical integration \cite{Mathematica}, the condition~(\ref{locvconstraint}) reads
\begin{widetext}
\begin{align}
     (\kappa d)^2\hspace{-1mm}\left[I_0( {\kappa}  {d})^2-I_1( {\kappa}  {d})^2\right]\hspace{-1mm}\left[\ln\left( \frac{\kappa a_\mathrm{BF}e^\gamma}{2}\right)\right]^2
    &+\frac{\sqrt{\pi}}{2 }G_{2,4}^{3,1}\left(\hspace{-2mm}
    \begin{matrix}
        &1,\frac{3}{2}\\
        &1,1,1,0
    \end{matrix}
    \hspace{1mm}\Bigg|\hspace{1mm} (  {\kappa}   {d})^2\right) +\frac{1}{\sqrt{\pi}}\ln\left( \frac{\kappa a_\mathrm{BF}e^\gamma}{2}\right)G_{1,3}^{2,1}\left(\hspace{-2mm}
    \begin{matrix}
        &\frac{3}{2}\\
        &1,1,0
    \end{matrix}\hspace{1mm}\Bigg|\hspace{1mm} (  {\kappa}   {d})^2\right) \nonumber\\
    &=\left(\frac{2 \kappa}{k_{\rm F}}\right)^2\left[K_0(  {\kappa}   {d})+\ln\left( \frac{\kappa a_\mathrm{BF}e^\gamma}{2}\right) I_0(  {\kappa}   {d})\right]^2
\end{align}
\begin{align}
(k d)^2\left[J_0(k d)^2+J_1(k d)^2\right]\left\{\frac{\pi^2}{4}+\left[\ln\left( \frac{k a_\mathrm{BF}e^\gamma}{2}\right)\right]^2\right\}&+\sqrt{\pi}\ln\left( \frac{k a_\mathrm{BF}e^\gamma}{2}\right)G_{1,3}^{2,0}\left(\hspace{-2mm}
   \begin{matrix}
        &\frac{3}{2}\\
        &1,1,0
    \end{matrix}\hspace{1mm}\Bigg|\hspace{1mm} (k d)^2\right) +\frac{\pi^{\frac{3}{2}}}{2}G_{3,5}^{3,1}\left(\hspace{-2mm}
    \begin{matrix}
        &1,\frac{3}{2},\frac{3}{2}\\
        &1,1,1,0,\frac{3}{2}
    \end{matrix}\hspace{1mm}\Bigg|\hspace{1mm} (k d)^2\right)\nonumber \\
&=  \left(\frac{2k}{k_{\rm F}}\right)^2\Big[\frac{\pi}{2} Y_0(k d) -\ln\left(\frac{k a_\mathrm{BF}e^\gamma}{2}\right) J_0(k d) \Big]^2,
\end{align}
\end{widetext}
for the attractive and repulsive branches, respectively, where the Meijer G-function\til\cite{Bateman} is used.


\begin{thebibliography}{146}%
\makeatletter
\providecommand \@ifxundefined [1]{%
 \@ifx{#1\undefined}
}%
\providecommand \@ifnum [1]{%
 \ifnum #1\expandafter \@firstoftwo
 \else \expandafter \@secondoftwo
 \fi
}%
\providecommand \@ifx [1]{%
 \ifx #1\expandafter \@firstoftwo
 \else \expandafter \@secondoftwo
 \fi
}%
\providecommand \natexlab [1]{#1}%
\providecommand \enquote  [1]{``#1''}%
\providecommand \bibnamefont  [1]{#1}%
\providecommand \bibfnamefont [1]{#1}%
\providecommand \citenamefont [1]{#1}%
\providecommand \href@noop [0]{\@secondoftwo}%
\providecommand \href [0]{\begingroup \@sanitize@url \@href}%
\providecommand \@href[1]{\@@startlink{#1}\@@href}%
\providecommand \@@href[1]{\endgroup#1\@@endlink}%
\providecommand \@sanitize@url [0]{\catcode `\\12\catcode `\$12\catcode `\&12\catcode `\#12\catcode `\^12\catcode `\_12\catcode `\%12\relax}%
\providecommand \@@startlink[1]{}%
\providecommand \@@endlink[0]{}%
\providecommand \url  [0]{\begingroup\@sanitize@url \@url }%
\providecommand \@url [1]{\endgroup\@href {#1}{\urlprefix }}%
\providecommand \urlprefix  [0]{URL }%
\providecommand \Eprint [0]{\href }%
\providecommand \doibase [0]{https://doi.org/}%
\providecommand \selectlanguage [0]{\@gobble}%
\providecommand \bibinfo  [0]{\@secondoftwo}%
\providecommand \bibfield  [0]{\@secondoftwo}%
\providecommand \translation [1]{[#1]}%
\providecommand \BibitemOpen [0]{}%
\providecommand \bibitemStop [0]{}%
\providecommand \bibitemNoStop [0]{.\EOS\space}%
\providecommand \EOS [0]{\spacefactor3000\relax}%
\providecommand \BibitemShut  [1]{\csname bibitem#1\endcsname}%
\let\auto@bib@innerbib\@empty
\bibitem [{\citenamefont {Bardeen}\ \emph {et~al.}(1967)\citenamefont {Bardeen}, \citenamefont {Baym},\ and\ \citenamefont {Pines}}]{BBP-1967}%
  \BibitemOpen
  \bibfield  {author} {\bibinfo {author} {\bibfnamefont {J.}~\bibnamefont {Bardeen}}, \bibinfo {author} {\bibfnamefont {G.}~\bibnamefont {Baym}},\ and\ \bibinfo {author} {\bibfnamefont {D.}~\bibnamefont {Pines}},\ }\bibfield  {title} {\bibinfo {title} {{Effective Interaction of $^{3}{\mathrm{He}}$ Atoms in Dilute Solutions of $^{3}{\mathrm{He}}$ in $^{4}{\mathrm{He}}$ at Low Temperatures}},\ }\href {https://doi.org/10.1103/PhysRev.156.207} {\bibfield  {journal} {\bibinfo  {journal} {Phys. Rev.}\ }\textbf {\bibinfo {volume} {156}},\ \bibinfo {pages} {207} (\bibinfo {year} {1967})}\BibitemShut {NoStop}%
\bibitem [{\citenamefont {Bardeen}\ \emph {et~al.}(1966)\citenamefont {Bardeen}, \citenamefont {Baym},\ and\ \citenamefont {Pines}}]{Bardeen-1966}%
  \BibitemOpen
  \bibfield  {author} {\bibinfo {author} {\bibfnamefont {J.}~\bibnamefont {Bardeen}}, \bibinfo {author} {\bibfnamefont {G.}~\bibnamefont {Baym}},\ and\ \bibinfo {author} {\bibfnamefont {D.}~\bibnamefont {Pines}},\ }\bibfield  {title} {\bibinfo {title} {{Interactions Between $^{3}$He Atoms in Dilute Solutions of $^{3}$He in Superfluid $^{4}$He}},\ }\href {https://doi.org/10.1103/PhysRevLett.17.372} {\bibfield  {journal} {\bibinfo  {journal} {Phys. Rev. Lett.}\ }\textbf {\bibinfo {volume} {17}},\ \bibinfo {pages} {372} (\bibinfo {year} {1966})}\BibitemShut {NoStop}%
\bibitem [{\citenamefont {Wheatley}(1968)}]{Wheatly-1968}%
  \BibitemOpen
  \bibfield  {author} {\bibinfo {author} {\bibfnamefont {J.~C.}\ \bibnamefont {Wheatley}},\ }\bibfield  {title} {\bibinfo {title} {{Dilute Solutions of $^{3}$He in $^{4}$He at Low Temperatures}},\ }\href {https://doi.org/10.1119/1.1974483} {\bibfield  {journal} {\bibinfo  {journal} {Am. J. Phys.}\ }\textbf {\bibinfo {volume} {36}},\ \bibinfo {pages} {181} (\bibinfo {year} {1968})}\BibitemShut {NoStop}%
\bibitem [{\citenamefont {Ebner}\ and\ \citenamefont {Edwards}(1971)}]{Edwards-1971}%
  \BibitemOpen
  \bibfield  {author} {\bibinfo {author} {\bibfnamefont {C.}~\bibnamefont {Ebner}}\ and\ \bibinfo {author} {\bibfnamefont {D.}~\bibnamefont {Edwards}},\ }\bibfield  {title} {\bibinfo {title} {{The low temperature thermodynamic properties of superfluid solutions of $^3$He in $^4$He}},\ }\href {https://doi.org/https://doi.org/10.1016/0370-1573(71)90003-2} {\bibfield  {journal} {\bibinfo  {journal} {Phys. Rep.}\ }\textbf {\bibinfo {volume} {2}},\ \bibinfo {pages} {77} (\bibinfo {year} {1971})}\BibitemShut {NoStop}%
\bibitem [{\citenamefont {Cohen}(1977)}]{Cohen-1977}%
  \BibitemOpen
  \bibfield  {author} {\bibinfo {author} {\bibfnamefont {E.~G.~D.}\ \bibnamefont {Cohen}},\ }\bibfield  {title} {\bibinfo {title} {{Quantum Statistics and Liquid Helium-3—Helium-4 Mixtures}},\ }\href {https://doi.org/10.1126/science.197.4298.11} {\bibfield  {journal} {\bibinfo  {journal} {Science}\ }\textbf {\bibinfo {volume} {197}},\ \bibinfo {pages} {11} (\bibinfo {year} {1977})}\BibitemShut {NoStop}%
\bibitem [{\citenamefont {Krotscheck}\ and\ \citenamefont {Saarela}(1993)}]{krotscheck-1993}%
  \BibitemOpen
  \bibfield  {author} {\bibinfo {author} {\bibfnamefont {E.}~\bibnamefont {Krotscheck}}\ and\ \bibinfo {author} {\bibfnamefont {M.}~\bibnamefont {Saarela}},\ }\bibfield  {title} {\bibinfo {title} {{Theory of $^3$He-$^4$He mixtures: energetics, structure, and stability}},\ }\href {https://doi.org/https://doi.org/10.1016/0370-1573(93)90004-W} {\bibfield  {journal} {\bibinfo  {journal} {Phys. Rep.}\ }\textbf {\bibinfo {volume} {232}},\ \bibinfo {pages} {1} (\bibinfo {year} {1993})}\BibitemShut {NoStop}%
\bibitem [{\citenamefont {Boronat}\ \emph {et~al.}(1989)\citenamefont {Boronat}, \citenamefont {Fabrocini},\ and\ \citenamefont {Polls}}]{Boronat-1989}%
  \BibitemOpen
  \bibfield  {author} {\bibinfo {author} {\bibfnamefont {J.}~\bibnamefont {Boronat}}, \bibinfo {author} {\bibfnamefont {A.}~\bibnamefont {Fabrocini}},\ and\ \bibinfo {author} {\bibfnamefont {A.}~\bibnamefont {Polls}},\ }\bibfield  {title} {\bibinfo {title} {{Variational calculation of the binding energy of one $^3$He impurity in liquid $^4$He}},\ }\href {https://doi.org/10.1007/BF00683379} {\bibfield  {journal} {\bibinfo  {journal} {J. Low. temp. Phys.}\ }\textbf {\bibinfo {volume} {74}},\ \bibinfo {pages} {347} (\bibinfo {year} {1989})}\BibitemShut {NoStop}%
\bibitem [{\citenamefont {Arias~de Saavedra}\ \emph {et~al.}(1994)\citenamefont {Arias~de Saavedra}, \citenamefont {Boronat}, \citenamefont {Polls},\ and\ \citenamefont {Fabrocini}}]{Boronat-1994}%
  \BibitemOpen
  \bibfield  {author} {\bibinfo {author} {\bibfnamefont {F.}~\bibnamefont {Arias~de Saavedra}}, \bibinfo {author} {\bibfnamefont {J.}~\bibnamefont {Boronat}}, \bibinfo {author} {\bibfnamefont {A.}~\bibnamefont {Polls}},\ and\ \bibinfo {author} {\bibfnamefont {A.}~\bibnamefont {Fabrocini}},\ }\bibfield  {title} {\bibinfo {title} {Effective mass of one $^{4}\mathrm{He}$ atom in liquid $^{3}\mathrm{He}$},\ }\href {https://doi.org/10.1103/PhysRevB.50.4248} {\bibfield  {journal} {\bibinfo  {journal} {Phys. Rev. B}\ }\textbf {\bibinfo {volume} {50}},\ \bibinfo {pages} {4248} (\bibinfo {year} {1994})}\BibitemShut {NoStop}%
\bibitem [{\citenamefont {Krotscheck}\ \emph {et~al.}(1988)\citenamefont {Krotscheck}, \citenamefont {Saarela},\ and\ \citenamefont {Epstein}}]{Krotscheck-1988}%
  \BibitemOpen
  \bibfield  {author} {\bibinfo {author} {\bibfnamefont {E.}~\bibnamefont {Krotscheck}}, \bibinfo {author} {\bibfnamefont {M.}~\bibnamefont {Saarela}},\ and\ \bibinfo {author} {\bibfnamefont {J.~L.}\ \bibnamefont {Epstein}},\ }\bibfield  {title} {\bibinfo {title} {{Impurity states in liquid-helium films}},\ }\href {https://doi.org/10.1103/PhysRevB.38.111} {\bibfield  {journal} {\bibinfo  {journal} {Phys. Rev. B}\ }\textbf {\bibinfo {volume} {38}},\ \bibinfo {pages} {111} (\bibinfo {year} {1988})}\BibitemShut {NoStop}%
\bibitem [{\citenamefont {Graf}\ \emph {et~al.}(1967)\citenamefont {Graf}, \citenamefont {Lee},\ and\ \citenamefont {Reppy}}]{Reppy-1967}%
  \BibitemOpen
  \bibfield  {author} {\bibinfo {author} {\bibfnamefont {E.~H.}\ \bibnamefont {Graf}}, \bibinfo {author} {\bibfnamefont {D.~M.}\ \bibnamefont {Lee}},\ and\ \bibinfo {author} {\bibfnamefont {J.~D.}\ \bibnamefont {Reppy}},\ }\bibfield  {title} {\bibinfo {title} {{Phase Separation and the Superfluid Transition in Liquid $^{3}$He-$^{4}$He Mixtures}},\ }\href {https://doi.org/10.1103/PhysRevLett.19.417} {\bibfield  {journal} {\bibinfo  {journal} {Phys. Rev. Lett.}\ }\textbf {\bibinfo {volume} {19}},\ \bibinfo {pages} {417} (\bibinfo {year} {1967})}\BibitemShut {NoStop}%
\bibitem [{\citenamefont {Edwards}\ and\ \citenamefont {Daunt}(1961)}]{Edwards-1961}%
  \BibitemOpen
  \bibfield  {author} {\bibinfo {author} {\bibfnamefont {D.~O.}\ \bibnamefont {Edwards}}\ and\ \bibinfo {author} {\bibfnamefont {J.~G.}\ \bibnamefont {Daunt}},\ }\bibfield  {title} {\bibinfo {title} {{Phase Separation in $^{3}{\mathrm{He}}$-$^{4}{\mathrm{He}}$ Mixtures near Absolute Zero}},\ }\href {https://doi.org/10.1103/PhysRev.124.640} {\bibfield  {journal} {\bibinfo  {journal} {Phys. Rev.}\ }\textbf {\bibinfo {volume} {124}},\ \bibinfo {pages} {640} (\bibinfo {year} {1961})}\BibitemShut {NoStop}%
\bibitem [{\citenamefont {Kagan}(2013)}]{Kagan-2013}%
  \BibitemOpen
  \bibfield  {author} {\bibinfo {author} {\bibfnamefont {M.~Y.}\ \bibnamefont {Kagan}},\ }\href {https://doi.org/10.1007/978-94-007-6961-8} {\emph {\bibinfo {title} {{Modern trends in Superconductivity and Superfluidity}}}}\ (\bibinfo  {publisher} {Springer Dordrecht},\ \bibinfo {year} {2013})\BibitemShut {NoStop}%
\bibitem [{\citenamefont {Iachello}\ and\ \citenamefont {van Isacker}(1991)}]{Iachello-1991}%
  \BibitemOpen
  \bibfield  {author} {\bibinfo {author} {\bibfnamefont {F.}~\bibnamefont {Iachello}}\ and\ \bibinfo {author} {\bibfnamefont {P.}~\bibnamefont {van Isacker}},\ }\href@noop {} {\emph {\bibinfo {title} {{The interacting boson-fermion model}}}}\ (\bibinfo  {publisher} {Cambridge University Press},\ \bibinfo {year} {1991})\BibitemShut {NoStop}%
\bibitem [{\citenamefont {Maeda}\ \emph {et~al.}(2009)\citenamefont {Maeda}, \citenamefont {Baym},\ and\ \citenamefont {Hatsuda}}]{Maeda-2009}%
  \BibitemOpen
  \bibfield  {author} {\bibinfo {author} {\bibfnamefont {K.}~\bibnamefont {Maeda}}, \bibinfo {author} {\bibfnamefont {G.}~\bibnamefont {Baym}},\ and\ \bibinfo {author} {\bibfnamefont {T.}~\bibnamefont {Hatsuda}},\ }\bibfield  {title} {\bibinfo {title} {{Simulating Dense QCD Matter with Ultracold Atomic Boson-Fermion Mixtures}},\ }\href {https://doi.org/10.1103/PhysRevLett.103.085301} {\bibfield  {journal} {\bibinfo  {journal} {Phys. Rev. Lett.}\ }\textbf {\bibinfo {volume} {103}},\ \bibinfo {pages} {085301} (\bibinfo {year} {2009})}\BibitemShut {NoStop}%
\bibitem [{\citenamefont {Stewart}(2017)}]{Stewart-2017}%
  \BibitemOpen
  \bibfield  {author} {\bibinfo {author} {\bibfnamefont {G.~R.}\ \bibnamefont {Stewart}},\ }\bibfield  {title} {\bibinfo {title} {{Unconventional superconductivity}},\ }\href {https://doi.org/10.1080/00018732.2017.1331615} {\bibfield  {journal} {\bibinfo  {journal} {Advances in Physics}\ }\textbf {\bibinfo {volume} {66}},\ \bibinfo {pages} {75} (\bibinfo {year} {2017})}\BibitemShut {NoStop}%
\bibitem [{\citenamefont {Zhang}\ \emph {et~al.}(2024{\natexlab{a}})\citenamefont {Zhang}, \citenamefont {Guo}, \citenamefont {Tajima},\ and\ \citenamefont {Liang}}]{Tajima-2024}%
  \BibitemOpen
  \bibfield  {author} {\bibinfo {author} {\bibfnamefont {T.}~\bibnamefont {Zhang}}, \bibinfo {author} {\bibfnamefont {Y.}~\bibnamefont {Guo}}, \bibinfo {author} {\bibfnamefont {H.}~\bibnamefont {Tajima}},\ and\ \bibinfo {author} {\bibfnamefont {H.}~\bibnamefont {Liang}},\ }\bibfield  {title} {\bibinfo {title} {{Probing the goldstino excitation through tunneling transport in a Bose-Fermi mixture with explicitly broken supersymmetry}},\ }\href {https://doi.org/10.1103/PhysRevB.110.064512} {\bibfield  {journal} {\bibinfo  {journal} {Phys. Rev. B}\ }\textbf {\bibinfo {volume} {110}},\ \bibinfo {pages} {064512} (\bibinfo {year} {2024}{\natexlab{a}})}\BibitemShut {NoStop}%
\bibitem [{\citenamefont {Bloch}\ \emph {et~al.}(2008)\citenamefont {Bloch}, \citenamefont {Dalibard},\ and\ \citenamefont {Zwerger}}]{Bloch-2008}%
  \BibitemOpen
  \bibfield  {author} {\bibinfo {author} {\bibfnamefont {I.}~\bibnamefont {Bloch}}, \bibinfo {author} {\bibfnamefont {J.}~\bibnamefont {Dalibard}},\ and\ \bibinfo {author} {\bibfnamefont {W.}~\bibnamefont {Zwerger}},\ }\bibfield  {title} {\bibinfo {title} {Many-body physics with ultracold gases},\ }\href {https://doi.org/10.1103/RevModPhys.80.885} {\bibfield  {journal} {\bibinfo  {journal} {Rev. Mod. Phys.}\ }\textbf {\bibinfo {volume} {80}},\ \bibinfo {pages} {885} (\bibinfo {year} {2008})}\BibitemShut {NoStop}%
\bibitem [{\citenamefont {Chin}\ \emph {et~al.}(2010)\citenamefont {Chin}, \citenamefont {Grimm}, \citenamefont {Julienne},\ and\ \citenamefont {Tiesinga}}]{Grimm-2010}%
  \BibitemOpen
  \bibfield  {author} {\bibinfo {author} {\bibfnamefont {C.}~\bibnamefont {Chin}}, \bibinfo {author} {\bibfnamefont {R.}~\bibnamefont {Grimm}}, \bibinfo {author} {\bibfnamefont {P.}~\bibnamefont {Julienne}},\ and\ \bibinfo {author} {\bibfnamefont {E.}~\bibnamefont {Tiesinga}},\ }\bibfield  {title} {\bibinfo {title} {{Feshbach resonances in ultracold gases}},\ }\href {https://doi.org/10.1103/RevModPhys.82.1225} {\bibfield  {journal} {\bibinfo  {journal} {Rev. Mod. Phys.}\ }\textbf {\bibinfo {volume} {82}},\ \bibinfo {pages} {1225} (\bibinfo {year} {2010})}\BibitemShut {NoStop}%
\bibitem [{\citenamefont {Baroni}\ \emph {et~al.}(2024{\natexlab{a}})\citenamefont {Baroni}, \citenamefont {Lamporesi},\ and\ \citenamefont {Zaccanti}}]{Zaccanti-2024}%
  \BibitemOpen
  \bibfield  {author} {\bibinfo {author} {\bibfnamefont {C.}~\bibnamefont {Baroni}}, \bibinfo {author} {\bibfnamefont {G.}~\bibnamefont {Lamporesi}},\ and\ \bibinfo {author} {\bibfnamefont {M.}~\bibnamefont {Zaccanti}},\ }\bibfield  {title} {\bibinfo {title} {{Quantum mixtures of ultracold gases of neutral atoms}},\ }\href {https://doi.org/10.1038/s42254-024-00773-6} {\bibfield  {journal} {\bibinfo  {journal} {Nat. Rev. Phys.}\ }\textbf {\bibinfo {volume} {6}},\ \bibinfo {pages} {736} (\bibinfo {year} {2024}{\natexlab{a}})}\BibitemShut {NoStop}%
\bibitem [{\citenamefont {Ni}\ \emph {et~al.}(2008)\citenamefont {Ni}, \citenamefont {Ospelkaus}, \citenamefont {de~Miranda}, \citenamefont {Pe'er}, \citenamefont {Neyenhuis}, \citenamefont {Zirbel}, \citenamefont {Kotochigova}, \citenamefont {Julienne}, \citenamefont {Jin},\ and\ \citenamefont {Ye}}]{Jin-2008}%
  \BibitemOpen
  \bibfield  {author} {\bibinfo {author} {\bibfnamefont {K.-K.}\ \bibnamefont {Ni}}, \bibinfo {author} {\bibfnamefont {S.}~\bibnamefont {Ospelkaus}}, \bibinfo {author} {\bibfnamefont {M.~H.~G.}\ \bibnamefont {de~Miranda}}, \bibinfo {author} {\bibfnamefont {A.}~\bibnamefont {Pe'er}}, \bibinfo {author} {\bibfnamefont {B.}~\bibnamefont {Neyenhuis}}, \bibinfo {author} {\bibfnamefont {J.~J.}\ \bibnamefont {Zirbel}}, \bibinfo {author} {\bibfnamefont {S.}~\bibnamefont {Kotochigova}}, \bibinfo {author} {\bibfnamefont {P.~S.}\ \bibnamefont {Julienne}}, \bibinfo {author} {\bibfnamefont {D.~S.}\ \bibnamefont {Jin}},\ and\ \bibinfo {author} {\bibfnamefont {J.}~\bibnamefont {Ye}},\ }\bibfield  {title} {\bibinfo {title} {{A High Phase-Space-Density Gas of Polar Molecules}},\ }\href {https://doi.org/10.1126/science.1163861} {\bibfield  {journal} {\bibinfo  {journal} {Science}\ }\textbf {\bibinfo {volume} {322}},\ \bibinfo {pages} {231} (\bibinfo {year} {2008})}\BibitemShut {NoStop}%
\bibitem [{\citenamefont {Wu}\ \emph {et~al.}(2011)\citenamefont {Wu}, \citenamefont {Santiago}, \citenamefont {Park}, \citenamefont {Ahmadi},\ and\ \citenamefont {Zwierlein}}]{Zwierlein-2011}%
  \BibitemOpen
  \bibfield  {author} {\bibinfo {author} {\bibfnamefont {C.-H.}\ \bibnamefont {Wu}}, \bibinfo {author} {\bibfnamefont {I.}~\bibnamefont {Santiago}}, \bibinfo {author} {\bibfnamefont {J.~W.}\ \bibnamefont {Park}}, \bibinfo {author} {\bibfnamefont {P.}~\bibnamefont {Ahmadi}},\ and\ \bibinfo {author} {\bibfnamefont {M.~W.}\ \bibnamefont {Zwierlein}},\ }\bibfield  {title} {\bibinfo {title} {{Strongly interacting isotopic Bose-Fermi mixture immersed in a Fermi sea}},\ }\href {https://doi.org/10.1103/PhysRevA.84.011601} {\bibfield  {journal} {\bibinfo  {journal} {Phys. Rev. A}\ }\textbf {\bibinfo {volume} {84}},\ \bibinfo {pages} {011601} (\bibinfo {year} {2011})}\BibitemShut {NoStop}%
\bibitem [{\citenamefont {Wu}\ \emph {et~al.}(2012)\citenamefont {Wu}, \citenamefont {Park}, \citenamefont {Ahmadi}, \citenamefont {Will},\ and\ \citenamefont {Zwierlein}}]{Zwierlein-2012}%
  \BibitemOpen
  \bibfield  {author} {\bibinfo {author} {\bibfnamefont {C.-H.}\ \bibnamefont {Wu}}, \bibinfo {author} {\bibfnamefont {J.~W.}\ \bibnamefont {Park}}, \bibinfo {author} {\bibfnamefont {P.}~\bibnamefont {Ahmadi}}, \bibinfo {author} {\bibfnamefont {S.}~\bibnamefont {Will}},\ and\ \bibinfo {author} {\bibfnamefont {M.~W.}\ \bibnamefont {Zwierlein}},\ }\bibfield  {title} {\bibinfo {title} {{Ultracold Fermionic Feshbach Molecules of $^{23}\mathrm{Na}^{40}\mathbf{K}$}},\ }\href {https://doi.org/10.1103/PhysRevLett.109.085301} {\bibfield  {journal} {\bibinfo  {journal} {Phys. Rev. Lett.}\ }\textbf {\bibinfo {volume} {109}},\ \bibinfo {pages} {085301} (\bibinfo {year} {2012})}\BibitemShut {NoStop}%
\bibitem [{\citenamefont {Park}\ \emph {et~al.}(2012)\citenamefont {Park}, \citenamefont {Wu}, \citenamefont {Santiago}, \citenamefont {Tiecke}, \citenamefont {Will}, \citenamefont {Ahmadi},\ and\ \citenamefont {Zwierlein}}]{Zwierlein-2012a}%
  \BibitemOpen
  \bibfield  {author} {\bibinfo {author} {\bibfnamefont {J.~W.}\ \bibnamefont {Park}}, \bibinfo {author} {\bibfnamefont {C.-H.}\ \bibnamefont {Wu}}, \bibinfo {author} {\bibfnamefont {I.}~\bibnamefont {Santiago}}, \bibinfo {author} {\bibfnamefont {T.~G.}\ \bibnamefont {Tiecke}}, \bibinfo {author} {\bibfnamefont {S.}~\bibnamefont {Will}}, \bibinfo {author} {\bibfnamefont {P.}~\bibnamefont {Ahmadi}},\ and\ \bibinfo {author} {\bibfnamefont {M.~W.}\ \bibnamefont {Zwierlein}},\ }\bibfield  {title} {\bibinfo {title} {{Quantum degenerate Bose-Fermi mixture of chemically different atomic species with widely tunable interactions}},\ }\href {https://doi.org/10.1103/PhysRevA.85.051602} {\bibfield  {journal} {\bibinfo  {journal} {Phys. Rev. A}\ }\textbf {\bibinfo {volume} {85}},\ \bibinfo {pages} {051602} (\bibinfo {year} {2012})}\BibitemShut {NoStop}%
\bibitem [{\citenamefont {Heo}\ \emph {et~al.}(2012)\citenamefont {Heo}, \citenamefont {Wang}, \citenamefont {Christensen}, \citenamefont {Rvachov}, \citenamefont {Cotta}, \citenamefont {Choi}, \citenamefont {Lee},\ and\ \citenamefont {Ketterle}}]{Ketterle-2012}%
  \BibitemOpen
  \bibfield  {author} {\bibinfo {author} {\bibfnamefont {M.-S.}\ \bibnamefont {Heo}}, \bibinfo {author} {\bibfnamefont {T.~T.}\ \bibnamefont {Wang}}, \bibinfo {author} {\bibfnamefont {C.~A.}\ \bibnamefont {Christensen}}, \bibinfo {author} {\bibfnamefont {T.~M.}\ \bibnamefont {Rvachov}}, \bibinfo {author} {\bibfnamefont {D.~A.}\ \bibnamefont {Cotta}}, \bibinfo {author} {\bibfnamefont {J.-H.}\ \bibnamefont {Choi}}, \bibinfo {author} {\bibfnamefont {Y.-R.}\ \bibnamefont {Lee}},\ and\ \bibinfo {author} {\bibfnamefont {W.}~\bibnamefont {Ketterle}},\ }\bibfield  {title} {\bibinfo {title} {{Formation of ultracold fermionic NaLi Feshbach molecules}},\ }\href {https://doi.org/10.1103/PhysRevA.86.021602} {\bibfield  {journal} {\bibinfo  {journal} {Phys. Rev. A}\ }\textbf {\bibinfo {volume} {86}},\ \bibinfo {pages} {021602} (\bibinfo {year} {2012})}\BibitemShut {NoStop}%
\bibitem [{\citenamefont {Cumby}\ \emph {et~al.}(2013)\citenamefont {Cumby}, \citenamefont {Shewmon}, \citenamefont {Hu}, \citenamefont {Perreault},\ and\ \citenamefont {Jin}}]{Jin-2013}%
  \BibitemOpen
  \bibfield  {author} {\bibinfo {author} {\bibfnamefont {T.~D.}\ \bibnamefont {Cumby}}, \bibinfo {author} {\bibfnamefont {R.~A.}\ \bibnamefont {Shewmon}}, \bibinfo {author} {\bibfnamefont {M.-G.}\ \bibnamefont {Hu}}, \bibinfo {author} {\bibfnamefont {J.~D.}\ \bibnamefont {Perreault}},\ and\ \bibinfo {author} {\bibfnamefont {D.~S.}\ \bibnamefont {Jin}},\ }\bibfield  {title} {\bibinfo {title} {{Feshbach-molecule formation in a Bose-Fermi mixture}},\ }\href {https://doi.org/10.1103/PhysRevA.87.012703} {\bibfield  {journal} {\bibinfo  {journal} {Phys. Rev. A}\ }\textbf {\bibinfo {volume} {87}},\ \bibinfo {pages} {012703} (\bibinfo {year} {2013})}\BibitemShut {NoStop}%
\bibitem [{\citenamefont {Bloom}\ \emph {et~al.}(2013)\citenamefont {Bloom}, \citenamefont {Hu}, \citenamefont {Cumby},\ and\ \citenamefont {Jin}}]{Jin-2013a}%
  \BibitemOpen
  \bibfield  {author} {\bibinfo {author} {\bibfnamefont {R.~S.}\ \bibnamefont {Bloom}}, \bibinfo {author} {\bibfnamefont {M.-G.}\ \bibnamefont {Hu}}, \bibinfo {author} {\bibfnamefont {T.~D.}\ \bibnamefont {Cumby}},\ and\ \bibinfo {author} {\bibfnamefont {D.~S.}\ \bibnamefont {Jin}},\ }\bibfield  {title} {\bibinfo {title} {{Tests of Universal Three-Body Physics in an Ultracold Bose-Fermi Mixture}},\ }\href {https://doi.org/10.1103/PhysRevLett.111.105301} {\bibfield  {journal} {\bibinfo  {journal} {Phys. Rev. Lett.}\ }\textbf {\bibinfo {volume} {111}},\ \bibinfo {pages} {105301} (\bibinfo {year} {2013})}\BibitemShut {NoStop}%
\bibitem [{\citenamefont {Vaidya}\ \emph {et~al.}(2015)\citenamefont {Vaidya}, \citenamefont {Tiamsuphat}, \citenamefont {Rolston},\ and\ \citenamefont {Porto}}]{Porto-2015}%
  \BibitemOpen
  \bibfield  {author} {\bibinfo {author} {\bibfnamefont {V.~D.}\ \bibnamefont {Vaidya}}, \bibinfo {author} {\bibfnamefont {J.}~\bibnamefont {Tiamsuphat}}, \bibinfo {author} {\bibfnamefont {S.~L.}\ \bibnamefont {Rolston}},\ and\ \bibinfo {author} {\bibfnamefont {J.~V.}\ \bibnamefont {Porto}},\ }\bibfield  {title} {\bibinfo {title} {{Degenerate Bose-Fermi mixtures of rubidium and ytterbium}},\ }\href {https://doi.org/10.1103/PhysRevA.92.043604} {\bibfield  {journal} {\bibinfo  {journal} {Phys. Rev. A}\ }\textbf {\bibinfo {volume} {92}},\ \bibinfo {pages} {043604} (\bibinfo {year} {2015})}\BibitemShut {NoStop}%
\bibitem [{\citenamefont {Ferrier-Barbut}\ \emph {et~al.}(2014)\citenamefont {Ferrier-Barbut}, \citenamefont {Delehaye}, \citenamefont {Laurent}, \citenamefont {Grier}, \citenamefont {Pierce}, \citenamefont {Rem}, \citenamefont {Chevy},\ and\ \citenamefont {Salomon}}]{Salomon-2014}%
  \BibitemOpen
  \bibfield  {author} {\bibinfo {author} {\bibfnamefont {I.}~\bibnamefont {Ferrier-Barbut}}, \bibinfo {author} {\bibfnamefont {M.}~\bibnamefont {Delehaye}}, \bibinfo {author} {\bibfnamefont {S.}~\bibnamefont {Laurent}}, \bibinfo {author} {\bibfnamefont {A.~T.}\ \bibnamefont {Grier}}, \bibinfo {author} {\bibfnamefont {M.}~\bibnamefont {Pierce}}, \bibinfo {author} {\bibfnamefont {B.~S.}\ \bibnamefont {Rem}}, \bibinfo {author} {\bibfnamefont {F.}~\bibnamefont {Chevy}},\ and\ \bibinfo {author} {\bibfnamefont {C.}~\bibnamefont {Salomon}},\ }\bibfield  {title} {\bibinfo {title} {{A mixture of Bose and Fermi superfluids}},\ }\href {https://doi.org/10.1126/science.1255380} {\bibfield  {journal} {\bibinfo  {journal} {Science}\ }\textbf {\bibinfo {volume} {345}},\ \bibinfo {pages} {1035} (\bibinfo {year} {2014})}\BibitemShut {NoStop}%
\bibitem [{\citenamefont {Delehaye}\ \emph {et~al.}(2015)\citenamefont {Delehaye}, \citenamefont {Laurent}, \citenamefont {Ferrier-Barbut}, \citenamefont {Jin}, \citenamefont {Chevy},\ and\ \citenamefont {Salomon}}]{Salomon-2015}%
  \BibitemOpen
  \bibfield  {author} {\bibinfo {author} {\bibfnamefont {M.}~\bibnamefont {Delehaye}}, \bibinfo {author} {\bibfnamefont {S.}~\bibnamefont {Laurent}}, \bibinfo {author} {\bibfnamefont {I.}~\bibnamefont {Ferrier-Barbut}}, \bibinfo {author} {\bibfnamefont {S.}~\bibnamefont {Jin}}, \bibinfo {author} {\bibfnamefont {F.}~\bibnamefont {Chevy}},\ and\ \bibinfo {author} {\bibfnamefont {C.}~\bibnamefont {Salomon}},\ }\bibfield  {title} {\bibinfo {title} {{Critical Velocity and Dissipation of an Ultracold Bose-Fermi Counterflow}},\ }\href {https://doi.org/10.1103/PhysRevLett.115.265303} {\bibfield  {journal} {\bibinfo  {journal} {Phys. Rev. Lett.}\ }\textbf {\bibinfo {volume} {115}},\ \bibinfo {pages} {265303} (\bibinfo {year} {2015})}\BibitemShut {NoStop}%
\bibitem [{\citenamefont {Hu}\ \emph {et~al.}(2016)\citenamefont {Hu}, \citenamefont {Van~de Graaff}, \citenamefont {Kedar}, \citenamefont {Corson}, \citenamefont {Cornell},\ and\ \citenamefont {Jin}}]{Jin-2016}%
  \BibitemOpen
  \bibfield  {author} {\bibinfo {author} {\bibfnamefont {M.-G.}\ \bibnamefont {Hu}}, \bibinfo {author} {\bibfnamefont {M.~J.}\ \bibnamefont {Van~de Graaff}}, \bibinfo {author} {\bibfnamefont {D.}~\bibnamefont {Kedar}}, \bibinfo {author} {\bibfnamefont {J.~P.}\ \bibnamefont {Corson}}, \bibinfo {author} {\bibfnamefont {E.~A.}\ \bibnamefont {Cornell}},\ and\ \bibinfo {author} {\bibfnamefont {D.~S.}\ \bibnamefont {Jin}},\ }\bibfield  {title} {\bibinfo {title} {{Bose Polarons in the Strongly Interacting Regime}},\ }\href {https://doi.org/10.1103/PhysRevLett.117.055301} {\bibfield  {journal} {\bibinfo  {journal} {Phys. Rev. Lett.}\ }\textbf {\bibinfo {volume} {117}},\ \bibinfo {pages} {055301} (\bibinfo {year} {2016})}\BibitemShut {NoStop}%
\bibitem [{\citenamefont {Wu}\ \emph {et~al.}(2017)\citenamefont {Wu}, \citenamefont {Yao}, \citenamefont {Chen}, \citenamefont {Liu}, \citenamefont {Wang}, \citenamefont {Chen},\ and\ \citenamefont {Pan}}]{Pan-2017}%
  \BibitemOpen
  \bibfield  {author} {\bibinfo {author} {\bibfnamefont {Y.-P.}\ \bibnamefont {Wu}}, \bibinfo {author} {\bibfnamefont {X.-C.}\ \bibnamefont {Yao}}, \bibinfo {author} {\bibfnamefont {H.-Z.}\ \bibnamefont {Chen}}, \bibinfo {author} {\bibfnamefont {X.-P.}\ \bibnamefont {Liu}}, \bibinfo {author} {\bibfnamefont {X.-Q.}\ \bibnamefont {Wang}}, \bibinfo {author} {\bibfnamefont {Y.-A.}\ \bibnamefont {Chen}},\ and\ \bibinfo {author} {\bibfnamefont {J.-W.}\ \bibnamefont {Pan}},\ }\bibfield  {title} {\bibinfo {title} {{A quantum degenerate Bose-Fermi mixture of 41K and 6Li}},\ }\href {https://doi.org/10.1088/1361-6455/aa658b} {\bibfield  {journal} {\bibinfo  {journal} {J. Phys. B At. Mol. Opt. Phys.}\ }\textbf {\bibinfo {volume} {50}},\ \bibinfo {pages} {094001} (\bibinfo {year} {2017})}\BibitemShut {NoStop}%
\bibitem [{\citenamefont {Roy}\ \emph {et~al.}(2017)\citenamefont {Roy}, \citenamefont {Green}, \citenamefont {Bowler},\ and\ \citenamefont {Gupta}}]{Gupta-2017}%
  \BibitemOpen
  \bibfield  {author} {\bibinfo {author} {\bibfnamefont {R.}~\bibnamefont {Roy}}, \bibinfo {author} {\bibfnamefont {A.}~\bibnamefont {Green}}, \bibinfo {author} {\bibfnamefont {R.}~\bibnamefont {Bowler}},\ and\ \bibinfo {author} {\bibfnamefont {S.}~\bibnamefont {Gupta}},\ }\bibfield  {title} {\bibinfo {title} {{Two-Element Mixture of Bose and Fermi Superfluids}},\ }\href {https://doi.org/10.1103/PhysRevLett.118.055301} {\bibfield  {journal} {\bibinfo  {journal} {Phys. Rev. Lett.}\ }\textbf {\bibinfo {volume} {118}},\ \bibinfo {pages} {055301} (\bibinfo {year} {2017})}\BibitemShut {NoStop}%
\bibitem [{\citenamefont {DeSalvo}\ \emph {et~al.}(2017)\citenamefont {DeSalvo}, \citenamefont {Patel}, \citenamefont {Johansen},\ and\ \citenamefont {Chin}}]{DeSalvo-2017}%
  \BibitemOpen
  \bibfield  {author} {\bibinfo {author} {\bibfnamefont {B.~J.}\ \bibnamefont {DeSalvo}}, \bibinfo {author} {\bibfnamefont {K.}~\bibnamefont {Patel}}, \bibinfo {author} {\bibfnamefont {J.}~\bibnamefont {Johansen}},\ and\ \bibinfo {author} {\bibfnamefont {C.}~\bibnamefont {Chin}},\ }\bibfield  {title} {\bibinfo {title} {{Observation of a Degenerate Fermi Gas Trapped by a Bose-Einstein Condensate}},\ }\href {https://doi.org/10.1103/PhysRevLett.119.233401} {\bibfield  {journal} {\bibinfo  {journal} {Phys. Rev. Lett.}\ }\textbf {\bibinfo {volume} {119}},\ \bibinfo {pages} {233401} (\bibinfo {year} {2017})}\BibitemShut {NoStop}%
\bibitem [{\citenamefont {Trautmann}\ \emph {et~al.}(2018)\citenamefont {Trautmann}, \citenamefont {Ilzh\"ofer}, \citenamefont {Durastante}, \citenamefont {Politi}, \citenamefont {Sohmen}, \citenamefont {Mark},\ and\ \citenamefont {Ferlaino}}]{Ferlaino-2018}%
  \BibitemOpen
  \bibfield  {author} {\bibinfo {author} {\bibfnamefont {A.}~\bibnamefont {Trautmann}}, \bibinfo {author} {\bibfnamefont {P.}~\bibnamefont {Ilzh\"ofer}}, \bibinfo {author} {\bibfnamefont {G.}~\bibnamefont {Durastante}}, \bibinfo {author} {\bibfnamefont {C.}~\bibnamefont {Politi}}, \bibinfo {author} {\bibfnamefont {M.}~\bibnamefont {Sohmen}}, \bibinfo {author} {\bibfnamefont {M.~J.}\ \bibnamefont {Mark}},\ and\ \bibinfo {author} {\bibfnamefont {F.}~\bibnamefont {Ferlaino}},\ }\bibfield  {title} {\bibinfo {title} {{Dipolar Quantum Mixtures of Erbium and Dysprosium Atoms}},\ }\href {https://doi.org/10.1103/PhysRevLett.121.213601} {\bibfield  {journal} {\bibinfo  {journal} {Phys. Rev. Lett.}\ }\textbf {\bibinfo {volume} {121}},\ \bibinfo {pages} {213601} (\bibinfo {year} {2018})}\BibitemShut {NoStop}%
\bibitem [{\citenamefont {Lous}\ \emph {et~al.}(2018{\natexlab{a}})\citenamefont {Lous}, \citenamefont {Fritsche}, \citenamefont {Jag}, \citenamefont {Lehmann}, \citenamefont {Kirilov}, \citenamefont {Huang},\ and\ \citenamefont {Grimm}}]{Grimm-2018}%
  \BibitemOpen
  \bibfield  {author} {\bibinfo {author} {\bibfnamefont {R.~S.}\ \bibnamefont {Lous}}, \bibinfo {author} {\bibfnamefont {I.}~\bibnamefont {Fritsche}}, \bibinfo {author} {\bibfnamefont {M.}~\bibnamefont {Jag}}, \bibinfo {author} {\bibfnamefont {F.}~\bibnamefont {Lehmann}}, \bibinfo {author} {\bibfnamefont {E.}~\bibnamefont {Kirilov}}, \bibinfo {author} {\bibfnamefont {B.}~\bibnamefont {Huang}},\ and\ \bibinfo {author} {\bibfnamefont {R.}~\bibnamefont {Grimm}},\ }\bibfield  {title} {\bibinfo {title} {{Probing the Interface of a Phase-Separated State in a Repulsive Bose-Fermi Mixture}},\ }\href {https://doi.org/10.1103/PhysRevLett.120.243403} {\bibfield  {journal} {\bibinfo  {journal} {Phys. Rev. Lett.}\ }\textbf {\bibinfo {volume} {120}},\ \bibinfo {pages} {243403} (\bibinfo {year} {2018}{\natexlab{a}})}\BibitemShut {NoStop}%
\bibitem [{\citenamefont {DeSalvo}\ \emph {et~al.}(2019)\citenamefont {DeSalvo}, \citenamefont {Patel}, \citenamefont {Cai},\ and\ \citenamefont {Chin}}]{DeSalVo-2019}%
  \BibitemOpen
  \bibfield  {author} {\bibinfo {author} {\bibfnamefont {B.~J.}\ \bibnamefont {DeSalvo}}, \bibinfo {author} {\bibfnamefont {K.}~\bibnamefont {Patel}}, \bibinfo {author} {\bibfnamefont {G.}~\bibnamefont {Cai}},\ and\ \bibinfo {author} {\bibfnamefont {C.}~\bibnamefont {Chin}},\ }\bibfield  {title} {\bibinfo {title} {Observation of fermion-mediated interactions between bosonic atoms},\ }\href {https://doi.org/10.1038/s41586-019-1055-0} {\bibfield  {journal} {\bibinfo  {journal} {Nature}\ }\textbf {\bibinfo {volume} {568}},\ \bibinfo {pages} {61} (\bibinfo {year} {2019})}\BibitemShut {NoStop}%
\bibitem [{\citenamefont {Marco}\ \emph {et~al.}(2019)\citenamefont {Marco}, \citenamefont {Valtolina}, \citenamefont {Matsuda}, \citenamefont {Tobias}, \citenamefont {Covey},\ and\ \citenamefont {Ye}}]{Valtolina-2019}%
  \BibitemOpen
  \bibfield  {author} {\bibinfo {author} {\bibfnamefont {L.~D.}\ \bibnamefont {Marco}}, \bibinfo {author} {\bibfnamefont {G.}~\bibnamefont {Valtolina}}, \bibinfo {author} {\bibfnamefont {K.}~\bibnamefont {Matsuda}}, \bibinfo {author} {\bibfnamefont {W.~G.}\ \bibnamefont {Tobias}}, \bibinfo {author} {\bibfnamefont {J.~P.}\ \bibnamefont {Covey}},\ and\ \bibinfo {author} {\bibfnamefont {J.}~\bibnamefont {Ye}},\ }\bibfield  {title} {\bibinfo {title} {{A degenerate Fermi gas of polar molecules}},\ }\href {https://doi.org/10.1126/science.aau7230} {\bibfield  {journal} {\bibinfo  {journal} {Science}\ }\textbf {\bibinfo {volume} {363}},\ \bibinfo {pages} {853} (\bibinfo {year} {2019})}\BibitemShut {NoStop}%
\bibitem [{\citenamefont {Yan}\ \emph {et~al.}(2020)\citenamefont {Yan}, \citenamefont {Ni}, \citenamefont {Robens},\ and\ \citenamefont {Zwierlein}}]{Zwierlein-2020}%
  \BibitemOpen
  \bibfield  {author} {\bibinfo {author} {\bibfnamefont {Z.~Z.}\ \bibnamefont {Yan}}, \bibinfo {author} {\bibfnamefont {Y.}~\bibnamefont {Ni}}, \bibinfo {author} {\bibfnamefont {C.}~\bibnamefont {Robens}},\ and\ \bibinfo {author} {\bibfnamefont {M.~W.}\ \bibnamefont {Zwierlein}},\ }\bibfield  {title} {\bibinfo {title} {{Bose polarons near quantum criticality}},\ }\href {https://doi.org/10.1126/science.aax5850} {\bibfield  {journal} {\bibinfo  {journal} {Science}\ }\textbf {\bibinfo {volume} {368}},\ \bibinfo {pages} {190} (\bibinfo {year} {2020})}\BibitemShut {NoStop}%
\bibitem [{\citenamefont {Ye}\ \emph {et~al.}(2020)\citenamefont {Ye}, \citenamefont {Xie}, \citenamefont {Guo}, \citenamefont {Ma}, \citenamefont {Wang}, \citenamefont {You},\ and\ \citenamefont {Tey}}]{Khoon-2020}%
  \BibitemOpen
  \bibfield  {author} {\bibinfo {author} {\bibfnamefont {Z.-X.}\ \bibnamefont {Ye}}, \bibinfo {author} {\bibfnamefont {L.-Y.}\ \bibnamefont {Xie}}, \bibinfo {author} {\bibfnamefont {Z.}~\bibnamefont {Guo}}, \bibinfo {author} {\bibfnamefont {X.-B.}\ \bibnamefont {Ma}}, \bibinfo {author} {\bibfnamefont {G.-R.}\ \bibnamefont {Wang}}, \bibinfo {author} {\bibfnamefont {L.}~\bibnamefont {You}},\ and\ \bibinfo {author} {\bibfnamefont {M.~K.}\ \bibnamefont {Tey}},\ }\bibfield  {title} {\bibinfo {title} {{Double-degenerate Bose-Fermi mixture of strontium and lithium}},\ }\href {https://doi.org/10.1103/PhysRevA.102.033307} {\bibfield  {journal} {\bibinfo  {journal} {Phys. Rev. A}\ }\textbf {\bibinfo {volume} {102}},\ \bibinfo {pages} {033307} (\bibinfo {year} {2020})}\BibitemShut {NoStop}%
\bibitem [{\citenamefont {Edri}\ \emph {et~al.}(2020)\citenamefont {Edri}, \citenamefont {Raz}, \citenamefont {Matzliah}, \citenamefont {Davidson},\ and\ \citenamefont {Ozeri}}]{Ozeri-2020}%
  \BibitemOpen
  \bibfield  {author} {\bibinfo {author} {\bibfnamefont {H.}~\bibnamefont {Edri}}, \bibinfo {author} {\bibfnamefont {B.}~\bibnamefont {Raz}}, \bibinfo {author} {\bibfnamefont {N.}~\bibnamefont {Matzliah}}, \bibinfo {author} {\bibfnamefont {N.}~\bibnamefont {Davidson}},\ and\ \bibinfo {author} {\bibfnamefont {R.}~\bibnamefont {Ozeri}},\ }\bibfield  {title} {\bibinfo {title} {{Observation of Spin-Spin Fermion-Mediated Interactions between Ultracold Bosons}},\ }\href {https://doi.org/10.1103/PhysRevLett.124.163401} {\bibfield  {journal} {\bibinfo  {journal} {Phys. Rev. Lett.}\ }\textbf {\bibinfo {volume} {124}},\ \bibinfo {pages} {163401} (\bibinfo {year} {2020})}\BibitemShut {NoStop}%
\bibitem [{\citenamefont {Fritsche}\ \emph {et~al.}(2021)\citenamefont {Fritsche}, \citenamefont {Baroni}, \citenamefont {Dobler}, \citenamefont {Kirilov}, \citenamefont {Huang}, \citenamefont {Grimm}, \citenamefont {Bruun},\ and\ \citenamefont {Massignan}}]{Fritsche-2021}%
  \BibitemOpen
  \bibfield  {author} {\bibinfo {author} {\bibfnamefont {I.}~\bibnamefont {Fritsche}}, \bibinfo {author} {\bibfnamefont {C.}~\bibnamefont {Baroni}}, \bibinfo {author} {\bibfnamefont {E.}~\bibnamefont {Dobler}}, \bibinfo {author} {\bibfnamefont {E.}~\bibnamefont {Kirilov}}, \bibinfo {author} {\bibfnamefont {B.}~\bibnamefont {Huang}}, \bibinfo {author} {\bibfnamefont {R.}~\bibnamefont {Grimm}}, \bibinfo {author} {\bibfnamefont {G.~M.}\ \bibnamefont {Bruun}},\ and\ \bibinfo {author} {\bibfnamefont {P.}~\bibnamefont {Massignan}},\ }\bibfield  {title} {\bibinfo {title} {{Stability and breakdown of Fermi polarons in a strongly interacting Fermi-Bose mixture}},\ }\href {https://doi.org/10.1103/PhysRevA.103.053314} {\bibfield  {journal} {\bibinfo  {journal} {Phys. Rev. A}\ }\textbf {\bibinfo {volume} {103}},\ \bibinfo {pages} {053314} (\bibinfo {year} {2021})}\BibitemShut {NoStop}%
\bibitem [{\citenamefont {Schindewolf}\ \emph {et~al.}(2022)\citenamefont {Schindewolf}, \citenamefont {Bause}, \citenamefont {Chen}, \citenamefont {Duda}, \citenamefont {Karman}, \citenamefont {Bloch},\ and\ \citenamefont {Luo}}]{Schindewolf-2022}%
  \BibitemOpen
  \bibfield  {author} {\bibinfo {author} {\bibfnamefont {A.}~\bibnamefont {Schindewolf}}, \bibinfo {author} {\bibfnamefont {R.}~\bibnamefont {Bause}}, \bibinfo {author} {\bibfnamefont {X.-Y.}\ \bibnamefont {Chen}}, \bibinfo {author} {\bibfnamefont {M.}~\bibnamefont {Duda}}, \bibinfo {author} {\bibfnamefont {T.}~\bibnamefont {Karman}}, \bibinfo {author} {\bibfnamefont {I.}~\bibnamefont {Bloch}},\ and\ \bibinfo {author} {\bibfnamefont {X.-Y.}\ \bibnamefont {Luo}},\ }\bibfield  {title} {\bibinfo {title} {{Evaporation of microwave-shielded polar molecules to quantum degeneracy}},\ }\href {https://doi.org/10.1038/s41586-022-04900-0} {\bibfield  {journal} {\bibinfo  {journal} {Nature}\ }\textbf {\bibinfo {volume} {607}},\ \bibinfo {pages} {677} (\bibinfo {year} {2022})}\BibitemShut {NoStop}%
\bibitem [{\citenamefont {Chen}\ \emph {et~al.}(2022)\citenamefont {Chen}, \citenamefont {Duda}, \citenamefont {Schindewolf}, \citenamefont {Bause}, \citenamefont {Bloch},\ and\ \citenamefont {Luo}}]{Bloch-2022}%
  \BibitemOpen
  \bibfield  {author} {\bibinfo {author} {\bibfnamefont {X.-Y.}\ \bibnamefont {Chen}}, \bibinfo {author} {\bibfnamefont {M.}~\bibnamefont {Duda}}, \bibinfo {author} {\bibfnamefont {A.}~\bibnamefont {Schindewolf}}, \bibinfo {author} {\bibfnamefont {R.}~\bibnamefont {Bause}}, \bibinfo {author} {\bibfnamefont {I.}~\bibnamefont {Bloch}},\ and\ \bibinfo {author} {\bibfnamefont {X.-Y.}\ \bibnamefont {Luo}},\ }\bibfield  {title} {\bibinfo {title} {{Suppression of Unitary Three-Body Loss in a Degenerate Bose-Fermi Mixture}},\ }\href {https://doi.org/10.1103/PhysRevLett.128.153401} {\bibfield  {journal} {\bibinfo  {journal} {Phys. Rev. Lett.}\ }\textbf {\bibinfo {volume} {128}},\ \bibinfo {pages} {153401} (\bibinfo {year} {2022})}\BibitemShut {NoStop}%
\bibitem [{\citenamefont {Duda}\ \emph {et~al.}(2023)\citenamefont {Duda}, \citenamefont {Chen}, \citenamefont {Schindewolf}, \citenamefont {Bause}, \citenamefont {von Milczewski}, \citenamefont {Schmidt}, \citenamefont {Bloch},\ and\ \citenamefont {Luo}}]{Duda-2023}%
  \BibitemOpen
  \bibfield  {author} {\bibinfo {author} {\bibfnamefont {M.}~\bibnamefont {Duda}}, \bibinfo {author} {\bibfnamefont {X.-Y.}\ \bibnamefont {Chen}}, \bibinfo {author} {\bibfnamefont {A.}~\bibnamefont {Schindewolf}}, \bibinfo {author} {\bibfnamefont {R.}~\bibnamefont {Bause}}, \bibinfo {author} {\bibfnamefont {J.}~\bibnamefont {von Milczewski}}, \bibinfo {author} {\bibfnamefont {R.}~\bibnamefont {Schmidt}}, \bibinfo {author} {\bibfnamefont {I.}~\bibnamefont {Bloch}},\ and\ \bibinfo {author} {\bibfnamefont {X.-Y.}\ \bibnamefont {Luo}},\ }\bibfield  {title} {\bibinfo {title} {{Transition from a polaronic condensate to a degenerate Fermi gas of heteronuclear molecules}},\ }\href {https://doi.org/10.1038/s41567-023-01948-1} {\bibfield  {journal} {\bibinfo  {journal} {Nat. Phys.}\ }\textbf {\bibinfo {volume} {19}},\ \bibinfo {pages} {720} (\bibinfo {year} {2023})}\BibitemShut {NoStop}%
\bibitem [{\citenamefont {Patel}\ \emph {et~al.}(2023)\citenamefont {Patel}, \citenamefont {Cai}, \citenamefont {Ando},\ and\ \citenamefont {Chin}}]{Patel-2023}%
  \BibitemOpen
  \bibfield  {author} {\bibinfo {author} {\bibfnamefont {K.}~\bibnamefont {Patel}}, \bibinfo {author} {\bibfnamefont {G.}~\bibnamefont {Cai}}, \bibinfo {author} {\bibfnamefont {H.}~\bibnamefont {Ando}},\ and\ \bibinfo {author} {\bibfnamefont {C.}~\bibnamefont {Chin}},\ }\bibfield  {title} {\bibinfo {title} {{Sound Propagation in a Bose-Fermi Mixture: From Weak to Strong Interactions}},\ }\href {https://doi.org/10.1103/PhysRevLett.131.083003} {\bibfield  {journal} {\bibinfo  {journal} {Phys. Rev. Lett.}\ }\textbf {\bibinfo {volume} {131}},\ \bibinfo {pages} {083003} (\bibinfo {year} {2023})}\BibitemShut {NoStop}%
\bibitem [{\citenamefont {Baroni}\ \emph {et~al.}(2024{\natexlab{b}})\citenamefont {Baroni}, \citenamefont {Huang}, \citenamefont {Fritsche}, \citenamefont {Dobler}, \citenamefont {Anich}, \citenamefont {Kirilov}, \citenamefont {Grimm}, \citenamefont {Bastarrachea-Magnani}, \citenamefont {Massignan},\ and\ \citenamefont {Bruun}}]{Baroni-2024}%
  \BibitemOpen
  \bibfield  {author} {\bibinfo {author} {\bibfnamefont {C.}~\bibnamefont {Baroni}}, \bibinfo {author} {\bibfnamefont {B.}~\bibnamefont {Huang}}, \bibinfo {author} {\bibfnamefont {I.}~\bibnamefont {Fritsche}}, \bibinfo {author} {\bibfnamefont {E.}~\bibnamefont {Dobler}}, \bibinfo {author} {\bibfnamefont {G.}~\bibnamefont {Anich}}, \bibinfo {author} {\bibfnamefont {E.}~\bibnamefont {Kirilov}}, \bibinfo {author} {\bibfnamefont {R.}~\bibnamefont {Grimm}}, \bibinfo {author} {\bibfnamefont {M.~A.}\ \bibnamefont {Bastarrachea-Magnani}}, \bibinfo {author} {\bibfnamefont {P.}~\bibnamefont {Massignan}},\ and\ \bibinfo {author} {\bibfnamefont {G.~M.}\ \bibnamefont {Bruun}},\ }\bibfield  {title} {\bibinfo {title} {{Mediated interactions between Fermi polarons and the role of impurity quantum statistics}},\ }\href {https://doi.org/10.1038/s41567-023-02248-4} {\bibfield  {journal} {\bibinfo  {journal} {Nat. Phys.}\ }\textbf {\bibinfo {volume} {20}},\ \bibinfo {pages} {68} (\bibinfo {year}
  {2024}{\natexlab{b}})}\BibitemShut {NoStop}%
\bibitem [{\citenamefont {Yan}\ \emph {et~al.}(2024)\citenamefont {Yan}, \citenamefont {Ni}, \citenamefont {Chuang}, \citenamefont {Dolgirev}, \citenamefont {Seetharam}, \citenamefont {Demler}, \citenamefont {Robens},\ and\ \citenamefont {Zwierlein}}]{Yan-2024}%
  \BibitemOpen
  \bibfield  {author} {\bibinfo {author} {\bibfnamefont {Z.~Z.}\ \bibnamefont {Yan}}, \bibinfo {author} {\bibfnamefont {Y.}~\bibnamefont {Ni}}, \bibinfo {author} {\bibfnamefont {A.}~\bibnamefont {Chuang}}, \bibinfo {author} {\bibfnamefont {P.~E.}\ \bibnamefont {Dolgirev}}, \bibinfo {author} {\bibfnamefont {K.}~\bibnamefont {Seetharam}}, \bibinfo {author} {\bibfnamefont {E.}~\bibnamefont {Demler}}, \bibinfo {author} {\bibfnamefont {C.}~\bibnamefont {Robens}},\ and\ \bibinfo {author} {\bibfnamefont {M.}~\bibnamefont {Zwierlein}},\ }\bibfield  {title} {\bibinfo {title} {{Collective flow of fermionic impurities immersed in a Bose--Einstein condensate}},\ }\href {https://doi.org/10.1038/s41567-024-02541-w} {\bibfield  {journal} {\bibinfo  {journal} {Nat. Phys.}\ }\textbf {\bibinfo {volume} {20}},\ \bibinfo {pages} {1395} (\bibinfo {year} {2024})}\BibitemShut {NoStop}%
\bibitem [{\citenamefont {Chuang}\ \emph {et~al.}(2025)\citenamefont {Chuang}, \citenamefont {Bui}, \citenamefont {Christianen}, \citenamefont {Zhang}, \citenamefont {Ni}, \citenamefont {Ahmed-Braun}, \citenamefont {Robens},\ and\ \citenamefont {Zwierlein}}]{Chuang-2024}%
  \BibitemOpen
  \bibfield  {author} {\bibinfo {author} {\bibfnamefont {A.~Y.}\ \bibnamefont {Chuang}}, \bibinfo {author} {\bibfnamefont {H.~Q.}\ \bibnamefont {Bui}}, \bibinfo {author} {\bibfnamefont {A.}~\bibnamefont {Christianen}}, \bibinfo {author} {\bibfnamefont {Y.}~\bibnamefont {Zhang}}, \bibinfo {author} {\bibfnamefont {Y.}~\bibnamefont {Ni}}, \bibinfo {author} {\bibfnamefont {D.}~\bibnamefont {Ahmed-Braun}}, \bibinfo {author} {\bibfnamefont {C.}~\bibnamefont {Robens}},\ and\ \bibinfo {author} {\bibfnamefont {M.}~\bibnamefont {Zwierlein}},\ }\bibfield  {title} {\bibinfo {title} {Observation of a halo trimer in an ultracold {Bose-Fermi} mixture},\ }\href {https://doi.org/10.1103/flty-9d72} {\bibfield  {journal} {\bibinfo  {journal} {Phys. Rev. X}\ }\textbf {\bibinfo {volume} {15}},\ \bibinfo {pages} {021098} (\bibinfo {year} {2025})}\BibitemShut {NoStop}%
\bibitem [{\citenamefont {Powell}\ \emph {et~al.}(2005)\citenamefont {Powell}, \citenamefont {Sachdev},\ and\ \citenamefont {B\"uchler}}]{Sachdev-2005}%
  \BibitemOpen
  \bibfield  {author} {\bibinfo {author} {\bibfnamefont {S.}~\bibnamefont {Powell}}, \bibinfo {author} {\bibfnamefont {S.}~\bibnamefont {Sachdev}},\ and\ \bibinfo {author} {\bibfnamefont {H.~P.}\ \bibnamefont {B\"uchler}},\ }\bibfield  {title} {\bibinfo {title} {{Depletion of the Bose-Einstein condensate in Bose-Fermi mixtures}},\ }\href {https://doi.org/10.1103/PhysRevB.72.024534} {\bibfield  {journal} {\bibinfo  {journal} {Phys. Rev. B}\ }\textbf {\bibinfo {volume} {72}},\ \bibinfo {pages} {024534} (\bibinfo {year} {2005})}\BibitemShut {NoStop}%
\bibitem [{\citenamefont {Dickerscheid}\ \emph {et~al.}(2005)\citenamefont {Dickerscheid}, \citenamefont {van Oosten}, \citenamefont {Tillema},\ and\ \citenamefont {Stoof}}]{Stoof-2005}%
  \BibitemOpen
  \bibfield  {author} {\bibinfo {author} {\bibfnamefont {D.~B.~M.}\ \bibnamefont {Dickerscheid}}, \bibinfo {author} {\bibfnamefont {D.}~\bibnamefont {van Oosten}}, \bibinfo {author} {\bibfnamefont {E.~J.}\ \bibnamefont {Tillema}},\ and\ \bibinfo {author} {\bibfnamefont {H.~T.~C.}\ \bibnamefont {Stoof}},\ }\bibfield  {title} {\bibinfo {title} {{Quantum Phases in a Resonantly Interacting Boson-Fermion Mixture}},\ }\href {https://doi.org/10.1103/PhysRevLett.94.230404} {\bibfield  {journal} {\bibinfo  {journal} {Phys. Rev. Lett.}\ }\textbf {\bibinfo {volume} {94}},\ \bibinfo {pages} {230404} (\bibinfo {year} {2005})}\BibitemShut {NoStop}%
\bibitem [{\citenamefont {Storozhenko}\ \emph {et~al.}(2005)\citenamefont {Storozhenko}, \citenamefont {Schuck}, \citenamefont {Suzuki}, \citenamefont {Yabu},\ and\ \citenamefont {Dukelsky}}]{Schuck-2005}%
  \BibitemOpen
  \bibfield  {author} {\bibinfo {author} {\bibfnamefont {A.}~\bibnamefont {Storozhenko}}, \bibinfo {author} {\bibfnamefont {P.}~\bibnamefont {Schuck}}, \bibinfo {author} {\bibfnamefont {T.}~\bibnamefont {Suzuki}}, \bibinfo {author} {\bibfnamefont {H.}~\bibnamefont {Yabu}},\ and\ \bibinfo {author} {\bibfnamefont {J.}~\bibnamefont {Dukelsky}},\ }\bibfield  {title} {\bibinfo {title} {{Boson-fermion pairing in a boson-fermion environment}},\ }\href {https://doi.org/10.1103/PhysRevA.71.063617} {\bibfield  {journal} {\bibinfo  {journal} {Phys. Rev. A}\ }\textbf {\bibinfo {volume} {71}},\ \bibinfo {pages} {063617} (\bibinfo {year} {2005})}\BibitemShut {NoStop}%
\bibitem [{\citenamefont {Avdeenkov}\ \emph {et~al.}(2006)\citenamefont {Avdeenkov}, \citenamefont {Bortolotti},\ and\ \citenamefont {Bohn}}]{Avdeenkov-2006}%
  \BibitemOpen
  \bibfield  {author} {\bibinfo {author} {\bibfnamefont {A.~V.}\ \bibnamefont {Avdeenkov}}, \bibinfo {author} {\bibfnamefont {D.~C.~E.}\ \bibnamefont {Bortolotti}},\ and\ \bibinfo {author} {\bibfnamefont {J.~L.}\ \bibnamefont {Bohn}},\ }\bibfield  {title} {\bibinfo {title} {{Stability of fermionic Feshbach molecules in a Bose-Fermi mixture}},\ }\href {https://doi.org/10.1103/PhysRevA.74.012709} {\bibfield  {journal} {\bibinfo  {journal} {Phys. Rev. A}\ }\textbf {\bibinfo {volume} {74}},\ \bibinfo {pages} {012709} (\bibinfo {year} {2006})}\BibitemShut {NoStop}%
\bibitem [{\citenamefont {Pollet}\ \emph {et~al.}(2006)\citenamefont {Pollet}, \citenamefont {Troyer}, \citenamefont {Van~Houcke},\ and\ \citenamefont {Rombouts}}]{Pollet-2006}%
  \BibitemOpen
  \bibfield  {author} {\bibinfo {author} {\bibfnamefont {L.}~\bibnamefont {Pollet}}, \bibinfo {author} {\bibfnamefont {M.}~\bibnamefont {Troyer}}, \bibinfo {author} {\bibfnamefont {K.}~\bibnamefont {Van~Houcke}},\ and\ \bibinfo {author} {\bibfnamefont {S.~M.~A.}\ \bibnamefont {Rombouts}},\ }\bibfield  {title} {\bibinfo {title} {{Phase Diagram of Bose-Fermi Mixtures in One-Dimensional Optical Lattices}},\ }\href {https://doi.org/10.1103/PhysRevLett.96.190402} {\bibfield  {journal} {\bibinfo  {journal} {Phys. Rev. Lett.}\ }\textbf {\bibinfo {volume} {96}},\ \bibinfo {pages} {190402} (\bibinfo {year} {2006})}\BibitemShut {NoStop}%
\bibitem [{\citenamefont {R\"othel}\ and\ \citenamefont {Pelster}(2007)}]{Rothel-2007}%
  \BibitemOpen
  \bibfield  {author} {\bibinfo {author} {\bibfnamefont {S.}~\bibnamefont {R\"othel}}\ and\ \bibinfo {author} {\bibfnamefont {A.}~\bibnamefont {Pelster}},\ }\bibfield  {title} {\bibinfo {title} {{Density and stability in ultracold dilute boson-fermion mixtures}},\ }\href {https://doi.org/10.1140/epjb/e2007-00288-x} {\bibfield  {journal} {\bibinfo  {journal} {Eur. Phys. Jour. B}\ }\textbf {\bibinfo {volume} {59}},\ \bibinfo {pages} {343} (\bibinfo {year} {2007})}\BibitemShut {NoStop}%
\bibitem [{\citenamefont {Barillier-Pertuisel}\ \emph {et~al.}(2008)\citenamefont {Barillier-Pertuisel}, \citenamefont {Pittel}, \citenamefont {Pollet},\ and\ \citenamefont {Schuck}}]{Barillier-2008}%
  \BibitemOpen
  \bibfield  {author} {\bibinfo {author} {\bibfnamefont {X.}~\bibnamefont {Barillier-Pertuisel}}, \bibinfo {author} {\bibfnamefont {S.}~\bibnamefont {Pittel}}, \bibinfo {author} {\bibfnamefont {L.}~\bibnamefont {Pollet}},\ and\ \bibinfo {author} {\bibfnamefont {P.}~\bibnamefont {Schuck}},\ }\bibfield  {title} {\bibinfo {title} {{Boson-fermion pairing in Bose-Fermi mixtures on one-dimensional optical lattices}},\ }\href {https://doi.org/10.1103/PhysRevA.77.012115} {\bibfield  {journal} {\bibinfo  {journal} {Phys. Rev. A}\ }\textbf {\bibinfo {volume} {77}},\ \bibinfo {pages} {012115} (\bibinfo {year} {2008})}\BibitemShut {NoStop}%
\bibitem [{\citenamefont {Pollet}\ \emph {et~al.}(2008)\citenamefont {Pollet}, \citenamefont {Kollath}, \citenamefont {Schollw\"ock},\ and\ \citenamefont {Troyer}}]{Pollet-2008}%
  \BibitemOpen
  \bibfield  {author} {\bibinfo {author} {\bibfnamefont {L.}~\bibnamefont {Pollet}}, \bibinfo {author} {\bibfnamefont {C.}~\bibnamefont {Kollath}}, \bibinfo {author} {\bibfnamefont {U.}~\bibnamefont {Schollw\"ock}},\ and\ \bibinfo {author} {\bibfnamefont {M.}~\bibnamefont {Troyer}},\ }\bibfield  {title} {\bibinfo {title} {{Mixture of bosonic and spin-polarized fermionic atoms in an optical lattice}},\ }\href {https://doi.org/10.1103/PhysRevA.77.023608} {\bibfield  {journal} {\bibinfo  {journal} {Phys. Rev. A}\ }\textbf {\bibinfo {volume} {77}},\ \bibinfo {pages} {023608} (\bibinfo {year} {2008})}\BibitemShut {NoStop}%
\bibitem [{\citenamefont {Bortolotti}\ \emph {et~al.}(2008)\citenamefont {Bortolotti}, \citenamefont {Avdeenkov},\ and\ \citenamefont {Bohn}}]{Bortolotti-2008}%
  \BibitemOpen
  \bibfield  {author} {\bibinfo {author} {\bibfnamefont {D.~C.~E.}\ \bibnamefont {Bortolotti}}, \bibinfo {author} {\bibfnamefont {A.~V.}\ \bibnamefont {Avdeenkov}},\ and\ \bibinfo {author} {\bibfnamefont {J.~L.}\ \bibnamefont {Bohn}},\ }\bibfield  {title} {\bibinfo {title} {{Generalized mean-field approach to a resonant Bose-Fermi mixture}},\ }\href {https://doi.org/10.1103/PhysRevA.78.063612} {\bibfield  {journal} {\bibinfo  {journal} {Phys. Rev. A}\ }\textbf {\bibinfo {volume} {78}},\ \bibinfo {pages} {063612} (\bibinfo {year} {2008})}\BibitemShut {NoStop}%
\bibitem [{\citenamefont {Marchetti}\ \emph {et~al.}(2008)\citenamefont {Marchetti}, \citenamefont {Mathy}, \citenamefont {Huse},\ and\ \citenamefont {Parish}}]{Parish-2008}%
  \BibitemOpen
  \bibfield  {author} {\bibinfo {author} {\bibfnamefont {F.~M.}\ \bibnamefont {Marchetti}}, \bibinfo {author} {\bibfnamefont {C.~J.~M.}\ \bibnamefont {Mathy}}, \bibinfo {author} {\bibfnamefont {D.~A.}\ \bibnamefont {Huse}},\ and\ \bibinfo {author} {\bibfnamefont {M.~M.}\ \bibnamefont {Parish}},\ }\bibfield  {title} {\bibinfo {title} {{Phase separation and collapse in Bose-Fermi mixtures with a Feshbach resonance}},\ }\href {https://doi.org/10.1103/PhysRevB.78.134517} {\bibfield  {journal} {\bibinfo  {journal} {Phys. Rev. B}\ }\textbf {\bibinfo {volume} {78}},\ \bibinfo {pages} {134517} (\bibinfo {year} {2008})}\BibitemShut {NoStop}%
\bibitem [{\citenamefont {Watanabe}\ \emph {et~al.}(2008)\citenamefont {Watanabe}, \citenamefont {Suzuki},\ and\ \citenamefont {Schuck}}]{Watanabe-2008}%
  \BibitemOpen
  \bibfield  {author} {\bibinfo {author} {\bibfnamefont {T.}~\bibnamefont {Watanabe}}, \bibinfo {author} {\bibfnamefont {T.}~\bibnamefont {Suzuki}},\ and\ \bibinfo {author} {\bibfnamefont {P.}~\bibnamefont {Schuck}},\ }\bibfield  {title} {\bibinfo {title} {{Bose-Fermi pair correlations in attractively interacting Bose-Fermi atomic mixtures}},\ }\href {https://doi.org/10.1103/PhysRevA.78.033601} {\bibfield  {journal} {\bibinfo  {journal} {Phys. Rev. A}\ }\textbf {\bibinfo {volume} {78}},\ \bibinfo {pages} {033601} (\bibinfo {year} {2008})}\BibitemShut {NoStop}%
\bibitem [{\citenamefont {Fratini}\ and\ \citenamefont {Pieri}(2010)}]{Fratini-2010}%
  \BibitemOpen
  \bibfield  {author} {\bibinfo {author} {\bibfnamefont {E.}~\bibnamefont {Fratini}}\ and\ \bibinfo {author} {\bibfnamefont {P.}~\bibnamefont {Pieri}},\ }\bibfield  {title} {\bibinfo {title} {{Pairing and condensation in a resonant Bose-Fermi mixture}},\ }\href {https://doi.org/10.1103/PhysRevA.81.051605} {\bibfield  {journal} {\bibinfo  {journal} {Phys. Rev. A}\ }\textbf {\bibinfo {volume} {81}},\ \bibinfo {pages} {051605} (\bibinfo {year} {2010})}\BibitemShut {NoStop}%
\bibitem [{\citenamefont {Yu}\ \emph {et~al.}(2011)\citenamefont {Yu}, \citenamefont {Zhang},\ and\ \citenamefont {Zhai}}]{Zhai-2011}%
  \BibitemOpen
  \bibfield  {author} {\bibinfo {author} {\bibfnamefont {Z.-Q.}\ \bibnamefont {Yu}}, \bibinfo {author} {\bibfnamefont {S.}~\bibnamefont {Zhang}},\ and\ \bibinfo {author} {\bibfnamefont {H.}~\bibnamefont {Zhai}},\ }\bibfield  {title} {\bibinfo {title} {{Stability condition of a strongly interacting boson-fermion mixture across an interspecies Feshbach resonance}},\ }\href {https://doi.org/10.1103/PhysRevA.83.041603} {\bibfield  {journal} {\bibinfo  {journal} {Phys. Rev. A}\ }\textbf {\bibinfo {volume} {83}},\ \bibinfo {pages} {041603} (\bibinfo {year} {2011})}\BibitemShut {NoStop}%
\bibitem [{\citenamefont {Song}\ and\ \citenamefont {Zhou}(2011)}]{Song-2011}%
  \BibitemOpen
  \bibfield  {author} {\bibinfo {author} {\bibfnamefont {J.-L.}\ \bibnamefont {Song}}\ and\ \bibinfo {author} {\bibfnamefont {F.}~\bibnamefont {Zhou}},\ }\bibfield  {title} {\bibinfo {title} {{Anomalous dimers in quantum mixtures near broad resonances: Pauli blocking, Fermi surface dynamics, and implications}},\ }\href {https://doi.org/10.1103/PhysRevA.84.013601} {\bibfield  {journal} {\bibinfo  {journal} {Phys. Rev. A}\ }\textbf {\bibinfo {volume} {84}},\ \bibinfo {pages} {013601} (\bibinfo {year} {2011})}\BibitemShut {NoStop}%
\bibitem [{\citenamefont {Ludwig}\ \emph {et~al.}(2011)\citenamefont {Ludwig}, \citenamefont {Floerchinger}, \citenamefont {Moroz},\ and\ \citenamefont {Wetterich}}]{Ludwig-2011}%
  \BibitemOpen
  \bibfield  {author} {\bibinfo {author} {\bibfnamefont {D.}~\bibnamefont {Ludwig}}, \bibinfo {author} {\bibfnamefont {S.}~\bibnamefont {Floerchinger}}, \bibinfo {author} {\bibfnamefont {S.}~\bibnamefont {Moroz}},\ and\ \bibinfo {author} {\bibfnamefont {C.}~\bibnamefont {Wetterich}},\ }\bibfield  {title} {\bibinfo {title} {{Quantum phase transition in Bose-Fermi mixtures}},\ }\href {https://doi.org/10.1103/PhysRevA.84.033629} {\bibfield  {journal} {\bibinfo  {journal} {Phys. Rev. A}\ }\textbf {\bibinfo {volume} {84}},\ \bibinfo {pages} {033629} (\bibinfo {year} {2011})}\BibitemShut {NoStop}%
\bibitem [{\citenamefont {Fratini}\ and\ \citenamefont {Pieri}(2012)}]{Fratini-2012}%
  \BibitemOpen
  \bibfield  {author} {\bibinfo {author} {\bibfnamefont {E.}~\bibnamefont {Fratini}}\ and\ \bibinfo {author} {\bibfnamefont {P.}~\bibnamefont {Pieri}},\ }\bibfield  {title} {\bibinfo {title} {{Mass imbalance effect in resonant Bose-Fermi mixtures}},\ }\href {https://doi.org/10.1103/PhysRevA.85.063618} {\bibfield  {journal} {\bibinfo  {journal} {Phys. Rev. A}\ }\textbf {\bibinfo {volume} {85}},\ \bibinfo {pages} {063618} (\bibinfo {year} {2012})}\BibitemShut {NoStop}%
\bibitem [{\citenamefont {Anders}\ \emph {et~al.}(2012)\citenamefont {Anders}, \citenamefont {Werner}, \citenamefont {Troyer}, \citenamefont {Sigrist},\ and\ \citenamefont {Pollet}}]{Anders-2012}%
  \BibitemOpen
  \bibfield  {author} {\bibinfo {author} {\bibfnamefont {P.}~\bibnamefont {Anders}}, \bibinfo {author} {\bibfnamefont {P.}~\bibnamefont {Werner}}, \bibinfo {author} {\bibfnamefont {M.}~\bibnamefont {Troyer}}, \bibinfo {author} {\bibfnamefont {M.}~\bibnamefont {Sigrist}},\ and\ \bibinfo {author} {\bibfnamefont {L.}~\bibnamefont {Pollet}},\ }\bibfield  {title} {\bibinfo {title} {{From the Cooper Problem to Canted Supersolids in Bose-Fermi Mixtures}},\ }\href {https://doi.org/10.1103/PhysRevLett.109.206401} {\bibfield  {journal} {\bibinfo  {journal} {Phys. Rev. Lett.}\ }\textbf {\bibinfo {volume} {109}},\ \bibinfo {pages} {206401} (\bibinfo {year} {2012})}\BibitemShut {NoStop}%
\bibitem [{\citenamefont {Bertaina}\ \emph {et~al.}(2013)\citenamefont {Bertaina}, \citenamefont {Fratini}, \citenamefont {Giorgini},\ and\ \citenamefont {Pieri}}]{Bertaina-2013}%
  \BibitemOpen
  \bibfield  {author} {\bibinfo {author} {\bibfnamefont {G.}~\bibnamefont {Bertaina}}, \bibinfo {author} {\bibfnamefont {E.}~\bibnamefont {Fratini}}, \bibinfo {author} {\bibfnamefont {S.}~\bibnamefont {Giorgini}},\ and\ \bibinfo {author} {\bibfnamefont {P.}~\bibnamefont {Pieri}},\ }\bibfield  {title} {\bibinfo {title} {{Quantum Monte Carlo Study of a Resonant Bose-Fermi Mixture}},\ }\href {https://doi.org/10.1103/PhysRevLett.110.115303} {\bibfield  {journal} {\bibinfo  {journal} {Phys. Rev. Lett.}\ }\textbf {\bibinfo {volume} {110}},\ \bibinfo {pages} {115303} (\bibinfo {year} {2013})}\BibitemShut {NoStop}%
\bibitem [{\citenamefont {Fratini}\ and\ \citenamefont {Pieri}(2013)}]{Fratini-2013}%
  \BibitemOpen
  \bibfield  {author} {\bibinfo {author} {\bibfnamefont {E.}~\bibnamefont {Fratini}}\ and\ \bibinfo {author} {\bibfnamefont {P.}~\bibnamefont {Pieri}},\ }\bibfield  {title} {\bibinfo {title} {{Single-particle spectral functions in the normal phase of a strongly attractive Bose-Fermi mixture}},\ }\href {https://doi.org/10.1103/PhysRevA.88.013627} {\bibfield  {journal} {\bibinfo  {journal} {Phys. Rev. A}\ }\textbf {\bibinfo {volume} {88}},\ \bibinfo {pages} {013627} (\bibinfo {year} {2013})}\BibitemShut {NoStop}%
\bibitem [{\citenamefont {Sogo}\ \emph {et~al.}(2013)\citenamefont {Sogo}, \citenamefont {Schuck},\ and\ \citenamefont {Urban}}]{Sogo-2013}%
  \BibitemOpen
  \bibfield  {author} {\bibinfo {author} {\bibfnamefont {T.}~\bibnamefont {Sogo}}, \bibinfo {author} {\bibfnamefont {P.}~\bibnamefont {Schuck}},\ and\ \bibinfo {author} {\bibfnamefont {M.}~\bibnamefont {Urban}},\ }\bibfield  {title} {\bibinfo {title} {{Bose-Fermi pairs in a mixture and the Luttinger theorem within a Nozi\`eres-Schmitt-Rink-like approach}},\ }\href {https://doi.org/10.1103/PhysRevA.88.023613} {\bibfield  {journal} {\bibinfo  {journal} {Phys. Rev. A}\ }\textbf {\bibinfo {volume} {88}},\ \bibinfo {pages} {023613} (\bibinfo {year} {2013})}\BibitemShut {NoStop}%
\bibitem [{\citenamefont {Guidini}\ \emph {et~al.}(2014)\citenamefont {Guidini}, \citenamefont {Bertaina}, \citenamefont {Fratini},\ and\ \citenamefont {Pieri}}]{Guidini-2014}%
  \BibitemOpen
  \bibfield  {author} {\bibinfo {author} {\bibfnamefont {A.}~\bibnamefont {Guidini}}, \bibinfo {author} {\bibfnamefont {G.}~\bibnamefont {Bertaina}}, \bibinfo {author} {\bibfnamefont {E.}~\bibnamefont {Fratini}},\ and\ \bibinfo {author} {\bibfnamefont {P.}~\bibnamefont {Pieri}},\ }\bibfield  {title} {\bibinfo {title} {{Bose-Fermi mixtures in the molecular limit}},\ }\href {https://doi.org/10.1103/PhysRevA.89.023634} {\bibfield  {journal} {\bibinfo  {journal} {Phys. Rev. A}\ }\textbf {\bibinfo {volume} {89}},\ \bibinfo {pages} {023634} (\bibinfo {year} {2014})}\BibitemShut {NoStop}%
\bibitem [{\citenamefont {Guidini}\ \emph {et~al.}(2015)\citenamefont {Guidini}, \citenamefont {Bertaina}, \citenamefont {Galli},\ and\ \citenamefont {Pieri}}]{Guidini-2015}%
  \BibitemOpen
  \bibfield  {author} {\bibinfo {author} {\bibfnamefont {A.}~\bibnamefont {Guidini}}, \bibinfo {author} {\bibfnamefont {G.}~\bibnamefont {Bertaina}}, \bibinfo {author} {\bibfnamefont {D.~E.}\ \bibnamefont {Galli}},\ and\ \bibinfo {author} {\bibfnamefont {P.}~\bibnamefont {Pieri}},\ }\bibfield  {title} {\bibinfo {title} {{Condensed phase of Bose-Fermi mixtures with a pairing interaction}},\ }\href {https://doi.org/10.1103/PhysRevA.91.023603} {\bibfield  {journal} {\bibinfo  {journal} {Phys. Rev. A}\ }\textbf {\bibinfo {volume} {91}},\ \bibinfo {pages} {023603} (\bibinfo {year} {2015})}\BibitemShut {NoStop}%
\bibitem [{\citenamefont {Kharga}\ \emph {et~al.}(2017)\citenamefont {Kharga}, \citenamefont {Inotani}, \citenamefont {Hanai},\ and\ \citenamefont {Ohashi}}]{Kharga-2017}%
  \BibitemOpen
  \bibfield  {author} {\bibinfo {author} {\bibfnamefont {D.}~\bibnamefont {Kharga}}, \bibinfo {author} {\bibfnamefont {D.}~\bibnamefont {Inotani}}, \bibinfo {author} {\bibfnamefont {R.}~\bibnamefont {Hanai}},\ and\ \bibinfo {author} {\bibfnamefont {Y.}~\bibnamefont {Ohashi}},\ }\bibfield  {title} {\bibinfo {title} {{Single-Particle Properties of a Strongly Interacting Bose--Fermi Mixture Above the BEC Phase Transition Temperature}},\ }\href {https://doi.org/10.1007/s10909-017-1742-x} {\bibfield  {journal} {\bibinfo  {journal} {J. Low. temp. Phys.}\ }\textbf {\bibinfo {volume} {187}},\ \bibinfo {pages} {661} (\bibinfo {year} {2017})}\BibitemShut {NoStop}%
\bibitem [{\citenamefont {Manabe}\ \emph {et~al.}(2019)\citenamefont {Manabe}, \citenamefont {Inotani},\ and\ \citenamefont {Ohashi}}]{Ohashi-2019}%
  \BibitemOpen
  \bibfield  {author} {\bibinfo {author} {\bibfnamefont {K.}~\bibnamefont {Manabe}}, \bibinfo {author} {\bibfnamefont {D.}~\bibnamefont {Inotani}},\ and\ \bibinfo {author} {\bibfnamefont {Y.}~\bibnamefont {Ohashi}},\ }\bibfield  {title} {\bibinfo {title} {{Single-particle properties of a strongly interacting Bose-Fermi mixture with mass and population imbalance}},\ }\href {https://doi.org/10.1103/PhysRevA.100.063609} {\bibfield  {journal} {\bibinfo  {journal} {Phys. Rev. A}\ }\textbf {\bibinfo {volume} {100}},\ \bibinfo {pages} {063609} (\bibinfo {year} {2019})}\BibitemShut {NoStop}%
\bibitem [{\citenamefont {Manabe}\ and\ \citenamefont {Ohashi}(2020)}]{Manabe-2020}%
  \BibitemOpen
  \bibfield  {author} {\bibinfo {author} {\bibfnamefont {K.}~\bibnamefont {Manabe}}\ and\ \bibinfo {author} {\bibfnamefont {Y.}~\bibnamefont {Ohashi}},\ }\bibfield  {title} {\bibinfo {title} {{Thermodynamic Stability and Effects of Bose--Bose Repulsion in an Ultracold Bose--Fermi Mixture with Strong Hetero-Pairing Fluctuations}},\ }\href {https://doi.org/10.1007/s10909-019-02321-4} {\bibfield  {journal} {\bibinfo  {journal} {J. Low. temp. Phys.}\ }\textbf {\bibinfo {volume} {201}},\ \bibinfo {pages} {65} (\bibinfo {year} {2020})}\BibitemShut {NoStop}%
\bibitem [{\citenamefont {Manabe}\ and\ \citenamefont {Ohashi}(2021)}]{Ohashi-2021}%
  \BibitemOpen
  \bibfield  {author} {\bibinfo {author} {\bibfnamefont {K.}~\bibnamefont {Manabe}}\ and\ \bibinfo {author} {\bibfnamefont {Y.}~\bibnamefont {Ohashi}},\ }\bibfield  {title} {\bibinfo {title} {{Thermodynamic stability, compressibility matrices, and effects of mediated interactions in a strongly interacting Bose-Fermi mixture}},\ }\href {https://doi.org/10.1103/PhysRevA.103.063317} {\bibfield  {journal} {\bibinfo  {journal} {Phys. Rev. A}\ }\textbf {\bibinfo {volume} {103}},\ \bibinfo {pages} {063317} (\bibinfo {year} {2021})}\BibitemShut {NoStop}%
\bibitem [{\citenamefont {von Milczewski}\ \emph {et~al.}(2022)\citenamefont {von Milczewski}, \citenamefont {Rose},\ and\ \citenamefont {Schmidt}}]{Schmidt-2022}%
  \BibitemOpen
  \bibfield  {author} {\bibinfo {author} {\bibfnamefont {J.}~\bibnamefont {von Milczewski}}, \bibinfo {author} {\bibfnamefont {F.}~\bibnamefont {Rose}},\ and\ \bibinfo {author} {\bibfnamefont {R.}~\bibnamefont {Schmidt}},\ }\bibfield  {title} {\bibinfo {title} {{Functional-renormalization-group approach to strongly coupled Bose-Fermi mixtures in two dimensions}},\ }\href {https://doi.org/10.1103/PhysRevA.105.013317} {\bibfield  {journal} {\bibinfo  {journal} {Phys. Rev. A}\ }\textbf {\bibinfo {volume} {105}},\ \bibinfo {pages} {013317} (\bibinfo {year} {2022})}\BibitemShut {NoStop}%
\bibitem [{\citenamefont {Shen}\ \emph {et~al.}(2024)\citenamefont {Shen}, \citenamefont {Davidson}, \citenamefont {Bruun}, \citenamefont {Sun},\ and\ \citenamefont {Wu}}]{Bruun-2024}%
  \BibitemOpen
  \bibfield  {author} {\bibinfo {author} {\bibfnamefont {X.}~\bibnamefont {Shen}}, \bibinfo {author} {\bibfnamefont {N.}~\bibnamefont {Davidson}}, \bibinfo {author} {\bibfnamefont {G.~M.}\ \bibnamefont {Bruun}}, \bibinfo {author} {\bibfnamefont {M.}~\bibnamefont {Sun}},\ and\ \bibinfo {author} {\bibfnamefont {Z.}~\bibnamefont {Wu}},\ }\bibfield  {title} {\bibinfo {title} {{Strongly Interacting Bose-Fermi Mixtures: Mediated Interaction, Phase Diagram, and Sound Propagation}},\ }\href {https://doi.org/10.1103/PhysRevLett.132.033401} {\bibfield  {journal} {\bibinfo  {journal} {Phys. Rev. Lett.}\ }\textbf {\bibinfo {volume} {132}},\ \bibinfo {pages} {033401} (\bibinfo {year} {2024})}\BibitemShut {NoStop}%
\bibitem [{\citenamefont {Zhang}\ \emph {et~al.}(2024{\natexlab{b}})\citenamefont {Zhang}, \citenamefont {Guo}, \citenamefont {Tajima},\ and\ \citenamefont {Liang}}]{Tajima-2024b}%
  \BibitemOpen
  \bibfield  {author} {\bibinfo {author} {\bibfnamefont {T.}~\bibnamefont {Zhang}}, \bibinfo {author} {\bibfnamefont {Y.}~\bibnamefont {Guo}}, \bibinfo {author} {\bibfnamefont {H.}~\bibnamefont {Tajima}},\ and\ \bibinfo {author} {\bibfnamefont {H.}~\bibnamefont {Liang}},\ }\bibfield  {title} {\bibinfo {title} {{Probing the goldstino excitation through tunneling transport in a Bose-Fermi mixture with explicitly broken supersymmetry}},\ }\href {https://doi.org/10.1103/PhysRevB.110.064512} {\bibfield  {journal} {\bibinfo  {journal} {Phys. Rev. B}\ }\textbf {\bibinfo {volume} {110}},\ \bibinfo {pages} {064512} (\bibinfo {year} {2024}{\natexlab{b}})}\BibitemShut {NoStop}%
\bibitem [{\citenamefont {Pisani}\ \emph {et~al.}(2025)\citenamefont {Pisani}, \citenamefont {Bovini}, \citenamefont {Pavan},\ and\ \citenamefont {Pieri}}]{Pisani-2025}%
  \BibitemOpen
  \bibfield  {author} {\bibinfo {author} {\bibfnamefont {L.}~\bibnamefont {Pisani}}, \bibinfo {author} {\bibfnamefont {P.}~\bibnamefont {Bovini}}, \bibinfo {author} {\bibfnamefont {F.}~\bibnamefont {Pavan}},\ and\ \bibinfo {author} {\bibfnamefont {P.}~\bibnamefont {Pieri}},\ }\bibfield  {title} {\bibinfo {title} {{Boson-fermion pairing and condensation in two-dimensional Bose-Fermi mixtures}},\ }\href {https://doi.org/10.21468/SciPostPhys.18.3.076} {\bibfield  {journal} {\bibinfo  {journal} {SciPost Phys.}\ }\textbf {\bibinfo {volume} {18}},\ \bibinfo {pages} {076} (\bibinfo {year} {2025})}\BibitemShut {NoStop}%
\bibitem [{\citenamefont {Gualerzi}\ \emph {et~al.}(2025)\citenamefont {Gualerzi}, \citenamefont {Pisani},\ and\ \citenamefont {Pieri}}]{Gualerzi-2025}%
  \BibitemOpen
  \bibfield  {author} {\bibinfo {author} {\bibfnamefont {C.}~\bibnamefont {Gualerzi}}, \bibinfo {author} {\bibfnamefont {L.}~\bibnamefont {Pisani}},\ and\ \bibinfo {author} {\bibfnamefont {P.}~\bibnamefont {Pieri}},\ }\bibfield  {title} {\bibinfo {title} {{Mechanical stability of resonant Bose-Fermi mixtures}},\ }\href {https://doi.org/10.21468/SciPostPhys.19.2.039} {\bibfield  {journal} {\bibinfo  {journal} {SciPost Phys.}\ }\textbf {\bibinfo {volume} {19}},\ \bibinfo {pages} {039} (\bibinfo {year} {2025})}\BibitemShut {NoStop}%
\bibitem [{\citenamefont {Foster}\ \emph {et~al.}(2026)\citenamefont {Foster}, \citenamefont {Bleu}, \citenamefont {Levinsen},\ and\ \citenamefont {Parish}}]{Foster-2026}%
  \BibitemOpen
  \bibfield  {author} {\bibinfo {author} {\bibfnamefont {S.}~\bibnamefont {Foster}}, \bibinfo {author} {\bibfnamefont {O.}~\bibnamefont {Bleu}}, \bibinfo {author} {\bibfnamefont {J.}~\bibnamefont {Levinsen}},\ and\ \bibinfo {author} {\bibfnamefont {M.~M.}\ \bibnamefont {Parish}},\ }\href@noop {} {\bibinfo {title} {{Quantum droplets in a resonant Bose-Fermi mixture}}} (\bibinfo {year} {2026}),\ \Eprint {https://arxiv.org/abs/2601.12777} {arXiv:2601.12777} \BibitemShut {NoStop}%
\bibitem [{\citenamefont {Modugno}\ \emph {et~al.}(2002)\citenamefont {Modugno}, \citenamefont {Roati}, \citenamefont {Riboli}, \citenamefont {Ferlaino}, \citenamefont {Brecha},\ and\ \citenamefont {Inguscio}}]{Modugno-2002}%
  \BibitemOpen
  \bibfield  {author} {\bibinfo {author} {\bibfnamefont {G.}~\bibnamefont {Modugno}}, \bibinfo {author} {\bibfnamefont {G.}~\bibnamefont {Roati}}, \bibinfo {author} {\bibfnamefont {F.}~\bibnamefont {Riboli}}, \bibinfo {author} {\bibfnamefont {F.}~\bibnamefont {Ferlaino}}, \bibinfo {author} {\bibfnamefont {R.~J.}\ \bibnamefont {Brecha}},\ and\ \bibinfo {author} {\bibfnamefont {M.}~\bibnamefont {Inguscio}},\ }\bibfield  {title} {\bibinfo {title} {{Collapse of a Degenerate Fermi Gas}},\ }\href {https://doi.org/10.1126/science.1077386} {\bibfield  {journal} {\bibinfo  {journal} {Science}\ }\textbf {\bibinfo {volume} {297}},\ \bibinfo {pages} {2240} (\bibinfo {year} {2002})}\BibitemShut {NoStop}%
\bibitem [{\citenamefont {Lous}\ \emph {et~al.}(2018{\natexlab{b}})\citenamefont {Lous}, \citenamefont {Fritsche}, \citenamefont {Jag}, \citenamefont {Lehmann}, \citenamefont {Kirilov}, \citenamefont {Huang},\ and\ \citenamefont {Grimm}}]{Lous-2018}%
  \BibitemOpen
  \bibfield  {author} {\bibinfo {author} {\bibfnamefont {R.~S.}\ \bibnamefont {Lous}}, \bibinfo {author} {\bibfnamefont {I.}~\bibnamefont {Fritsche}}, \bibinfo {author} {\bibfnamefont {M.}~\bibnamefont {Jag}}, \bibinfo {author} {\bibfnamefont {F.}~\bibnamefont {Lehmann}}, \bibinfo {author} {\bibfnamefont {E.}~\bibnamefont {Kirilov}}, \bibinfo {author} {\bibfnamefont {B.}~\bibnamefont {Huang}},\ and\ \bibinfo {author} {\bibfnamefont {R.}~\bibnamefont {Grimm}},\ }\bibfield  {title} {\bibinfo {title} {Probing the interface of a phase-separated state in a repulsive {Bose-Fermi} mixture},\ }\href {https://doi.org/10.1103/PhysRevLett.120.243403} {\bibfield  {journal} {\bibinfo  {journal} {Phys. Rev. Lett.}\ }\textbf {\bibinfo {volume} {120}},\ \bibinfo {pages} {243403} (\bibinfo {year} {2018}{\natexlab{b}})}\BibitemShut {NoStop}%
\bibitem [{\citenamefont {M\o{}lmer}(1998)}]{Molmer-1998}%
  \BibitemOpen
  \bibfield  {author} {\bibinfo {author} {\bibfnamefont {K.}~\bibnamefont {M\o{}lmer}},\ }\bibfield  {title} {\bibinfo {title} {{Bose Condensates and Fermi Gases at Zero Temperature}},\ }\href {https://doi.org/10.1103/PhysRevLett.80.1804} {\bibfield  {journal} {\bibinfo  {journal} {Phys. Rev. Lett.}\ }\textbf {\bibinfo {volume} {80}},\ \bibinfo {pages} {1804} (\bibinfo {year} {1998})}\BibitemShut {NoStop}%
\bibitem [{\citenamefont {Viverit}\ \emph {et~al.}(2000)\citenamefont {Viverit}, \citenamefont {Pethick},\ and\ \citenamefont {Smith}}]{Viverit-2000}%
  \BibitemOpen
  \bibfield  {author} {\bibinfo {author} {\bibfnamefont {L.}~\bibnamefont {Viverit}}, \bibinfo {author} {\bibfnamefont {C.~J.}\ \bibnamefont {Pethick}},\ and\ \bibinfo {author} {\bibfnamefont {H.}~\bibnamefont {Smith}},\ }\bibfield  {title} {\bibinfo {title} {Zero-temperature phase diagram of binary boson-fermion mixtures},\ }\href {https://doi.org/10.1103/PhysRevA.61.053605} {\bibfield  {journal} {\bibinfo  {journal} {Phys. Rev. A}\ }\textbf {\bibinfo {volume} {61}},\ \bibinfo {pages} {053605} (\bibinfo {year} {2000})}\BibitemShut {NoStop}%
\bibitem [{\citenamefont {Yi}\ and\ \citenamefont {Sun}(2001)}]{Yi-2001}%
  \BibitemOpen
  \bibfield  {author} {\bibinfo {author} {\bibfnamefont {X.~X.}\ \bibnamefont {Yi}}\ and\ \bibinfo {author} {\bibfnamefont {C.~P.}\ \bibnamefont {Sun}},\ }\bibfield  {title} {\bibinfo {title} {{Phase separation of a trapped Bose-Fermi gas mixture: Beyond the Thomas-Fermi approximation}},\ }\href {https://doi.org/10.1103/PhysRevA.64.043608} {\bibfield  {journal} {\bibinfo  {journal} {Phys. Rev. A}\ }\textbf {\bibinfo {volume} {64}},\ \bibinfo {pages} {043608} (\bibinfo {year} {2001})}\BibitemShut {NoStop}%
\bibitem [{\citenamefont {Roth}\ and\ \citenamefont {Feldmeier}(2002)}]{Feldmeier-2002}%
  \BibitemOpen
  \bibfield  {author} {\bibinfo {author} {\bibfnamefont {R.}~\bibnamefont {Roth}}\ and\ \bibinfo {author} {\bibfnamefont {H.}~\bibnamefont {Feldmeier}},\ }\bibfield  {title} {\bibinfo {title} {Mean-field instability of trapped dilute boson-fermion mixtures},\ }\href {https://doi.org/10.1103/PhysRevA.65.021603} {\bibfield  {journal} {\bibinfo  {journal} {Phys. Rev. A}\ }\textbf {\bibinfo {volume} {65}},\ \bibinfo {pages} {021603} (\bibinfo {year} {2002})}\BibitemShut {NoStop}%
\bibitem [{\citenamefont {Viverit}\ and\ \citenamefont {Giorgini}(2002)}]{Viverit-2002}%
  \BibitemOpen
  \bibfield  {author} {\bibinfo {author} {\bibfnamefont {L.}~\bibnamefont {Viverit}}\ and\ \bibinfo {author} {\bibfnamefont {S.}~\bibnamefont {Giorgini}},\ }\bibfield  {title} {\bibinfo {title} {{Ground-state properties of a dilute Bose-Fermi mixture}},\ }\href {https://doi.org/10.1103/PhysRevA.66.063604} {\bibfield  {journal} {\bibinfo  {journal} {Phys. Rev. A}\ }\textbf {\bibinfo {volume} {66}},\ \bibinfo {pages} {063604} (\bibinfo {year} {2002})}\BibitemShut {NoStop}%
\bibitem [{\citenamefont {Capuzzi}\ \emph {et~al.}(2003)\citenamefont {Capuzzi}, \citenamefont {Minguzzi},\ and\ \citenamefont {Tosi}}]{Capuzzi-2003}%
  \BibitemOpen
  \bibfield  {author} {\bibinfo {author} {\bibfnamefont {P.}~\bibnamefont {Capuzzi}}, \bibinfo {author} {\bibfnamefont {A.}~\bibnamefont {Minguzzi}},\ and\ \bibinfo {author} {\bibfnamefont {M.~P.}\ \bibnamefont {Tosi}},\ }\bibfield  {title} {\bibinfo {title} {Collective excitations in trapped boson-fermion mixtures: From demixing to collapse},\ }\href {https://doi.org/10.1103/PhysRevA.68.033605} {\bibfield  {journal} {\bibinfo  {journal} {Phys. Rev. A}\ }\textbf {\bibinfo {volume} {68}},\ \bibinfo {pages} {033605} (\bibinfo {year} {2003})}\BibitemShut {NoStop}%
\bibitem [{\citenamefont {Chui}\ and\ \citenamefont {Ryzhov}(2004)}]{Chui-2004}%
  \BibitemOpen
  \bibfield  {author} {\bibinfo {author} {\bibfnamefont {S.~T.}\ \bibnamefont {Chui}}\ and\ \bibinfo {author} {\bibfnamefont {V.~N.}\ \bibnamefont {Ryzhov}},\ }\bibfield  {title} {\bibinfo {title} {Collapse transition in mixtures of bosons and fermions},\ }\href {https://doi.org/10.1103/PhysRevA.69.043607} {\bibfield  {journal} {\bibinfo  {journal} {Phys. Rev. A}\ }\textbf {\bibinfo {volume} {69}},\ \bibinfo {pages} {043607} (\bibinfo {year} {2004})}\BibitemShut {NoStop}%
\bibitem [{\citenamefont {Shirasaki}\ \emph {et~al.}(2014)\citenamefont {Shirasaki}, \citenamefont {Nakano},\ and\ \citenamefont {Yabu}}]{Yabu-2014}%
  \BibitemOpen
  \bibfield  {author} {\bibinfo {author} {\bibfnamefont {K.}~\bibnamefont {Shirasaki}}, \bibinfo {author} {\bibfnamefont {E.}~\bibnamefont {Nakano}},\ and\ \bibinfo {author} {\bibfnamefont {H.}~\bibnamefont {Yabu}},\ }\bibfield  {title} {\bibinfo {title} {{Bose-Einstein condensation and density collapse in a weakly coupled boson-fermion mixture}},\ }\href {https://doi.org/10.1103/PhysRevA.90.063629} {\bibfield  {journal} {\bibinfo  {journal} {Phys. Rev. A}\ }\textbf {\bibinfo {volume} {90}},\ \bibinfo {pages} {063629} (\bibinfo {year} {2014})}\BibitemShut {NoStop}%
\bibitem [{\citenamefont {D'Alberto}\ \emph {et~al.}(2024)\citenamefont {D'Alberto}, \citenamefont {Cardarelli}, \citenamefont {Galli}, \citenamefont {Bertaina},\ and\ \citenamefont {Pieri}}]{DAlberto-2024}%
  \BibitemOpen
  \bibfield  {author} {\bibinfo {author} {\bibfnamefont {J.}~\bibnamefont {D'Alberto}}, \bibinfo {author} {\bibfnamefont {L.}~\bibnamefont {Cardarelli}}, \bibinfo {author} {\bibfnamefont {D.~E.}\ \bibnamefont {Galli}}, \bibinfo {author} {\bibfnamefont {G.}~\bibnamefont {Bertaina}},\ and\ \bibinfo {author} {\bibfnamefont {P.}~\bibnamefont {Pieri}},\ }\bibfield  {title} {\bibinfo {title} {{Quantum Monte Carlo and perturbative study of two-dimensional Bose-Fermi mixtures}},\ }\href {https://doi.org/10.1103/PhysRevA.109.053302} {\bibfield  {journal} {\bibinfo  {journal} {Phys. Rev. A}\ }\textbf {\bibinfo {volume} {109}},\ \bibinfo {pages} {053302} (\bibinfo {year} {2024})}\BibitemShut {NoStop}%
\bibitem [{\citenamefont {Averbuch}(1986)}]{Averbuch-1986}%
  \BibitemOpen
  \bibfield  {author} {\bibinfo {author} {\bibfnamefont {P.~G.}\ \bibnamefont {Averbuch}},\ }\bibfield  {title} {\bibinfo {title} {{Zero energy divergence of scattering cross sections in two dimensions}},\ }\href {https://doi.org/10.1088/0305-4470/19/12/018} {\bibfield  {journal} {\bibinfo  {journal} {J. Phys. A: Math. Gen.}\ }\textbf {\bibinfo {volume} {19}},\ \bibinfo {pages} {2325} (\bibinfo {year} {1986})}\BibitemShut {NoStop}%
\bibitem [{\citenamefont {Hammer}\ and\ \citenamefont {Lee}(2010)}]{Hammer-2010}%
  \BibitemOpen
  \bibfield  {author} {\bibinfo {author} {\bibfnamefont {H.-W.}\ \bibnamefont {Hammer}}\ and\ \bibinfo {author} {\bibfnamefont {D.}~\bibnamefont {Lee}},\ }\bibfield  {title} {\bibinfo {title} {Causality and the effective range expansion},\ }\href {https://doi.org/10.1016/j.aop.2010.06.006} {\bibfield  {journal} {\bibinfo  {journal} {Ann. Phys. (N.Y.)}\ }\textbf {\bibinfo {volume} {325}},\ \bibinfo {pages} {2212} (\bibinfo {year} {2010})}\BibitemShut {NoStop}%
\bibitem [{\citenamefont {Bloom}(1975)}]{Bloom-1975}%
  \BibitemOpen
  \bibfield  {author} {\bibinfo {author} {\bibfnamefont {P.}~\bibnamefont {Bloom}},\ }\bibfield  {title} {\bibinfo {title} {{Two-dimensional Fermi gas}},\ }\href {https://doi.org/10.1103/PhysRevB.12.125} {\bibfield  {journal} {\bibinfo  {journal} {Phys. Rev. B}\ }\textbf {\bibinfo {volume} {12}},\ \bibinfo {pages} {125} (\bibinfo {year} {1975})}\BibitemShut {NoStop}%
\bibitem [{\citenamefont {Olshanii}(1998)}]{Olshanii-1998}%
  \BibitemOpen
  \bibfield  {author} {\bibinfo {author} {\bibfnamefont {M.}~\bibnamefont {Olshanii}},\ }\bibfield  {title} {\bibinfo {title} {{Atomic Scattering in the Presence of an External Confinement and a Gas of Impenetrable Bosons}},\ }\href {https://doi.org/10.1103/PhysRevLett.81.938} {\bibfield  {journal} {\bibinfo  {journal} {Phys. Rev. Lett.}\ }\textbf {\bibinfo {volume} {81}},\ \bibinfo {pages} {938} (\bibinfo {year} {1998})}\BibitemShut {NoStop}%
\bibitem [{\citenamefont {Haller}\ \emph {et~al.}(2010)\citenamefont {Haller}, \citenamefont {Mark}, \citenamefont {Hart}, \citenamefont {Danzl}, \citenamefont {Reichs\"ollner}, \citenamefont {Melezhik}, \citenamefont {Schmelcher},\ and\ \citenamefont {N\"agerl}}]{Haller-2010}%
  \BibitemOpen
  \bibfield  {author} {\bibinfo {author} {\bibfnamefont {E.}~\bibnamefont {Haller}}, \bibinfo {author} {\bibfnamefont {M.~J.}\ \bibnamefont {Mark}}, \bibinfo {author} {\bibfnamefont {R.}~\bibnamefont {Hart}}, \bibinfo {author} {\bibfnamefont {J.~G.}\ \bibnamefont {Danzl}}, \bibinfo {author} {\bibfnamefont {L.}~\bibnamefont {Reichs\"ollner}}, \bibinfo {author} {\bibfnamefont {V.}~\bibnamefont {Melezhik}}, \bibinfo {author} {\bibfnamefont {P.}~\bibnamefont {Schmelcher}},\ and\ \bibinfo {author} {\bibfnamefont {H.-C.}\ \bibnamefont {N\"agerl}},\ }\bibfield  {title} {\bibinfo {title} {Confinement-induced resonances in low-dimensional quantum systems},\ }\href {https://doi.org/10.1103/PhysRevLett.104.153203} {\bibfield  {journal} {\bibinfo  {journal} {Phys. Rev. Lett.}\ }\textbf {\bibinfo {volume} {104}},\ \bibinfo {pages} {153203} (\bibinfo {year} {2010})}\BibitemShut {NoStop}%
\bibitem [{\citenamefont {Mathey}\ \emph {et~al.}(2006)\citenamefont {Mathey}, \citenamefont {Tsai},\ and\ \citenamefont {Neto}}]{Mathey-2006}%
  \BibitemOpen
  \bibfield  {author} {\bibinfo {author} {\bibfnamefont {L.}~\bibnamefont {Mathey}}, \bibinfo {author} {\bibfnamefont {S.-W.}\ \bibnamefont {Tsai}},\ and\ \bibinfo {author} {\bibfnamefont {A.~H.~C.}\ \bibnamefont {Neto}},\ }\bibfield  {title} {\bibinfo {title} {{Competing Types of Order in Two-Dimensional Bose-Fermi Mixtures}},\ }\href {https://doi.org/10.1103/PhysRevLett.97.030601} {\bibfield  {journal} {\bibinfo  {journal} {Phys. Rev. Lett.}\ }\textbf {\bibinfo {volume} {97}},\ \bibinfo {pages} {030601} (\bibinfo {year} {2006})}\BibitemShut {NoStop}%
\bibitem [{\citenamefont {Suba{\c{s}}i}\ \emph {et~al.}(2010)\citenamefont {Suba{\c{s}}i}, \citenamefont {Sevin{\c{c}}li}, \citenamefont {Vignolo},\ and\ \citenamefont {Tanatar}}]{Subasi-2010}%
  \BibitemOpen
  \bibfield  {author} {\bibinfo {author} {\bibfnamefont {A.~L.}\ \bibnamefont {Suba{\c{s}}i}}, \bibinfo {author} {\bibfnamefont {S.}~\bibnamefont {Sevin{\c{c}}li}}, \bibinfo {author} {\bibfnamefont {P.}~\bibnamefont {Vignolo}},\ and\ \bibinfo {author} {\bibfnamefont {B.}~\bibnamefont {Tanatar}},\ }\bibfield  {title} {\bibinfo {title} {{Dimensional crossover in two-dimensional Bose-Fermi mixtures}},\ }\href {https://doi.org/10.1134/S1054660X1005018X} {\bibfield  {journal} {\bibinfo  {journal} {Las. Phys.}\ }\textbf {\bibinfo {volume} {20}},\ \bibinfo {pages} {683} (\bibinfo {year} {2010})}\BibitemShut {NoStop}%
\bibitem [{\citenamefont {Noda}\ \emph {et~al.}(2011)\citenamefont {Noda}, \citenamefont {Peters}, \citenamefont {Kawakami},\ and\ \citenamefont {Pruschke}}]{Noda-2011}%
  \BibitemOpen
  \bibfield  {author} {\bibinfo {author} {\bibfnamefont {K.}~\bibnamefont {Noda}}, \bibinfo {author} {\bibfnamefont {R.}~\bibnamefont {Peters}}, \bibinfo {author} {\bibfnamefont {N.}~\bibnamefont {Kawakami}},\ and\ \bibinfo {author} {\bibfnamefont {T.}~\bibnamefont {Pruschke}},\ }\bibfield  {title} {\bibinfo {title} {{Quantum Phases of Bose\textendash Fermi Mixtures in Optical Lattices}},\ }\href {https://doi.org/10.1088/1742-6596/273/1/012146} {\bibfield  {journal} {\bibinfo  {journal} {J. Phys.: Conf. Ser.}\ }\textbf {\bibinfo {volume} {273}},\ \bibinfo {pages} {012146} (\bibinfo {year} {2011})}\BibitemShut {NoStop}%
\bibitem [{\citenamefont {Krotscheck}\ \emph {et~al.}(2000)\citenamefont {Krotscheck}, \citenamefont {Paaso}, \citenamefont {Saarela},\ and\ \citenamefont {Sch\"orkhuber}}]{Krotscheck-2000}%
  \BibitemOpen
  \bibfield  {author} {\bibinfo {author} {\bibfnamefont {E.}~\bibnamefont {Krotscheck}}, \bibinfo {author} {\bibfnamefont {J.}~\bibnamefont {Paaso}}, \bibinfo {author} {\bibfnamefont {M.}~\bibnamefont {Saarela}},\ and\ \bibinfo {author} {\bibfnamefont {K.}~\bibnamefont {Sch\"orkhuber}},\ }\bibfield  {title} {\bibinfo {title} {Dimers in two-dimensional {${}^{3}{\mathrm{H}\mathrm{e}\ensuremath{-}}^{4}\mathrm{He}$} mixtures},\ }\href {https://doi.org/10.1103/PhysRevLett.85.2344} {\bibfield  {journal} {\bibinfo  {journal} {Phys. Rev. Lett.}\ }\textbf {\bibinfo {volume} {85}},\ \bibinfo {pages} {2344} (\bibinfo {year} {2000})}\BibitemShut {NoStop}%
\bibitem [{\citenamefont {Krotscheck}\ \emph {et~al.}(2001)\citenamefont {Krotscheck}, \citenamefont {Paaso}, \citenamefont {Saarela},\ and\ \citenamefont {Sch\"orkhuber}}]{Krotscheck-2001}%
  \BibitemOpen
  \bibfield  {author} {\bibinfo {author} {\bibfnamefont {E.}~\bibnamefont {Krotscheck}}, \bibinfo {author} {\bibfnamefont {J.}~\bibnamefont {Paaso}}, \bibinfo {author} {\bibfnamefont {M.}~\bibnamefont {Saarela}},\ and\ \bibinfo {author} {\bibfnamefont {K.}~\bibnamefont {Sch\"orkhuber}},\ }\bibfield  {title} {\bibinfo {title} {{Phases of ${}^{3}{\mathrm{H}\mathrm{e}\ensuremath{-}}^{4}\mathrm{He}$ mixtures in two dimensions}},\ }\href {https://doi.org/10.1103/PhysRevB.64.054504} {\bibfield  {journal} {\bibinfo  {journal} {Phys. Rev. B}\ }\textbf {\bibinfo {volume} {64}},\ \bibinfo {pages} {054504} (\bibinfo {year} {2001})}\BibitemShut {NoStop}%
\bibitem [{\citenamefont {Pandharipande}(1971{\natexlab{a}})}]{Pandharipande-1971}%
  \BibitemOpen
  \bibfield  {author} {\bibinfo {author} {\bibfnamefont {V.}~\bibnamefont {Pandharipande}},\ }\bibfield  {title} {\bibinfo {title} {{Dense neutron matter with realistic interactions}},\ }\href {https://doi.org/https://doi.org/10.1016/0375-9474(71)90413-1} {\bibfield  {journal} {\bibinfo  {journal} {Nucl. Phys. A}\ }\textbf {\bibinfo {volume} {174}},\ \bibinfo {pages} {641} (\bibinfo {year} {1971}{\natexlab{a}})}\BibitemShut {NoStop}%
\bibitem [{\citenamefont {Pandharipande}(1971{\natexlab{b}})}]{Pandharipande2-1971}%
  \BibitemOpen
  \bibfield  {author} {\bibinfo {author} {\bibfnamefont {V.}~\bibnamefont {Pandharipande}},\ }\bibfield  {title} {\bibinfo {title} {Hyperonic matter},\ }\href {https://doi.org/https://doi.org/10.1016/0375-9474(71)90193-X} {\bibfield  {journal} {\bibinfo  {journal} {Nucl. Phys. A}\ }\textbf {\bibinfo {volume} {178}},\ \bibinfo {pages} {123} (\bibinfo {year} {1971}{\natexlab{b}})}\BibitemShut {NoStop}%
\bibitem [{\citenamefont {Pandharipande}\ and\ \citenamefont {Bethe}(1973)}]{Pandharipande-1973}%
  \BibitemOpen
  \bibfield  {author} {\bibinfo {author} {\bibfnamefont {V.~R.}\ \bibnamefont {Pandharipande}}\ and\ \bibinfo {author} {\bibfnamefont {H.~A.}\ \bibnamefont {Bethe}},\ }\bibfield  {title} {\bibinfo {title} {{Variational Method for Dense Systems}},\ }\href {https://doi.org/10.1103/PhysRevC.7.1312} {\bibfield  {journal} {\bibinfo  {journal} {Phys. Rev. C}\ }\textbf {\bibinfo {volume} {7}},\ \bibinfo {pages} {1312} (\bibinfo {year} {1973})}\BibitemShut {NoStop}%
\bibitem [{\citenamefont {Cowell}\ \emph {et~al.}(2002)\citenamefont {Cowell}, \citenamefont {Heiselberg}, \citenamefont {Mazets}, \citenamefont {Morales}, \citenamefont {Pandharipande},\ and\ \citenamefont {Pethick}}]{Pethick-2002}%
  \BibitemOpen
  \bibfield  {author} {\bibinfo {author} {\bibfnamefont {S.}~\bibnamefont {Cowell}}, \bibinfo {author} {\bibfnamefont {H.}~\bibnamefont {Heiselberg}}, \bibinfo {author} {\bibfnamefont {I.~E.}\ \bibnamefont {Mazets}}, \bibinfo {author} {\bibfnamefont {J.}~\bibnamefont {Morales}}, \bibinfo {author} {\bibfnamefont {V.~R.}\ \bibnamefont {Pandharipande}},\ and\ \bibinfo {author} {\bibfnamefont {C.~J.}\ \bibnamefont {Pethick}},\ }\bibfield  {title} {\bibinfo {title} {{Cold Bose Gases with Large Scattering Lengths}},\ }\href {https://doi.org/10.1103/PhysRevLett.88.210403} {\bibfield  {journal} {\bibinfo  {journal} {Phys. Rev. Lett.}\ }\textbf {\bibinfo {volume} {88}},\ \bibinfo {pages} {210403} (\bibinfo {year} {2002})}\BibitemShut {NoStop}%
\bibitem [{\citenamefont {Heiselberg}(2011)}]{Heiselberg-2011}%
  \BibitemOpen
  \bibfield  {author} {\bibinfo {author} {\bibfnamefont {H.}~\bibnamefont {Heiselberg}},\ }\bibfield  {title} {\bibinfo {title} {{Itinerant ferromagnetism in ultracold Fermi gases}},\ }\href {https://doi.org/10.1103/PhysRevA.83.053635} {\bibfield  {journal} {\bibinfo  {journal} {Phys. Rev. A}\ }\textbf {\bibinfo {volume} {83}},\ \bibinfo {pages} {053635} (\bibinfo {year} {2011})}\BibitemShut {NoStop}%
\bibitem [{\citenamefont {Taylor}\ \emph {et~al.}(2011)\citenamefont {Taylor}, \citenamefont {Zhang}, \citenamefont {Schneider},\ and\ \citenamefont {Randeria}}]{Taylor-2011}%
  \BibitemOpen
  \bibfield  {author} {\bibinfo {author} {\bibfnamefont {E.}~\bibnamefont {Taylor}}, \bibinfo {author} {\bibfnamefont {S.}~\bibnamefont {Zhang}}, \bibinfo {author} {\bibfnamefont {W.}~\bibnamefont {Schneider}},\ and\ \bibinfo {author} {\bibfnamefont {M.}~\bibnamefont {Randeria}},\ }\bibfield  {title} {\bibinfo {title} {Colliding clouds of strongly interacting spin-polarized fermions},\ }\href {https://doi.org/10.1103/PhysRevA.84.063622} {\bibfield  {journal} {\bibinfo  {journal} {Phys. Rev. A}\ }\textbf {\bibinfo {volume} {84}},\ \bibinfo {pages} {063622} (\bibinfo {year} {2011})}\BibitemShut {NoStop}%
\bibitem [{\citenamefont {Grochowski}\ \emph {et~al.}(2020)\citenamefont {Grochowski}, \citenamefont {Karpiuk}, \citenamefont {Brewczyk},\ and\ \citenamefont {Rz\k{a}żewski}}]{Grochowski-2020}%
  \BibitemOpen
  \bibfield  {author} {\bibinfo {author} {\bibfnamefont {P.~T.}\ \bibnamefont {Grochowski}}, \bibinfo {author} {\bibfnamefont {T.}~\bibnamefont {Karpiuk}}, \bibinfo {author} {\bibfnamefont {M.}~\bibnamefont {Brewczyk}},\ and\ \bibinfo {author} {\bibfnamefont {K.}~\bibnamefont {Rz\k{a}żewski}},\ }\bibfield  {title} {\bibinfo {title} {Breathing mode of a {Bose-Einstein} condensate immersed in a {Fermi} sea},\ }\href {https://doi.org/10.1103/PhysRevLett.125.103401} {\bibfield  {journal} {\bibinfo  {journal} {Phys. Rev. Lett.}\ }\textbf {\bibinfo {volume} {125}},\ \bibinfo {pages} {103401} (\bibinfo {year} {2020})}\BibitemShut {NoStop}%
\bibitem [{\citenamefont {Gawryluk}\ and\ \citenamefont {Brewczyk}(2024)}]{Gawryluk-2024}%
  \BibitemOpen
  \bibfield  {author} {\bibinfo {author} {\bibfnamefont {K.}~\bibnamefont {Gawryluk}}\ and\ \bibinfo {author} {\bibfnamefont {M.}~\bibnamefont {Brewczyk}},\ }\bibfield  {title} {\bibinfo {title} {{Mechanism for sound dissipation in a two-dimensional degenerate Fermi gas}},\ }\href {https://doi.org/10.1038/s41598-024-61521-5} {\bibfield  {journal} {\bibinfo  {journal} {Sci. Rep.}\ }\textbf {\bibinfo {volume} {14}},\ \bibinfo {pages} {10815} (\bibinfo {year} {2024})}\BibitemShut {NoStop}%
\bibitem [{\citenamefont {Massignan}\ \emph {et~al.}(2014)\citenamefont {Massignan}, \citenamefont {Zaccanti},\ and\ \citenamefont {Bruun}}]{Massignan-2014}%
  \BibitemOpen
  \bibfield  {author} {\bibinfo {author} {\bibfnamefont {P.}~\bibnamefont {Massignan}}, \bibinfo {author} {\bibfnamefont {M.}~\bibnamefont {Zaccanti}},\ and\ \bibinfo {author} {\bibfnamefont {G.~M.}\ \bibnamefont {Bruun}},\ }\bibfield  {title} {\bibinfo {title} {Polarons, dressed molecules and itinerant ferromagnetism in ultracold fermi gases},\ }\href {https://doi.org/10.1088/0034-4885/77/3/034401} {\bibfield  {journal} {\bibinfo  {journal} {Rep. Progr. Phys.}\ }\textbf {\bibinfo {volume} {77}},\ \bibinfo {pages} {034401} (\bibinfo {year} {2014})}\BibitemShut {NoStop}%
\bibitem [{\citenamefont {Jastrow}(1955)}]{Jastrow-1955}%
  \BibitemOpen
  \bibfield  {author} {\bibinfo {author} {\bibfnamefont {R.}~\bibnamefont {Jastrow}},\ }\bibfield  {title} {\bibinfo {title} {Many-body problem with strong forces},\ }\href {https://doi.org/10.1103/PhysRev.98.1479} {\bibfield  {journal} {\bibinfo  {journal} {Phys. Rev.}\ }\textbf {\bibinfo {volume} {98}},\ \bibinfo {pages} {1479} (\bibinfo {year} {1955})}\BibitemShut {NoStop}%
\bibitem [{\citenamefont {de~Boer}(1949)}]{deBoer-1949}%
  \BibitemOpen
  \bibfield  {author} {\bibinfo {author} {\bibfnamefont {J.}~\bibnamefont {de~Boer}},\ }\bibfield  {title} {\bibinfo {title} {{Molecular distribution and equation of state of gases}},\ }\href {https://doi.org/10.1088/0034-4885/12/1/314} {\bibfield  {journal} {\bibinfo  {journal} {Rep. Progr. Phys.}\ }\textbf {\bibinfo {volume} {12}},\ \bibinfo {pages} {305} (\bibinfo {year} {1949})}\BibitemShut {NoStop}%
\bibitem [{\citenamefont {Chang}(2006)}]{Chang-2006}%
  \BibitemOpen
  \bibfield  {author} {\bibinfo {author} {\bibfnamefont {S.~Y.}\ \bibnamefont {Chang}},\ }\emph {\bibinfo {title} {{Study of the properties of dilute Fermi gases in the strongly interacting regime}}},\ \href {https://www.ideals.illinois.edu/items/32371} {\bibinfo {type} {{Ph.D. thesis}}},\ \bibinfo  {school} {University of Illinois at Urbana-Champaign} (\bibinfo {year} {2006})\BibitemShut {NoStop}%
\bibitem [{\citenamefont {Cordioli}(2025)}]{Cordioli-2025}%
  \BibitemOpen
  \bibfield  {author} {\bibinfo {author} {\bibfnamefont {P.}~\bibnamefont {Cordioli}},\ }\emph {\bibinfo {title} {Lowest order constrained variational approximation for ultracold Bose-Fermi mixtures}},\ \href {https://amslaurea.unibo.it/id/eprint/35871/} {\bibinfo {type} {Master's thesis}},\ \bibinfo  {school} {University of Bologna} (\bibinfo {year} {2025})\BibitemShut {NoStop}%
\bibitem [{\citenamefont {Emery}(1958)}]{Emery-1958}%
  \BibitemOpen
  \bibfield  {author} {\bibinfo {author} {\bibfnamefont {V.}~\bibnamefont {Emery}},\ }\bibfield  {title} {\bibinfo {title} {A variational approach to the nuclear many-body problem},\ }\href {https://doi.org/https://doi.org/10.1016/0029-5582(58)90211-6} {\bibfield  {journal} {\bibinfo  {journal} {Nucl. Phys.}\ }\textbf {\bibinfo {volume} {6}},\ \bibinfo {pages} {585} (\bibinfo {year} {1958})}\BibitemShut {NoStop}%
\bibitem [{\citenamefont {Clark}\ and\ \citenamefont {Ristig}(1972)}]{Clark-1972}%
  \BibitemOpen
  \bibfield  {author} {\bibinfo {author} {\bibfnamefont {J.~W.}\ \bibnamefont {Clark}}\ and\ \bibinfo {author} {\bibfnamefont {M.~L.}\ \bibnamefont {Ristig}},\ }\bibfield  {title} {\bibinfo {title} {{Subsidiary Conditions on Nuclear Many-Body Theories}},\ }\href {https://doi.org/10.1103/PhysRevC.5.1553} {\bibfield  {journal} {\bibinfo  {journal} {Phys. Rev. C}\ }\textbf {\bibinfo {volume} {5}},\ \bibinfo {pages} {1553} (\bibinfo {year} {1972})}\BibitemShut {NoStop}%
\bibitem [{\citenamefont {Krotscheck}(1977)}]{Krotscheck-1977}%
  \BibitemOpen
  \bibfield  {author} {\bibinfo {author} {\bibfnamefont {E.}~\bibnamefont {Krotscheck}},\ }\bibfield  {title} {\bibinfo {title} {{Variational problem in Jastrow theory}},\ }\href {https://doi.org/10.1103/PhysRevA.15.397} {\bibfield  {journal} {\bibinfo  {journal} {Phys. Rev. A}\ }\textbf {\bibinfo {volume} {15}},\ \bibinfo {pages} {397} (\bibinfo {year} {1977})}\BibitemShut {NoStop}%
\bibitem [{\citenamefont {Pilati}\ \emph {et~al.}(2005)\citenamefont {Pilati}, \citenamefont {Boronat}, \citenamefont {Casulleras},\ and\ \citenamefont {Giorgini}}]{Boronat-2005}%
  \BibitemOpen
  \bibfield  {author} {\bibinfo {author} {\bibfnamefont {S.}~\bibnamefont {Pilati}}, \bibinfo {author} {\bibfnamefont {J.}~\bibnamefont {Boronat}}, \bibinfo {author} {\bibfnamefont {J.}~\bibnamefont {Casulleras}},\ and\ \bibinfo {author} {\bibfnamefont {S.}~\bibnamefont {Giorgini}},\ }\bibfield  {title} {\bibinfo {title} {{Quantum Monte Carlo simulation of a two-dimensional Bose gas}},\ }\href {https://doi.org/10.1103/PhysRevA.71.023605} {\bibfield  {journal} {\bibinfo  {journal} {Phys. Rev. A}\ }\textbf {\bibinfo {volume} {71}},\ \bibinfo {pages} {023605} (\bibinfo {year} {2005})}\BibitemShut {NoStop}%
\bibitem [{\citenamefont {Whitehead}\ \emph {et~al.}(2016)\citenamefont {Whitehead}, \citenamefont {Schonenberg}, \citenamefont {Kongsuwan}, \citenamefont {Needs},\ and\ \citenamefont {Conduit}}]{Whitehead-2016}%
  \BibitemOpen
  \bibfield  {author} {\bibinfo {author} {\bibfnamefont {T.~M.}\ \bibnamefont {Whitehead}}, \bibinfo {author} {\bibfnamefont {L.~M.}\ \bibnamefont {Schonenberg}}, \bibinfo {author} {\bibfnamefont {N.}~\bibnamefont {Kongsuwan}}, \bibinfo {author} {\bibfnamefont {R.~J.}\ \bibnamefont {Needs}},\ and\ \bibinfo {author} {\bibfnamefont {G.~J.}\ \bibnamefont {Conduit}},\ }\bibfield  {title} {\bibinfo {title} {Pseudopotential for the two-dimensional contact interaction},\ }\href {https://doi.org/10.1103/PhysRevA.93.042702} {\bibfield  {journal} {\bibinfo  {journal} {Phys. Rev. A}\ }\textbf {\bibinfo {volume} {93}},\ \bibinfo {pages} {042702} (\bibinfo {year} {2016})}\BibitemShut {NoStop}%
\bibitem [{\citenamefont {Bethe}\ and\ \citenamefont {Peierls}(1935)}]{Bethe-1935}%
  \BibitemOpen
  \bibfield  {author} {\bibinfo {author} {\bibfnamefont {H.}~\bibnamefont {Bethe}}\ and\ \bibinfo {author} {\bibfnamefont {R.}~\bibnamefont {Peierls}},\ }\bibfield  {title} {\bibinfo {title} {Quantum theory of the diplon},\ }\href {https://doi.org/10.1098/rspa.1935.0010} {\bibfield  {journal} {\bibinfo  {journal} {Proc. R. Soc. Lond. A Math. Phys. Sci.}\ }\textbf {\bibinfo {volume} {148}},\ \bibinfo {pages} {146} (\bibinfo {year} {1935})}\BibitemShut {NoStop}%
\bibitem [{\citenamefont {Bertaina}\ and\ \citenamefont {Giorgini}(2011)}]{Bertaina-2011}%
  \BibitemOpen
  \bibfield  {author} {\bibinfo {author} {\bibfnamefont {G.}~\bibnamefont {Bertaina}}\ and\ \bibinfo {author} {\bibfnamefont {S.}~\bibnamefont {Giorgini}},\ }\bibfield  {title} {\bibinfo {title} {{BCS-BEC Crossover in a Two-Dimensional Fermi Gas}},\ }\href {https://doi.org/10.1103/PhysRevLett.106.110403} {\bibfield  {journal} {\bibinfo  {journal} {Phys. Rev. Lett.}\ }\textbf {\bibinfo {volume} {106}},\ \bibinfo {pages} {110403} (\bibinfo {year} {2011})}\BibitemShut {NoStop}%
\bibitem [{\citenamefont {Chang}\ \emph {et~al.}(2004)\citenamefont {Chang}, \citenamefont {Pandharipande}, \citenamefont {Carlson},\ and\ \citenamefont {Schmidt}}]{Chang-2004}%
  \BibitemOpen
  \bibfield  {author} {\bibinfo {author} {\bibfnamefont {S.~Y.}\ \bibnamefont {Chang}}, \bibinfo {author} {\bibfnamefont {V.~R.}\ \bibnamefont {Pandharipande}}, \bibinfo {author} {\bibfnamefont {J.}~\bibnamefont {Carlson}},\ and\ \bibinfo {author} {\bibfnamefont {K.~E.}\ \bibnamefont {Schmidt}},\ }\bibfield  {title} {\bibinfo {title} {{Quantum Monte Carlo studies of superfluid Fermi gases}},\ }\href {https://doi.org/10.1103/PhysRevA.70.043602} {\bibfield  {journal} {\bibinfo  {journal} {Phys. Rev. A}\ }\textbf {\bibinfo {volume} {70}},\ \bibinfo {pages} {043602} (\bibinfo {year} {2004})}\BibitemShut {NoStop}%
\bibitem [{\citenamefont {Astrakharchik}\ \emph {et~al.}(2004)\citenamefont {Astrakharchik}, \citenamefont {Boronat}, \citenamefont {Casulleras},\ and\ \citenamefont {Giorgini}}]{Astra-2004}%
  \BibitemOpen
  \bibfield  {author} {\bibinfo {author} {\bibfnamefont {G.~E.}\ \bibnamefont {Astrakharchik}}, \bibinfo {author} {\bibfnamefont {J.}~\bibnamefont {Boronat}}, \bibinfo {author} {\bibfnamefont {J.}~\bibnamefont {Casulleras}},\ and\ \bibinfo {author} {\bibfnamefont {S.}~\bibnamefont {Giorgini}},\ }\bibfield  {title} {\bibinfo {title} {{Equation of State of a Fermi Gas in the BEC-BCS Crossover: A Quantum Monte Carlo Study}},\ }\href {https://doi.org/10.1103/PhysRevLett.93.200404} {\bibfield  {journal} {\bibinfo  {journal} {Phys. Rev. Lett.}\ }\textbf {\bibinfo {volume} {93}},\ \bibinfo {pages} {200404} (\bibinfo {year} {2004})}\BibitemShut {NoStop}%
\bibitem [{\citenamefont {Chang}\ \emph {et~al.}(2011)\citenamefont {Chang}, \citenamefont {Randeria},\ and\ \citenamefont {Trivedi}}]{Randeria-2011}%
  \BibitemOpen
  \bibfield  {author} {\bibinfo {author} {\bibfnamefont {S.-Y.}\ \bibnamefont {Chang}}, \bibinfo {author} {\bibfnamefont {M.}~\bibnamefont {Randeria}},\ and\ \bibinfo {author} {\bibfnamefont {N.}~\bibnamefont {Trivedi}},\ }\bibfield  {title} {\bibinfo {title} {{Ferromagnetism in the upper branch of the Feshbach resonance and the hard-sphere Fermi gas}},\ }\href {https://doi.org/10.1073/pnas.1011990108} {\bibfield  {journal} {\bibinfo  {journal} {PNAS}\ }\textbf {\bibinfo {volume} {108}},\ \bibinfo {pages} {51} (\bibinfo {year} {2011})}\BibitemShut {NoStop}%
\bibitem [{\citenamefont {Pilati}\ \emph {et~al.}(2021)\citenamefont {Pilati}, \citenamefont {Orso},\ and\ \citenamefont {Bertaina}}]{Bertaina-2021}%
  \BibitemOpen
  \bibfield  {author} {\bibinfo {author} {\bibfnamefont {S.}~\bibnamefont {Pilati}}, \bibinfo {author} {\bibfnamefont {G.}~\bibnamefont {Orso}},\ and\ \bibinfo {author} {\bibfnamefont {G.}~\bibnamefont {Bertaina}},\ }\bibfield  {title} {\bibinfo {title} {{Quantum Monte Carlo simulations of two-dimensional repulsive Fermi gases with population imbalance}},\ }\href {https://doi.org/10.1103/PhysRevA.103.063314} {\bibfield  {journal} {\bibinfo  {journal} {Phys. Rev. A}\ }\textbf {\bibinfo {volume} {103}},\ \bibinfo {pages} {063314} (\bibinfo {year} {2021})}\BibitemShut {NoStop}%
\bibitem [{\citenamefont {Werner}\ and\ \citenamefont {Castin}(2012)}]{Werner-2012}%
  \BibitemOpen
  \bibfield  {author} {\bibinfo {author} {\bibfnamefont {F.}~\bibnamefont {Werner}}\ and\ \bibinfo {author} {\bibfnamefont {Y.}~\bibnamefont {Castin}},\ }\bibfield  {title} {\bibinfo {title} {{General relations for quantum gases in two and three dimensions: Two-component fermions}},\ }\href {https://doi.org/10.1103/PhysRevA.86.013626} {\bibfield  {journal} {\bibinfo  {journal} {Phys. Rev. A}\ }\textbf {\bibinfo {volume} {86}},\ \bibinfo {pages} {013626} (\bibinfo {year} {2012})}\BibitemShut {NoStop}%
\bibitem [{\citenamefont {Chang}\ and\ \citenamefont {Pandharipande}(2005)}]{Chang-2005}%
  \BibitemOpen
  \bibfield  {author} {\bibinfo {author} {\bibfnamefont {S.~Y.}\ \bibnamefont {Chang}}\ and\ \bibinfo {author} {\bibfnamefont {V.~R.}\ \bibnamefont {Pandharipande}},\ }\bibfield  {title} {\bibinfo {title} {{Ground-State Properties of Fermi Gases in the Strongly Interacting Regime}},\ }\href {https://doi.org/10.1103/PhysRevLett.95.080402} {\bibfield  {journal} {\bibinfo  {journal} {Phys. Rev. Lett.}\ }\textbf {\bibinfo {volume} {95}},\ \bibinfo {pages} {080402} (\bibinfo {year} {2005})}\BibitemShut {NoStop}%
\bibitem [{\citenamefont {Pilati}\ \emph {et~al.}(2010)\citenamefont {Pilati}, \citenamefont {Bertaina}, \citenamefont {Giorgini},\ and\ \citenamefont {Troyer}}]{Pilati-2010}%
  \BibitemOpen
  \bibfield  {author} {\bibinfo {author} {\bibfnamefont {S.}~\bibnamefont {Pilati}}, \bibinfo {author} {\bibfnamefont {G.}~\bibnamefont {Bertaina}}, \bibinfo {author} {\bibfnamefont {S.}~\bibnamefont {Giorgini}},\ and\ \bibinfo {author} {\bibfnamefont {M.}~\bibnamefont {Troyer}},\ }\bibfield  {title} {\bibinfo {title} {{Itinerant Ferromagnetism of a Repulsive Atomic Fermi Gas: A Quantum Monte Carlo Study}},\ }\href {https://doi.org/10.1103/PhysRevLett.105.030405} {\bibfield  {journal} {\bibinfo  {journal} {Phys. Rev. Lett.}\ }\textbf {\bibinfo {volume} {105}},\ \bibinfo {pages} {030405} (\bibinfo {year} {2010})}\BibitemShut {NoStop}%
\bibitem [{\citenamefont {Koschorreck}\ \emph {et~al.}(2012)\citenamefont {Koschorreck}, \citenamefont {Pertot}, \citenamefont {Vogt}, \citenamefont {Fröhlich}, \citenamefont {Feld},\ and\ \citenamefont {Köhl}}]{Koehl-2012}%
  \BibitemOpen
  \bibfield  {author} {\bibinfo {author} {\bibfnamefont {M.}~\bibnamefont {Koschorreck}}, \bibinfo {author} {\bibfnamefont {D.}~\bibnamefont {Pertot}}, \bibinfo {author} {\bibfnamefont {E.}~\bibnamefont {Vogt}}, \bibinfo {author} {\bibfnamefont {B.}~\bibnamefont {Fröhlich}}, \bibinfo {author} {\bibfnamefont {M.}~\bibnamefont {Feld}},\ and\ \bibinfo {author} {\bibfnamefont {M.}~\bibnamefont {Köhl}},\ }\bibfield  {title} {\bibinfo {title} {{Attractive and repulsive Fermi polarons in two dimensions}},\ }\href {https://doi.org/10.1038/nature11151} {\bibfield  {journal} {\bibinfo  {journal} {Nature}\ }\textbf {\bibinfo {volume} {485}},\ \bibinfo {pages} {619–622} (\bibinfo {year} {2012})}\BibitemShut {NoStop}%
\bibitem [{\citenamefont {Ngampruetikorn}\ \emph {et~al.}(2012)\citenamefont {Ngampruetikorn}, \citenamefont {Levinsen},\ and\ \citenamefont {Parish}}]{Parish-2012}%
  \BibitemOpen
  \bibfield  {author} {\bibinfo {author} {\bibfnamefont {V.}~\bibnamefont {Ngampruetikorn}}, \bibinfo {author} {\bibfnamefont {J.}~\bibnamefont {Levinsen}},\ and\ \bibinfo {author} {\bibfnamefont {M.~M.}\ \bibnamefont {Parish}},\ }\bibfield  {title} {\bibinfo {title} {{Repulsive polarons in two-dimensional Fermi gases}},\ }\href {https://doi.org/10.1209/0295-5075/98/30005} {\bibfield  {journal} {\bibinfo  {journal} {Europhys. Lett.}\ }\textbf {\bibinfo {volume} {98}},\ \bibinfo {pages} {30005} (\bibinfo {year} {2012})}\BibitemShut {NoStop}%
\bibitem [{\citenamefont {Chevy}(2006)}]{Chevy-2006}%
  \BibitemOpen
  \bibfield  {author} {\bibinfo {author} {\bibfnamefont {F.}~\bibnamefont {Chevy}},\ }\bibfield  {title} {\bibinfo {title} {Universal phase diagram of a strongly interacting {Fermi} gas with unbalanced spin populations},\ }\href {https://doi.org/10.1103/PhysRevA.74.063628} {\bibfield  {journal} {\bibinfo  {journal} {Physical Review A}\ }\textbf {\bibinfo {volume} {74}},\ \bibinfo {pages} {063628} (\bibinfo {year} {2006})}\BibitemShut {NoStop}%
\bibitem [{\citenamefont {Christianen}\ \emph {et~al.}(2022)\citenamefont {Christianen}, \citenamefont {Cirac},\ and\ \citenamefont {Schmidt}}]{Christianen-2022}%
  \BibitemOpen
  \bibfield  {author} {\bibinfo {author} {\bibfnamefont {A.}~\bibnamefont {Christianen}}, \bibinfo {author} {\bibfnamefont {J.~I.}\ \bibnamefont {Cirac}},\ and\ \bibinfo {author} {\bibfnamefont {R.}~\bibnamefont {Schmidt}},\ }\bibfield  {title} {\bibinfo {title} {{Bose polaron and the Efimov effect: A Gaussian-state approach}},\ }\href {https://doi.org/10.1103/PhysRevA.105.053302} {\bibfield  {journal} {\bibinfo  {journal} {Phys. Rev. A}\ }\textbf {\bibinfo {volume} {105}},\ \bibinfo {pages} {053302} (\bibinfo {year} {2022})}\BibitemShut {NoStop}%
\bibitem [{\citenamefont {Christianen}\ \emph {et~al.}(2024)\citenamefont {Christianen}, \citenamefont {Cirac},\ and\ \citenamefont {Schmidt}}]{Christianen-2024}%
  \BibitemOpen
  \bibfield  {author} {\bibinfo {author} {\bibfnamefont {A.}~\bibnamefont {Christianen}}, \bibinfo {author} {\bibfnamefont {J.~I.}\ \bibnamefont {Cirac}},\ and\ \bibinfo {author} {\bibfnamefont {R.}~\bibnamefont {Schmidt}},\ }\bibfield  {title} {\bibinfo {title} {{Phase diagram for strong-coupling Bose polarons}},\ }\href {https://doi.org/10.21468/SciPostPhys.16.3.067} {\bibfield  {journal} {\bibinfo  {journal} {SciPost Phys.}\ }\textbf {\bibinfo {volume} {16}},\ \bibinfo {pages} {067} (\bibinfo {year} {2024})}\BibitemShut {NoStop}%
\bibitem [{\citenamefont {Yu}\ and\ \citenamefont {Pethick}(2012)}]{Yu-2012}%
  \BibitemOpen
  \bibfield  {author} {\bibinfo {author} {\bibfnamefont {Z.}~\bibnamefont {Yu}}\ and\ \bibinfo {author} {\bibfnamefont {C.~J.}\ \bibnamefont {Pethick}},\ }\bibfield  {title} {\bibinfo {title} {Induced interactions in dilute atomic gases and liquid helium mixtures},\ }\href {https://doi.org/10.1103/PhysRevA.85.063616} {\bibfield  {journal} {\bibinfo  {journal} {Phys. Rev. A}\ }\textbf {\bibinfo {volume} {85}},\ \bibinfo {pages} {063616} (\bibinfo {year} {2012})}\BibitemShut {NoStop}%
\bibitem [{\citenamefont {Schick}(1971)}]{Schick-1971}%
  \BibitemOpen
  \bibfield  {author} {\bibinfo {author} {\bibfnamefont {M.}~\bibnamefont {Schick}},\ }\bibfield  {title} {\bibinfo {title} {{Two-Dimensional System of Hard-Core Bosons}},\ }\href {https://doi.org/10.1103/PhysRevA.3.1067} {\bibfield  {journal} {\bibinfo  {journal} {Phys. Rev. A}\ }\textbf {\bibinfo {volume} {3}},\ \bibinfo {pages} {1067} (\bibinfo {year} {1971})}\BibitemShut {NoStop}%
\bibitem [{\citenamefont {Yu}\ \emph {et~al.}(2012)\citenamefont {Yu}, \citenamefont {Zhang},\ and\ \citenamefont {Zhai}}]{Yu-2012err}%
  \BibitemOpen
  \bibfield  {author} {\bibinfo {author} {\bibfnamefont {Z.-Q.}\ \bibnamefont {Yu}}, \bibinfo {author} {\bibfnamefont {S.}~\bibnamefont {Zhang}},\ and\ \bibinfo {author} {\bibfnamefont {H.}~\bibnamefont {Zhai}},\ }\bibfield  {title} {\bibinfo {title} {{Erratum: Stability condition of a strongly interacting boson-fermion mixture across an interspecies Feshbach resonance [Phys. Rev. A 83, 041603(R) (2011)]}},\ }\href {https://doi.org/10.1103/PhysRevA.86.069904} {\bibfield  {journal} {\bibinfo  {journal} {Phys. Rev. A}\ }\textbf {\bibinfo {volume} {86}},\ \bibinfo {pages} {069904} (\bibinfo {year} {2012})}\BibitemShut {NoStop}%
\bibitem [{\citenamefont {Ristig}\ \emph {et~al.}(1975)\citenamefont {Ristig}, \citenamefont {Lam},\ and\ \citenamefont {Clark}}]{Ristig-1975}%
  \BibitemOpen
  \bibfield  {author} {\bibinfo {author} {\bibfnamefont {M.}~\bibnamefont {Ristig}}, \bibinfo {author} {\bibfnamefont {P.}~\bibnamefont {Lam}},\ and\ \bibinfo {author} {\bibfnamefont {J.}~\bibnamefont {Clark}},\ }\bibfield  {title} {\bibinfo {title} {{Condensate fraction and momentum distribution of liquid Helium}},\ }\href {https://doi.org/https://doi.org/10.1016/0375-9601(75)90142-5} {\bibfield  {journal} {\bibinfo  {journal} {Phys. Lett. A}\ }\textbf {\bibinfo {volume} {55}},\ \bibinfo {pages} {101} (\bibinfo {year} {1975})}\BibitemShut {NoStop}%
\bibitem [{\citenamefont {Levinsen}\ \emph {et~al.}(2025)\citenamefont {Levinsen}, \citenamefont {Bleu},\ and\ \citenamefont {Parish}}]{Levinsen-2025}%
  \BibitemOpen
  \bibfield  {author} {\bibinfo {author} {\bibfnamefont {J.}~\bibnamefont {Levinsen}}, \bibinfo {author} {\bibfnamefont {O.}~\bibnamefont {Bleu}},\ and\ \bibinfo {author} {\bibfnamefont {M.~M.}\ \bibnamefont {Parish}},\ }\bibfield  {title} {\bibinfo {title} {Medium-enhanced polaron repulsion in a dilute {Bos}e mixture},\ }\href {https://doi.org/10.1103/x2xb-6hzv} {\bibfield  {journal} {\bibinfo  {journal} {Phys. Rev. Lett.}\ }\textbf {\bibinfo {volume} {135}},\ \bibinfo {pages} {113402} (\bibinfo {year} {2025})}\BibitemShut {NoStop}%
\bibitem [{\citenamefont {Levinsen}\ \emph {et~al.}(2026)\citenamefont {Levinsen}, \citenamefont {Marchetti}, \citenamefont {Bleu},\ and\ \citenamefont {Parish}}]{Levinsen-2026}%
  \BibitemOpen
  \bibfield  {author} {\bibinfo {author} {\bibfnamefont {J.}~\bibnamefont {Levinsen}}, \bibinfo {author} {\bibfnamefont {F.~M.}\ \bibnamefont {Marchetti}}, \bibinfo {author} {\bibfnamefont {O.}~\bibnamefont {Bleu}},\ and\ \bibinfo {author} {\bibfnamefont {M.~M.}\ \bibnamefont {Parish}},\ }\bibfield  {title} {\bibinfo {title} {Role of impurity statistics and medium constraints in polaron-polaron interactions},\ }\href {https://doi.org/10.21468/SciPostPhys.21.1.025} {\bibfield  {journal} {\bibinfo  {journal} {SciPost Phys.}\ }\textbf {\bibinfo {volume} {21}},\ \bibinfo {pages} {025} (\bibinfo {year} {2026})}\BibitemShut {NoStop}%
\bibitem [{\citenamefont {Rakshit}\ \emph {et~al.}(2019{\natexlab{a}})\citenamefont {Rakshit}, \citenamefont {Karpiuk}, \citenamefont {Brewczyk},\ and\ \citenamefont {Gajda}}]{Rakshit3D-2019}%
  \BibitemOpen
  \bibfield  {author} {\bibinfo {author} {\bibfnamefont {D.}~\bibnamefont {Rakshit}}, \bibinfo {author} {\bibfnamefont {T.}~\bibnamefont {Karpiuk}}, \bibinfo {author} {\bibfnamefont {M.}~\bibnamefont {Brewczyk}},\ and\ \bibinfo {author} {\bibfnamefont {M.}~\bibnamefont {Gajda}},\ }\bibfield  {title} {\bibinfo {title} {{Quantum Bose-Fermi droplets}},\ }\href {https://doi.org/10.21468/SciPostPhys.6.6.079} {\bibfield  {journal} {\bibinfo  {journal} {SciPost Physics}\ }\textbf {\bibinfo {volume} {6}},\ \bibinfo {pages} {079} (\bibinfo {year} {2019}{\natexlab{a}})}\BibitemShut {NoStop}%
\bibitem [{\citenamefont {Rakshit}\ \emph {et~al.}(2019{\natexlab{b}})\citenamefont {Rakshit}, \citenamefont {Karpiuk}, \citenamefont {Zi{\'n}}, \citenamefont {Brewczyk}, \citenamefont {Lewenstein},\ and\ \citenamefont {Gajda}}]{Rakshit2D-2019}%
  \BibitemOpen
  \bibfield  {author} {\bibinfo {author} {\bibfnamefont {D.}~\bibnamefont {Rakshit}}, \bibinfo {author} {\bibfnamefont {T.}~\bibnamefont {Karpiuk}}, \bibinfo {author} {\bibfnamefont {P.}~\bibnamefont {Zi{\'n}}}, \bibinfo {author} {\bibfnamefont {M.}~\bibnamefont {Brewczyk}}, \bibinfo {author} {\bibfnamefont {M.}~\bibnamefont {Lewenstein}},\ and\ \bibinfo {author} {\bibfnamefont {M.}~\bibnamefont {Gajda}},\ }\bibfield  {title} {\bibinfo {title} {Self-bound {Bose--Fermi} liquids in lower dimensions},\ }\href {https://doi.org/10.1088/1367-2630/ab2ce3} {\bibfield  {journal} {\bibinfo  {journal} {New Jour. Phys.}\ }\textbf {\bibinfo {volume} {21}},\ \bibinfo {pages} {073027} (\bibinfo {year} {2019}{\natexlab{b}})}\BibitemShut {NoStop}%
\bibitem [{\citenamefont {Bajdich}\ \emph {et~al.}(2006)\citenamefont {Bajdich}, \citenamefont {Mitas}, \citenamefont {Drobn\'y}, \citenamefont {Wagner},\ and\ \citenamefont {Schmidt}}]{Bajdich-2006}%
  \BibitemOpen
  \bibfield  {author} {\bibinfo {author} {\bibfnamefont {M.}~\bibnamefont {Bajdich}}, \bibinfo {author} {\bibfnamefont {L.}~\bibnamefont {Mitas}}, \bibinfo {author} {\bibfnamefont {G.}~\bibnamefont {Drobn\'y}}, \bibinfo {author} {\bibfnamefont {L.~K.}\ \bibnamefont {Wagner}},\ and\ \bibinfo {author} {\bibfnamefont {K.~E.}\ \bibnamefont {Schmidt}},\ }\bibfield  {title} {\bibinfo {title} {Pfaffian pairing wave functions in electronic-structure {Quantum Monte Carlo} simulations},\ }\href {https://doi.org/10.1103/PhysRevLett.96.130201} {\bibfield  {journal} {\bibinfo  {journal} {Phys. Rev. Lett.}\ }\textbf {\bibinfo {volume} {96}},\ \bibinfo {pages} {130201} (\bibinfo {year} {2006})}\BibitemShut {NoStop}%
\bibitem [{\citenamefont {Carlson}\ \emph {et~al.}(2003)\citenamefont {Carlson}, \citenamefont {Chang}, \citenamefont {Pandharipande},\ and\ \citenamefont {Schmidt}}]{Carlson-2003}%
  \BibitemOpen
  \bibfield  {author} {\bibinfo {author} {\bibfnamefont {J.}~\bibnamefont {Carlson}}, \bibinfo {author} {\bibfnamefont {S.-Y.}\ \bibnamefont {Chang}}, \bibinfo {author} {\bibfnamefont {V.~R.}\ \bibnamefont {Pandharipande}},\ and\ \bibinfo {author} {\bibfnamefont {K.~E.}\ \bibnamefont {Schmidt}},\ }\bibfield  {title} {\bibinfo {title} {Superfluid {Fermi} gases with large scattering length},\ }\href {https://doi.org/10.1103/PhysRevLett.91.050401} {\bibfield  {journal} {\bibinfo  {journal} {Phys. Rev. Lett.}\ }\textbf {\bibinfo {volume} {91}},\ \bibinfo {pages} {050401} (\bibinfo {year} {2003})}\BibitemShut {NoStop}%
\bibitem [{\citenamefont {Cordioli}\ \emph {et~al.}(2026)\citenamefont {Cordioli}, \citenamefont {Pisani},\ and\ \citenamefont {Pieri}}]{DatasetZenodo}%
  \BibitemOpen
  \bibfield  {author} {\bibinfo {author} {\bibfnamefont {P.}~\bibnamefont {Cordioli}}, \bibinfo {author} {\bibfnamefont {L.}~\bibnamefont {Pisani}},\ and\ \bibinfo {author} {\bibfnamefont {P.}~\bibnamefont {Pieri}},\ }\href {https://doi.org/10.5281/zenodo.18668994} {\bibinfo {title} {{Data for "Stability of Bose-Fermi mixtures in two dimensions: a lowest-order constrained variational approach"}}} (\bibinfo {year} {2026}),\ \bibinfo {note} {{Zenodo Repository}}\BibitemShut {NoStop}%
\bibitem [{\citenamefont {{Wolfram Research, Inc.}}(2024)}]{Mathematica}%
  \BibitemOpen
  \bibfield  {author} {\bibinfo {author} {\bibnamefont {{Wolfram Research, Inc.}}},\ }\href@noop {} {\emph {\bibinfo {title} {{Mathematica, Version 14.2}}}},\ \bibinfo {organization} {Wolfram Research, Inc.},\ \bibinfo {address} {Champaign, Illinois, USA} (\bibinfo {year} {2024}),\ \bibinfo {note} {\url{https://www.wolfram.com/mathematica/}}\BibitemShut {NoStop}%
\bibitem [{\citenamefont {Erd{\'e}lyi}\ \emph {et~al.}(1953)\citenamefont {Erd{\'e}lyi}, \citenamefont {Magnus}, \citenamefont {Oberhettinger},\ and\ \citenamefont {Tricomi}}]{Bateman}%
  \BibitemOpen
  \bibinfo {editor} {\bibfnamefont {A.}~\bibnamefont {Erd{\'e}lyi}}, \bibinfo {editor} {\bibfnamefont {W.}~\bibnamefont {Magnus}}, \bibinfo {editor} {\bibfnamefont {F.}~\bibnamefont {Oberhettinger}},\ and\ \bibinfo {editor} {\bibfnamefont {F.~G.}\ \bibnamefont {Tricomi}},\ eds.,\ \href@noop {} {\emph {\bibinfo {title} {Higher Transcendental Functions}}},\ Vol.\ \bibinfo {volume} {I--III}\ (\bibinfo  {publisher} {McGraw-Hill Book Company},\ \bibinfo {address} {New York},\ \bibinfo {year} {1953})\BibitemShut {NoStop}%
\end{thebibliography}

%

\end{document}